\documentclass[11pt]{article}
\usepackage{amssymb}
\usepackage{graphics}
\usepackage{epsfig}
\usepackage{a4wide}
\begin{document}
\title {Full one-loop QCD and electroweak corrections to sfermion pair production in $\gamma
\gamma$ collisions \footnote{Supported by National Natural Science
Foundation of China.}} \vspace{3mm}
\author{{Xing Li-Rong$^{b}$, Ma Wen-Gan$^{a,b}$, Zhang Ren-You$^{b}$, Jiang Yi$^{b}$, Han Liang$^{b}$, Li Gang$^{b}$}\\
{\small $^{a}$ CCAST (World Laboratory), P.O.Box 8730, Beijing
100080, P.R.China}\\
{\small $^{b}$ Department of Modern Physics, University of Science and Technology}\\
{\small of China (USTC), Hefei, Anhui 230027, P.R.China}}
\date{}
\maketitle \vskip 12mm

\begin{abstract}
We have calculated the full one-loop electroweak (EW) and QCD
corrections to the third generation scalar-fermion pair production
processes $e^+e^- \to \gamma \gamma \rightarrow
\tilde{f_i}\bar{\tilde{f_i}} (f=t,b,\tau)$ at an electron-positron
linear collider(LC) in the minimal supersymmetric standard model
(MSSM). We analyze the dependence of the radiative corrections on
the parameters such as the colliding energy $\sqrt{\hat s}$ and
the SUSY fundamental parameters $A_f$, $\tan \beta$, $\mu$,
$M_{SUSY}$ and so forth. The numerical results show that the EW
corrections to the squark-, stau-pair production processes and QCD
corrections to the squark-pair production processes give
substantial contributions in some parameter space. The EW relative
corrections to squark-pair production processes can be comparable
with QCD corrections at high energies. Therefore, these EW and QCD
corrections cannot be neglected in precise measurement of sfermion
pair productions via $\gamma\gamma$ collision at future linear
colliders.
\end{abstract}


\vfill \eject \baselineskip=0.5cm      
\renewcommand{\theequation}{\arabic{section}.\arabic{equation}}
\newcommand{\nb}{\nonumber}
\makeatletter      
\@addtoreset{equation}{section}
\makeatother       

\section{Introduction}
\par
The standard model (SM) has been successful in describing the
strong, weak and electromagnetic interaction phenomena at the
energy scale up to $10^2$ GeV. At the higher energy scale, it is
likely that the minimal supersymmetric standard model (MSSM) is
the most attractive candidate among various extensions of the SM.
In the MSSM, the existence of scalar partners of all fermions in
the SM, namely, two chiral scalar fermions $\tilde f_L$ and
$\tilde f_R$ are required. At future colliders running in TeV
energy region, the supersymmetric scalar particle
$\tilde{f}\bar{\tilde{f}}$ pair production processes are very
promising channels in probing directly the existence of these
scalar fermions, since their production cross sections can be
comparatively large, if the scalar fermions are not too heavy.
\par
The two chiral SUSY states $\tilde f_L$ and $\tilde f_R$ of each
fermion $f$ turn to their mass eigenstates by mixing with each
other. The mixing size is proportional to the mass of the
corresponding SM fermion \cite{ellis}. Generally, people believe
that the sfermions of the third generation are more important in
direct SUSY discovery than those of the former two generations,
because the sfermions $\tilde{f}_{L}$ and $\tilde{f}_{R}$ of the
third generation mix strongly to form the two mass eigenstates
$\tilde{f}_{1}$ and $\tilde{f}_{2}$. We assume that the mass
eigenstates $\tilde{f}_1(f=t,b,\tau)$ have lower masses than
$\tilde{f}_2$. Therefore, $\tilde{f}_1$ is very probably to be
discovered in a relatively lower colliding energy range. Another
significance of the sfermion pair production is that it gives
access to one of the SUSY fundamental parameters $A_f$, the
trilinear coupling parameter.
\par
The future higher energy $e^+e^-$ linear colliders (LC) is
designed to look for the evidences of Higgs boson and other new
particles beyond the SM. There have been already some detailed
designs of linear colliders, such as NLC\cite{NLC}, JLC\cite{JLC},
TESLA\cite{TESLA} and CLIC\cite{CLIC}. Because of the cleaner
background of $e^+e^-$ collision than $pp (\bar{p})$ collision, LC
can produce more distinctive experimental signature of new
physics. The slepton pair production at LC are intensively
discussed in Refs.\cite{zerwas, huitu, howard, freitas}. The squark pair
produced by $e^+e^-$ annihilation has been studied thoroughly,
both at tree level and at next-to-leading order\cite{hikasa}
\cite{been}. In Ref.\cite{han} the QCD correction to stop pair
production via $\gamma \gamma$ fusion at $e^+e^-$ linear collider
is investigated. The scalar fermion pair production via $e^+e^-$
collisions $e^+e^- \to \tilde{f}_i \bar{\tilde{f}_j}
(f=\tau,t,b,~i,j = 1,2)$ at one-loop level, has been studied in
detail in \cite{eberl, hollik}. They have considered the complete
SUSY-QCD and electroweak (EW) one-loop corrections. Their results
show that at the energy of $\sqrt s = 500 \sim 1000$ GeV, the QCD
corrections are dominated while the EW corrections are of the same
magnitude as the SUSY-QCD corrections at the higher energy scale.
\par
However, the future $e^{+}e^{-}$ linear colliders are designed to
give other facilities operating in $e^- e^-$, $\gamma \gamma$ and
other collision modes at the energy of $500 \sim 5000$ GeV with a
luminosity of the order $10^{33} cm^{-2} s^{-1}$ \cite{a1}. The
future LC's can turn the high energy electron-positron beams into
the Compton backscattering energetic photon beams with high
efficiency in the scattering of intense laser photons. With the
help of the new experimental techniques, it is feasible to yield a
scaler fermion pair production directly via the high energy photon
collision. Different options of the colliding mode are
complementary to each other and will add essential new information
to that obtained from the CERN Large Hadron Collider (LHC).
Therefore, the sfermion pair production via $\gamma \gamma$ fusion
provides another important mechanism in producing sfermion pair.
Moreover, their production rates should be larger than those by
the $e^+e^-$ annihilation because of the existing of the s-channel
suppression in the latter. At the tree-level, the two final
sfermions produced in $\gamma \gamma$ collisions should be the
same sfermion mass eigenstate, since only the electromagnetic
interaction is involved. Although there are some studies on the
$e^+e^- \to \gamma\gamma \to \tilde{f}_i \bar{\tilde{f}}_i
(f=t,b,\tau,~i=1,2)$ at tree level\cite{a2}, the complete one-loop
level effects of the EW and QCD in the sfermion pair production
via $\gamma \gamma$ collisions are still absent at present. In a
word, the process of scalar fermion pair production via
photon-photon collisions $e^+e^- \to \gamma \gamma \to \tilde{f}_i
\bar{\tilde{f}}_i ~(f=t,b,\tau,i=1,2)$ will be worthwhile to
investigate precisely and can be accessible in accurate
experiments.
\par
In this paper, we will calculate the full one-loop EW and QCD
corrections to this process. The paper is organized as follows: In
Section 2, we give the definitions of the notations and the
analytical calculations of the cross sections involving the ${\cal
O}(\alpha_{ew})$ EW and ${\cal O}(\alpha_{s})$ QCD corrections.
The numerical results and discussions are presented in Section 3.
Finally, we give a short summery in Section 4.

\vskip 5mm
\section{Analytical calculations}
\par
In this section, we present the analytical calculations for the
subprocesses $\gamma \gamma \to \tilde{f}_i \bar{\tilde{f_i}}
(f=\tau, t, b,i=1,2)$ and their parent processes $e^+e^- \to
\gamma \gamma \to \tilde{f}_i \bar{\tilde{f_i}}$ at the lowest
order and the one-loop level in the MSSM. We adopt the 't
Hooft-Feynman gauge and the definitions of one-loop integral
functions in Ref.\cite{s14}. As we know that for the subprocesses
$\gamma \gamma \to \tilde{q}_i \bar{\tilde{q_i}} (q=t, b,i=1,2)$
there exist both QCD and EW quantum corrections, while for $\gamma
\gamma \to \tilde{\tau}_i \bar{\tilde{\tau_i}}(i=1,2)$
subprocesses they have only EW quantum contributions.

\vskip 5mm
\subsection{The sfermion sector and the lowest order cross section of the subprocess
$\gamma \gamma \to \tilde{f}_i
\bar{\tilde{f_i}}(f=\tau,t,b,~i=1,2)$} In the MSSM, the Lagrangian
mass term of the scalar fermion $\tilde f$ can be written as
\begin{eqnarray}
 -{\cal L}_{\tilde{f}}^{mass} &=&
 \left(
    \begin{array}{cc}
    \tilde{f}^{\ast}_{L} & \tilde{f}^{\ast}_{R}
     \end{array}
    \right)
    {\cal M}^2_{\tilde{f}}
      \left(
        \begin{array}{c}
       \tilde{f}_{L} \\
       \tilde{f}_{R}
     \end{array}
        \right), ~ ~ ~ ~ ~ ~ ~(f= \tau, t, b),
\end{eqnarray}
where ${\cal M}^2_{\tilde{f}}$ is the mass matrix of $\tilde f$, expressed as
\begin{eqnarray}
\nonumber
{\cal M}^2_{\tilde{f}} &=&
        \left( \begin{array}{cc}
    m^2_{\tilde f_L} & m_f a_f \\
    a^{\dag}_f m_f & m^2_{{\tilde f}_R}
    \end{array} \right) \\
\end{eqnarray}
and
\begin{eqnarray}
\nonumber
  m_{\tilde f_L}^{2} &=& M_{\{\tilde Q, \tilde L \}}^2
       + (I^{3L}_f - Q_f \sin^2 \theta_W)\cos2\beta m_{Z}^2
       + m_{f}^2, \\ \nonumber
m_{\tilde f_R}^{2} &=& M_{\{\tilde U, \tilde D, \tilde E \}}^2
       + Q_{f} \sin^2 \theta_W \cos2 \beta m_{Z}^{2}
       + m_f^2, \\
  a_f &=& A_f - \mu (\tan \beta)^{-2 I^{3L}_f} .
\end{eqnarray}
where $M_{\tilde Q}, M_{\tilde L}, M_{\tilde U}, M_{\tilde D}$ and
$M_{\tilde E}$ are the soft SUSY breaking masses, $I_f^{3L}$ is
the third component of the weak isospin of the fermion, $Q_f$ the
electric charge of the scalar fermion, $\theta_W$ the Weinberg
angle, and $A_f$ is the trilinear scalar coupling parameters of
Higgs boson with scalar quarks, $\mu$ the higgsino mass parameter.
\par
The mass matrix ${\cal M}_{\tilde {f}}$ can be diagonalized by introducing an unitary
matrix ${\cal R}^{\tilde f}$. The mass eigenstates $\tilde f_1$, $\tilde f_2$ are
defined as
\begin{eqnarray}
       \left(
       \begin{array}{c}
       \tilde{f}_{1} \\
       \tilde{f}_{2}
       \end{array}
        \right) =
    {\cal R}^{\tilde f}
    \left(
       \begin{array}{c}
       \tilde{f}_{L} \\
       \tilde{f}_{R}
       \end{array}
        \right)=
\left(
    \begin{array}{cc}
    \cos \theta_{\tilde{f}} &  \sin \theta_{\tilde{f}} \\
    -\sin \theta_{\tilde{f}} &  \cos \theta_{\tilde{f}}
     \end{array}
    \right)
    \left(
       \begin{array}{c}
       \tilde{f}_{L} \\
       \tilde{f}_{R}
       \end{array}
        \right)
\end{eqnarray}
Then the mass term of sfermion $\tilde f$ can be expressed
\begin{eqnarray}
 -{\cal L}_{\tilde{f}}^{mass} &=&
 \left(
    \begin{array}{cc}
    \tilde{f}^{\ast}_{1} & \tilde{f}^{\ast}_{2}
     \end{array}
    \right)
    {\cal M}_D^{\tilde{f}~2 }
      \left(
        \begin{array}{c}
       \tilde{f}_{1} \\
       \tilde{f}_{2}
     \end{array}
        \right) ~ ~ ~ ~ ~ ~ ~(f= \tau, t, b),
\end{eqnarray}
where
\begin{eqnarray}
{\cal M}_D^{\tilde{f}~2 } =
{\cal R}^{\tilde f} {\cal M}^2_{\tilde {f}} {\cal R}^{\tilde f~ \dag}=
   \left(
    \begin{array}{cc}
    m^2_{\tilde{f}_{1}} &  0 \\
    0 &  m^2_{\tilde{f}_{2}}
     \end{array}
    \right).
\end{eqnarray}
The masses of $\tilde f_1, \tilde f_2$ and the angle $\theta_{\tilde f}$
are fixed by the following equation
\begin{equation}
(m^2_{\tilde{f}_{1}},m^2_{\tilde{f}_{2}})=\frac{1}{2}\{m^2_{\tilde{f}_{L}}+m^2_{\tilde{f}_{R}}
\mp [(m^2_{\tilde{f}_{L}}-m^2_{\tilde{f}_{R}})^2 +
4|a_f|^2m_f^2]^{1/2}\},
\end{equation}
\begin{eqnarray}
\tan{2\theta_{\tilde f}}
=\frac{2|a_f|m_f}{m^2_{\tilde{f}_{L}}-m^2_{\tilde{f}_{R}}} ~ ~ ~ (0 < \theta_f < \pi )
\end{eqnarray}
\par
\begin{figure}[htb]
\centering
\includegraphics{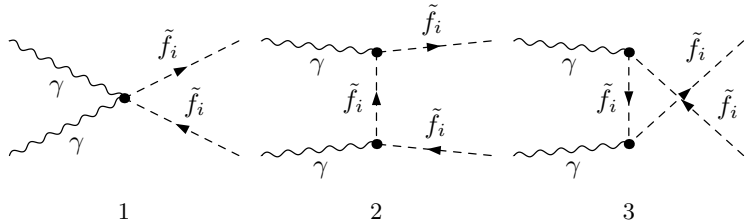}
\caption{The lowest order diagrams for the $\gamma\gamma \to
\tilde f_i \bar{\tilde f_i} ~(f= \tau, t, b)$ subprocess.}
\label{fig:feyn_born}
\end{figure}

We denote the subprocess $\gamma \gamma \to \tilde{f}_i \bar{\tilde{f_i}}$ as
\begin{equation}
\gamma(p_1)+\gamma(p_2) \to
\tilde{f}_i(p_3)+\bar{\tilde{f_i}}(p_4) ~ ~(f= \tau, t, b,~i=1,2).
\end{equation}
where $p_1$ and $p_2$ represent the four-momenta of the two
incoming photons, $p_3$ and $p_4$ denote the four-momenta of the
outgoing scalar fermion and its anti-particle, respectively. The
momenta $p_i (i=1,\cdots,4)$ obey the on-shell equations, namely,
$p^2_1 = p^2_2 = 0$ and $p^2_3 = p^2_4 = m_{\tilde{f}_i}^2$. There
are three Feynman diagrams for this subprocess at the tree level,
which are shown in Fig.\ref{fig:feyn_born}.
The corresponding tree level amplitudes of this subprocess $\gamma
\gamma \to \tilde{f}_i \bar{\tilde{f_i}}$ are represented as
\begin{equation}
 M_0 = M_0^{\hat{t}} + M_0^{\hat{u}} + M_0^{\hat{q}}
\end{equation}
where $M_0^{\hat{t}}$, $M_0^{\hat{u}}$ and $M_0^{\hat{q}}$
represent the amplitudes of the t-channel, u-channel and quartic
coupling diagrams respectively. The explicit expressions can be
written as
\begin{equation}
M_0^{\hat{t}} = \frac{4 i e^2 Q_{f}^2}{\hat{t} -
m_{\tilde{f}_i}^2}
                \epsilon_{\mu}(p_1)
                \epsilon_{\nu}(p_2)~p_3^{\mu}~p_4^{\nu},
\end{equation}
\begin{equation}
M_0^{\hat{u}} = M_0^{\hat{t}} ~ (p_1 \leftrightarrow p_2),~~~
M_0^{\hat{q}} = 2 i e^2 Q_{f}^2  g^{\mu\nu}
\epsilon_{\mu}(p_1) \epsilon_{\nu}(p_2).
\end{equation}
The Mandelstam variables $\hat{t}$, $\hat{u}$ and $\hat{s}$ are
defined as $\hat{t}=(p_1 - p_3)^2,~\hat{u}=(p_1 - p_4)^2,~
\hat{s}=(p_1 + p_2)^2 = (p_3 + p_4)^2$. $m_{\tilde{f}_i}~ (i=1,2)$
denotes the masses of the mass eigenstates of scalar fermions.
\par
The cross section at tree-level can be expressed as
\begin{equation}
\label{sigma0} \hat{\sigma}_0(\hat{s}) = \frac{1}{16 \pi
\hat{s}^2} \int_{t_{min}}^{t_{max}} dt~ \bar{\Sigma} |M_0|^2,
\end{equation}
with
\begin{equation}
\label{Mandlstein}
 t_{max,min}=\frac{1}{2}\left[
(2m^{2}_{\tilde{f}_i} - \hat{s})\pm \sqrt{(2m^{2}_{\tilde{f}_i} -
\hat{s})^2 -4m^{4}_{\tilde{f}_i}} \right].
\end{equation}
The summation is taken over the spins and colors of initial and
final states, and the bar over the summation recalls averaging
over the initial spins. After integration of Eq.(\ref{sigma0}) we
get the analytical expressions of the cross section of $\gamma
\gamma \rightarrow \tilde{f}_i \bar{\tilde{f}}_i$ subprocess at
the tree level as
\begin{equation}
\label{SigmaT}
\hat{\sigma}_{0}(\hat{s}) =
   \frac{2 \pi \alpha^2}{\hat{s}}Q_{f}^4~N_C^f~\beta~
   \{ 1 + \frac{16 \mu^4}{1 - \beta^2} +
      \frac{4 \mu^2 (1 - 2 \mu^2)}{\beta} \log{v} \}.
\end{equation}
Here $\mu^2=m_{\tilde{f}_i}^2/\hat{s}$ and $\beta=\sqrt{1 - 4
m_{\tilde{f}_i}^2/\hat{s}}$ is the velocity of the
produced scalar fermion. The kinematical variable $v$ is
defined as $v = (1 - \beta)/(1 + \beta)$. For squarks, we have $N_C^f
= 3$, while for sleptons $N_C^f = 1$.

\vskip 5mm
\subsection{${\cal O}(\alpha_{ew})$ EW corrections to subprocess
$\gamma \gamma \to \tilde{f}_i \bar{\tilde{f_i}} ~ ~ (f= \tau, t,
b,~i=1,2)$ }
\par
In the calculation of the one-loop EW corrections, we adopt the
dimensional reduction ($\overline {DR}$) regularization scheme,
which is supersymmetric invariant at least at one-loop level. We
assume that there is no quark mixing, i.e., the CKM-matrix is
identity matrix, and use the complete on-mass-shell (COMS)
renormalization scheme \cite{COMS}. We use ${\it FeynArts}~3$
\cite{FA3} package to generate the ${\cal O}(\alpha_{ew})$ Feynman
diagrams and amplitudes of the ${\cal O}(\alpha_{ew})$ EW virtual
contributions to $\gamma \gamma \to \tilde{f}_i \bar{\tilde{f_i}}
~ ~ (f= \tau, t, b)$ subprocess. There are total 469 EW one-loop
Feynman diagrams, and we classified them into four groups:
self-energy, vertex, box diagrams and counter-term diagrams. The
relevant renormalization constants are defined as
$$
 e_0=(1+\delta Z_e)e,~~~ m_{W,0}^2=m_W^2+\delta m_W^2,
$$
$$
 m_{Z,0}^2=m_Z^2+\delta m_Z^2, ~~~ A_0=\frac{1}{2} \delta
Z_{AZ}Z+(1+\frac{1}{2}\delta Z_{AA})A
$$
\begin{equation}
\label{counterterm}
 m^2_{\tilde{f_i}, 0}= m^2_{\tilde{f_i}} +
\delta m^2_{\tilde{f_i}}, ~ ~ ~ \tilde{f}_{1,0} =
Z_{11}^{\tilde{f}^{1/2}}\tilde{f}_1 +
Z_{12}^{\tilde{f}^{1/2}}\tilde{f}_2,~~~ \tilde{f}_{2,0} =
Z_{22}^{\tilde{f}^{1/2}}\tilde{f}_2 +
Z_{21}^{\tilde{f}^{1/2}}\tilde{f}_1.
\end{equation}
where
$$
Z_{ij}^{\tilde{f}^{1/2}}=\delta_{ij} + \frac{1}{2}\delta
Z_{ij}^{\tilde{f}}.
$$
With the on-mass-shell conditions, we can obtain the
renormalized constants expressed as
$$
\delta m_W^2 = \tilde{Re} \Sigma_T^W(m_W^2),~~~
\delta m_Z^2 = \tilde{Re} \Sigma_T^{ZZ}(m_Z^2),
$$
$$
\delta Z_{AA}= -\tilde{Re}\frac{\partial \Sigma_T^{AA}(p^2)}{\partial p^2}|_{p^2=0},~~~
\delta Z_{ZA}= 2\frac{\tilde{Re} \Sigma_T^{ZA}(0)}{m_Z^2},
$$
\begin{equation}
\label{CT electric charge} \delta Z_e = -\frac{1}{2} \delta
Z_{AA}+ \frac{s_W}{c_W}\frac{1}{2} \delta Z_{ZA}
           = \frac{1}{2} \tilde{Re}\frac{\partial \Sigma_T^{AA}(p^2)}{\partial p^2}|_{p^2=0}
         + \frac{\sin \theta_W}{\cos \theta_W} \frac{\tilde{Re}\Sigma_T^{ZA}(0)}{m_Z^2},
\end{equation}
$$
\delta m^2_{\tilde{f}_i}= \tilde{Re}\Sigma_{ii}^{\tilde{f}}(m^2_{\tilde{f}_i}),~ ~ ~
\delta Z^{\tilde f}_{ii}= -\tilde{Re}
\frac{\partial \Sigma^{\tilde{f}}_{ii}(p^2)}{\partial p^2}|_{p^2=m^2_{\tilde{f}_i}},
$$
\begin{equation}
\label{CT expressions}
\delta Z^{\tilde f}_{ij}= -\tilde{Re}
\frac{2
\Sigma^{\tilde{f}}_{ij}(m^2_{\tilde{f}_j})}{m^2_{\tilde{f}_j} -
m^2_{\tilde{f}_i}} ~ ~ ~ ~ (i,j=1,2 ~ ~ ~i\neq j) .
\end{equation}
$\tilde{Re}$ means taking the real part of the loop integrals
appearing in the self-energy. The
$\Sigma^{\tilde{f}}_{ij},(i,j=1,2)$ appeared in Eqs.(\ref{CT
expressions}) denote the unrenormalized sfermion self-energy
involving only the EW interactions. The ${\cal O}(\alpha_{ew})$
one-loop virtual corrections to $\gamma \gamma \to \tilde{f}_i
\bar{\tilde{f_i}}$ is represented as
\begin{equation}
\Delta \hat{\sigma}_{vir}^{EW} = \hat{\sigma}_0
\hat{\delta}^{EW}_{vir}= \frac{1}{16 \pi \hat{s}^2}
\int_{\hat{t}_{min}}^{\hat{t}_{max}} d\hat{t}~ 2 Re
\overline{\sum} (M^{EW}_{vir} M_0^{\dag}) .
\end{equation}
The expressions of $\hat{t}_{max,min}$ have been presented in
Eq.(\ref{Mandlstein}) and the summation with bar over head
represents same operation as that appeared in Eq.(\ref{sigma0}).
$M^{EW}_{vir}$ is the renormalized amplitude of the EW one-loop
Feynman diagrams, which include self-energy, vertex, box and
counter-term diagrams.
\par
\begin{figure}[htb]
\centering
\includegraphics{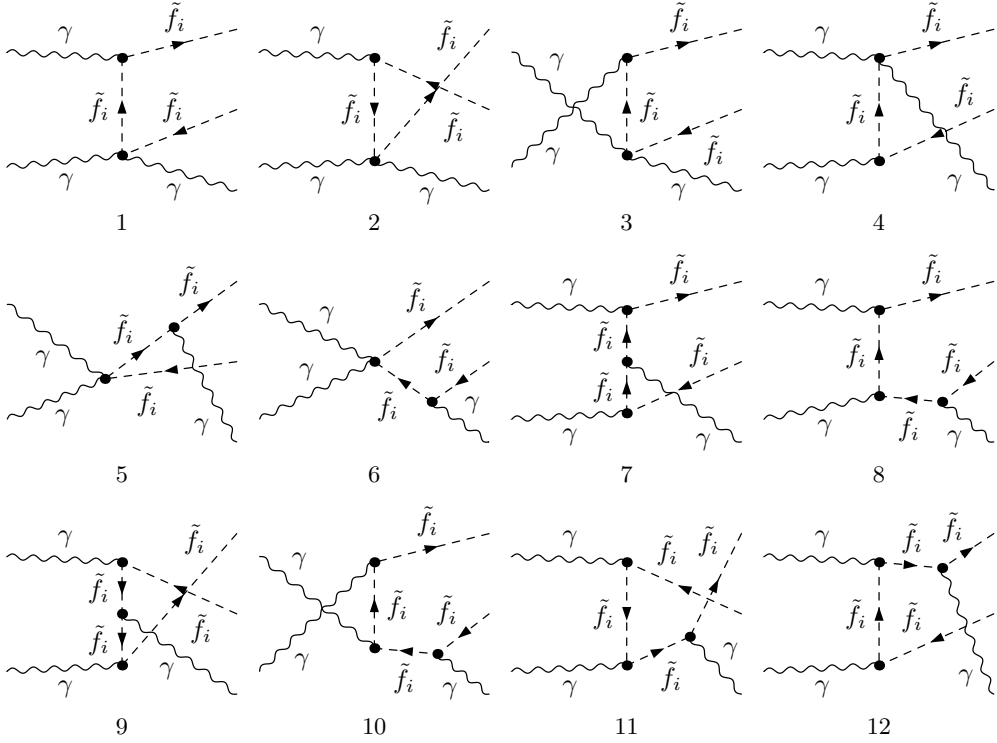}
\caption{The real photon emission diagrams for the process
$\gamma \gamma \to \tilde{f}_i \bar{\tilde{f_i}} \gamma ~(f= \tau, t, b)$} \label{fig:feyn_realphoton}
\end{figure}

The ${\cal O}(\alpha_{ew})$ virtual corrections contain both
ultraviolet (UV) and infrared (IR) divergences. After
renormalization procedure, the UV divergence should vanish. We
have checked the cancellation of the UV divergence both
analytically and numerically, and confirmed that we got a UV
finite amplitude at the ${\cal O}(\alpha_{ew})$ order. The IR
singularity in the $M^{EW}_{vir}$ is originated from virtual
photonic loop correction. It can be cancelled by the contribution
of the real photon emission process. We denote the real photon
emission as
\begin{equation}
\label{process2}
 \gamma (p_1) + \gamma(p_2) \to \tilde{f}_i(p_3) +
\bar{\tilde{f_i}}(p_4) + \gamma (k) ~~~(f = \tau, t ,b,~ i=1,2),
\end{equation}
where $k = (k^0, \vec{k})$ is the four momentum of the radiated
photon, and $p_1$, $p_2$, $p_3$, $p_4$ are the four momenta of two
initial photons and final state particles
$\tilde{f}_i\bar{\tilde{f_i}}$, respectively. The real photon
emission Feynman diagrams for the process $\gamma \gamma \to
\tilde{f}_i \bar{\tilde{f_i}} \gamma$ are displayed in
Fig.\ref{fig:feyn_realphoton}.
In our paper,
we adopt the general phase-space-slicing method \cite{PSS} to
separate the soft photon emission singularity from the real photon
emission process. By using this method, the bremsstrahlung phase
space is divided into singular and non-singular regions. Then the
correction of the real photon emission is broken down into
corresponding soft and hard terms
\begin{equation}
\Delta \hat{\sigma}_{real}^{EW} =\Delta
\hat{\sigma}_{soft}^{EW}+\Delta \hat{\sigma}_{hard}^{EW}
=\hat{\sigma}_0(\hat{\delta}_{soft}^{EW}+\hat{\delta}_{hard}^{EW}).
\end{equation}
In the c.m.s. frame, the radiated photon energy $k^0 =
\sqrt{|\vec{k}|^2 + m_{\gamma}^2}$ is called `soft' if $k^0 \leq
\Delta E_{\gamma}$ or `hard' if $k^0 > \Delta E_{\gamma}$. Here,
$m_{\gamma}$ is a small photon mass, which is used to regulate the
infrared divergences existing in the soft term. Although both
$\Delta \hat{\sigma}_{soft}^{EW}$ and $\Delta
\hat{\sigma}_{hard}^{EW}$ depend on the soft photon cutoff $\Delta
E_{\gamma}/E_b$, where $E_b = \frac{\sqrt{\hat s}}{2}$ is the
electron beam energy in the c.m.s. frame, the real correction
$\Delta \hat{\sigma}_{real}^{EW}$ is cutoff independent. In the
calculation of soft term, we use the soft photon approximation.
Since the diagrams in Fig.\ref{fig:feyn_realphoton} with real
photon radiation from the internal sfermion line or
photon-sfermions vertex do not lead to IR-singularity, we can
neglect them in the calculation of soft photon emission
subprocesses (\ref{process2}) by using the soft photon
approximation method.  In this approach the contribution of the
soft photon emission subprocess is expressed as\cite{COMS,Velt}
\begin{equation}
{\rm d} \Delta \hat{\sigma}^{EW}_{{\rm soft}} = -{\rm d}
\hat{\sigma}_0 \frac{\alpha_{{\rm ew}}Q_{\tilde{f}}^2}{2\pi^2}
 \int_{|\vec{k}| \leq \Delta E_{\gamma}}\frac{{\rm d}^3 k}{2 k^0} \left[
 \frac{p_3}{p_3 \cdot k}-\frac{p_4}{p_4 \cdot k} \right]^2
\end{equation}
where the soft photon cutoff $\Delta E_{\gamma}$ satisfies $k^0
\leq \Delta E_{\gamma} \ll \sqrt{\hat{s}}$. The integral over the
soft photon phase space has been implemented in Ref.\cite{COMS},
then one can obtain the analytical result of the soft real photon
emission correction to $\gamma\gamma \to \tilde{f}_i \tilde{f}_i$.
\par
As mentioned above, the IR divergence of the virtual photonic
corrections can be exactly cancelled by that of soft real
correction. Therefore, $\Delta \hat{\sigma}^{EW}_{{\rm vir} + {\rm
soft}}$, the sum of the ${\cal O}(\alpha_{ew})$ virtual and soft
contributions, is independent of the IR regulator $m_{\gamma}$. In
the following numerical calculations, we have checked the
cancellation of IR divergencies and verified that the total
contributions of soft photon emission and the virtual corrections
are numerically independent of $m_{\gamma}$. In addition, we
present the numerical verification of that the total one-loop
level EW correction to the cross section of $\gamma \gamma \to
\tilde{f}_i \bar{\tilde{f}_i}$, defined as $\Delta
\hat{\sigma}^{EW}=\Delta \hat{\sigma}_{vir}^{EW} +\Delta
\hat{\sigma}_{real}^{EW}$, is independent of the cutoff $\Delta
E_{\gamma}$.
\par
Finally, we get an UV and IR finite ${\cal O}(\alpha_{ew})$
EW correction $\Delta \hat{\sigma}^{EW}$:
$$
\Delta \hat{\sigma}^{EW}
= \Delta \hat{\sigma}_{vir}^{EW} +\Delta \hat{\sigma}_{real}^{EW}
=\hat{\sigma}_0 \hat{\delta}^{EW}
$$
where
$\hat{\delta}^{EW}=\hat{\delta}^{EW}_{vir}+\hat{\delta}^{EW}_{soft}+\hat{\delta}^{EW}_{hard}$
is the ${\cal O}(\alpha_{ew})$ EW relative correction.

\vskip 5mm
\subsection{${\cal O}(\alpha_{s})$ QCD correction to subprocess
$\gamma \gamma \to \tilde{q}_i \bar{\tilde{q_i}}~(q= t, b,~i=1,2)$
}
\par
In this subsection, we calculate the supersymmetric ${\cal
O}(\alpha_{s})$ QCD corrections. The relevant Feynman diagrams and
the corresponding amplitudes of the subprocess $\gamma \gamma \to
\tilde{q}_i \bar{\tilde{q_i}},~(q= t, b,~i=1,2)$ both at
tree-level and at one-loop level, are again generated by the
package FeynArts 3 \cite{FA3}. The Feynman diagrams of the
one-loop ${\cal O}(\alpha_{s})$ QCD corrections also can be
classified into self-energy, vertex, box and counter-term
diagrams. The relevant renormalized constants used in the
calculation are similar with those in the calculation of the
one-loop ${\cal O}(\alpha_{ew})$ EW correction, which are defined
and expressed as in Eqs.(\ref{counterterm}) and Eqs.(\ref{CT
expressions}) respectively, except all the EW one-loop
self-energies are replaced by the corresponding QCD ones. The SUSY
QCD unrenormalized self-energy of the scalar quark
$\tilde{q}_i~(q=t,b,i=1,2)$ can be written as a summation of three
parts as the follows:
\begin{equation}
\Sigma_{ii}^{\tilde{q}} (p^2) = \Sigma_{ii}^{\tilde{q}(g)}(p^2) +
\Sigma_{ii}^{\tilde{q}(\tilde{g})}(p^2)
          + \Sigma_{ii}^{\tilde{q}(\tilde{q})}(p^2),
\end{equation}
where $\Sigma_{ii}^{\tilde{q}(g)}$,
$\Sigma_{ii}^{\tilde{q}(\tilde{g})}$ and
$\Sigma_{ii}^{\tilde{q}(\tilde{q})}$ denote the scalar quark
self-energy parts corresponding to the diagrams with virtual
gluon, virtual gluino exchanges and the squark quartic
interactions respectively. The squark quartic interactions are
introduced by the superpotential of the SUSY model. The three
parts from the squark $\tilde{q}_i$ self-energy can be written
explicitly as
\begin{equation}
\Sigma_{ii}^{\tilde{q}(g)}(p^2) = - \frac{g_s^2 C_F}{16 \pi^2}
             \left \{
                      A_0[m_g] +
                      4 p^2 (B_0 + B_1)[p^2, m_g, m_{\tilde{q}_i}] +
                      m_g^2 B_0[p^2, m_g, m_{\tilde{q}_i}]-A_0[m_{\tilde{q}_i}] \right \},
\end{equation}
\begin{equation}
\Sigma_{ii}^{\tilde{q}(\tilde{g})}(p^2) = - \frac{g_s^2 C_F}{16
\pi^2} D
  \{ A_0[m_q] +
     (m_{\tilde{g}}^2 + m_q m_{\tilde{g}} \sin{2 \theta_{\tilde{q}})}
     B_0[p^2, m_{\tilde{g}}, m_q] + p^2 B_1[p^2, m_{\tilde{g}}, m_q]
  \},
\end{equation}
\begin{equation}
\Sigma_{ii}^{\tilde{q}(\tilde{q})}(p^2) =  \frac{g_s^2}{12 \pi^2}
        \{ A_0[m_{\tilde{q}_1}^2]~cos^2 2 \theta_{\tilde{q}} +
           A_0[m_{\tilde{q}_2}^2]~sin^2 2 \theta_{\tilde{q}} \},
\end{equation}
where $m_g$ denotes the small gluon mass, $D=4-2\epsilon$ is the
space-time dimension and the group Casimir operator has
$C_{F}=\frac{4}{3}$.
\par
The one-loop ${\cal O}(\alpha_{s})$ QCD virtual correction of
subprocess $\gamma \gamma \to \tilde{q}_i \bar{\tilde{q_i}}~(q= t,
b,~i=1,2)$ can be expressed as
\begin{equation}
\Delta \hat{\sigma}_{vir}^{QCD} = \hat{\sigma}_0
\hat{\delta}^{QCD}_{vir}= \frac{1}{16 \pi \hat{s}^2}
\int_{\hat{t}_{min}}^{\hat{t}_{max}} d\hat{t}~ 2 Re
\overline{\sum} (M^{QCD}_{vir} M_0^{\dag})
\end{equation}
where $M^{QCD}_{vir}$ is the
renormalized amplitude of the one-loop ${\cal O}(\alpha_{s})$ QCD
Feynman diagrams, which include self-energy, vertex, box and
counter-term diagrams.
\par
\begin{figure}[htb]
\centering
\includegraphics{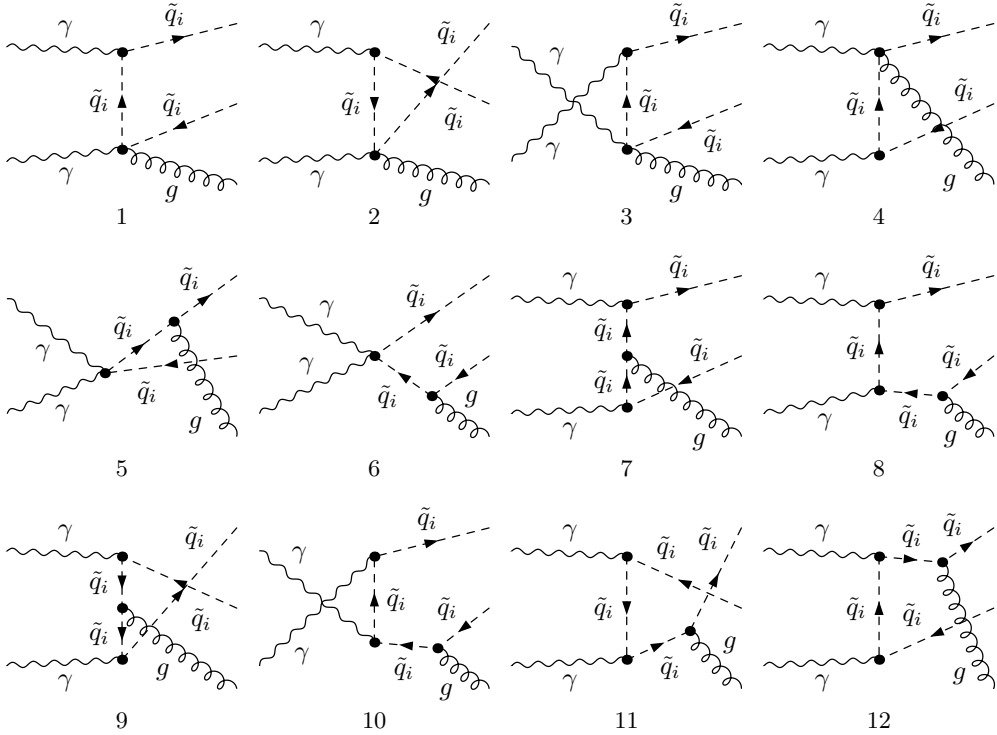}
\caption{The real gluon emission diagrams for the process $\gamma
\gamma \to \tilde{q}_i \bar{\tilde{q_i}} g ~ (q=t,b,~i=1,2)$}
\label{fig:feyn_realgluon}
\end{figure}

The virtual QCD corrections contain both ultraviolet (UV) and
infrared (IR) divergences in general. To regularize the UV
divergences in loop integrals, we adopt the dimensional
regularization in which the dimensions of spinor and space-time
manifolds are extended to $D = 4 - 2 \epsilon$. We have verified
the cancellation of the UV divergence both analytically and numerically. Then
we get a UV finite amplitude including ${\cal O}(\alpha_{s})$
virtual radiative corrections. The IR divergence of the QCD
virtual corrections of the subprocess $\gamma \gamma \to
\tilde{q}_i \bar{\tilde{q_i}} ~ ~ (q= t, b,~i=1,2)$ coming from
virtual gluonic correction can be cancelled by the real soft
gluonic bremsstrahlung, which is analogous to the real soft
photonic one. The real gluon emission diagrams of the process
$\gamma + \gamma \to \tilde{q}_i \bar{\tilde{q_i}} g$ are shown in
Fig.\ref{fig:feyn_realgluon}. We denote the real gluon emission as
\begin{equation}
\gamma (p_1) + \gamma (p_2) \to \tilde{q}_i (p_3) +
\bar{\tilde{q_i}}(p_4) + g (k),~(q=t,b,~i=1,2).
\end{equation}
\par
Analogously, we use
again the general phase-space-slicing method to separate the soft
gluon emission singularity from the real gluon emission process.
Therefore, the correction of the real gluon emission is divided
into soft and hard terms
\begin{equation}
\Delta \hat{\sigma}^{QCD}_{real} = \Delta
\hat{\sigma}^{QCD}_{soft} + \Delta \hat{\sigma}^{QCD}_{hard} =
\hat{\sigma}_0 (\hat{\delta}^{QCD}_{soft} +
\hat{\delta}^{QCD}_{hard})
\end{equation}
By using the soft gluon approximation, we get the contribution of the
soft gluon emission sunbprocess expressed as
\begin{equation}
{\rm d} \Delta \hat{\sigma}^{QCD}_{{ soft}} = -{\rm d}
\hat{\sigma}_0 \frac{\alpha_{s}C_F} {2\pi^2}
 \int_{|\vec{k}| \leq \Delta E_g}\frac{{\rm d}^3 k}{2 k^0} \left[
 \frac{p_3}{p_3 \cdot k}-\frac{p_4}{p_4 \cdot k} \right]^2
\end{equation}
in which $\Delta E_g$ is the energy cutoff of the soft gluon and
$k^0 \leq \Delta E_g \ll \sqrt {\hat s}$. $k^0 = \sqrt{|\vec{k}|^2
+ m_{g}^2}$ is the gluon energy. $p_3$ and $p_4$ are the four
momenta of two final state particles
$\tilde{q}_i$ and $\bar{\tilde{q_i}}$. In this approach, we may again
refer to Ref.\cite{COMS} to get the analytical expression of the
soft gluon correction.
\par
Finally we obtain an UV and IR finite ${\cal O}(\alpha_{s})$ QCD
correction $ \Delta \hat{\sigma}^{QCD}$
to the subprocess $\gamma \gamma \to \tilde{q}_i
\bar{\tilde{q_i}}$ containing one-loop ${\cal O}(\alpha_{s})$ QCD
correction
\begin{equation}
\Delta \hat{\sigma}^{QCD} =
\Delta \hat{\sigma}_{vir}^{QCD} +\Delta
\hat{\sigma}_{real}^{QCD} =\hat{\sigma}_0 \hat{\delta}^{QCD}
\end{equation}
where
$\hat{\delta}^{QCD}=\hat{\delta}^{QCD}_{vir}+\hat{\delta}^{QCD}_{soft}+\hat{\delta}^{QCD}_{hard}$
is the ${\cal O}(\alpha_{s})$ QCD relative correction.

\vskip 5mm
\subsection{The cross sections of parent processes $e^+ e^- \to \gamma \gamma \to
\tilde{f}_i \bar{\tilde{f}_i} ~ ~ (f= \tau, t, b,~i=1,2)$}

The $\tilde {f}_i \bar{\tilde {f}_i}$ pair production via
photon-photon fusion is only a subprocess of the parent process $e^+e^-
\to \gamma \gamma \to \tilde{f}_i \bar{\tilde{f}_i}$. The laser
back-scattering technique on electron beam can transform $e^+e^-$
beams into photon beams \cite{Com1,Com2,Com3}. After integrating
over the photon luminosity in an $e^+e^-$ linear collider, we obtain
the total cross section of the process $e^+ e^- \to \gamma \gamma \to
\tilde{f}_i \bar{\tilde{f}_i}$ expressed
\begin{equation}
\sigma(s)= \int_{2m_{\tilde{f}_i} / \sqrt{s}} ^{x_{max}} d z
\frac{dL_{\gamma\gamma}}{d z} \hat{\sigma}(\gamma\gamma \to
\tilde{f}_i \bar{\tilde{f}}_i  ~ at ~ \hat{s}=z^{2} s),
\end{equation}
where $\sqrt{s}$ and $\sqrt{\hat{s}}$ are the $e^{+}e^{-}$ and
$\gamma\gamma$ c.m.s. energies respectively and
$dL_{\gamma\gamma}/dz$ is the distribution function of photon
luminosity, which is expressed as
\begin{equation}
\frac{dL_{\gamma\gamma}}{dz}=2z\int_{z^2/x_{max}}^{x_{max}}
\frac{dx}{x} f_{\gamma/e}(x)f_{\gamma/e}(z^2/x)
\end{equation}
where $f_{\gamma/e}$ is the photon structure function of the
electron beam \cite {sd,rb}. For the initial unpolarized electrons
and laser photon beams, the photon structure function is given by
the most promising Compton backscattering as \cite {sd,vt,phospec}
\begin{equation}
f_{\gamma/e}=\frac{1}{D(\xi)}\left[1-x+\frac{1}{1-x}-
\frac{4x}{\xi(1-x)}+\frac{4x^{2}}{\xi^{2}(1-x)^2}\right],
\end{equation}
where
$$
D(\xi)=(1-\frac{4}{\xi}-\frac{8}{{\xi}^2})\ln{(1+\xi)}+\frac{1}{2}+
  \frac{8}{\xi}-\frac{1}{2{(1+\xi)}^2},
$$
\begin{equation}
\xi=\frac{2\sqrt{s} \omega_0}{{m_e}^2}.
\end{equation}
$m_e$ and $\sqrt s /2$ represent the mass and energy of the
electron respectively. $\omega_0$ is the laser-photon energy and
$x$ is the fraction of the energy of the incident electron carried
by the backscattered photon. The maximum fraction of energy
carried by the backscattered photon is $x_{max}=2
\omega_{max}/\sqrt{s}=\xi/(1+\xi)$. In our calculations, we choose
$\omega_0$ to maximize the backscattered photon energy without
spoiling the luminosity through $e^{+}e^{-}$ pair creation. Then
we have ${\xi}=2(1+\sqrt{2})$, $x_{max}\simeq 0.83$, and $D(\xi)
\approx 1.8397$\cite{photon para}.

\vskip 5mm
\section{Numerical results}
\par
In this section, we present some numerical results for the one
loop ${\cal O}(\alpha_{s})$ QCD and ${\cal O}(\alpha_{ew})$ EW
corrections to subprocesses $\gamma \gamma \to \tilde{f}_i
\bar{\tilde{f_i}}$ and the parent processes $e^+e^- \to \gamma
\gamma \to \tilde{f}_i \bar{\tilde{f_i}}$. In our numerical
calculation, the SM parameters are set to be $\alpha_{s}(m_Z^2)=0.1190$,
$m_{e}=0.5110998902$ MeV, $m_{\mu}=105.658357$ MeV, $m_{\tau}=1.77699$ GeV, $m_u=66$
MeV, $m_d=66$ MeV, $m_c=1.2$ GeV, $m_s=150$ MeV, $m_b=4.3$ GeV,
$m_t=174.3$ GeV, $m_Z=91.1876$ GeV, $m_W=80.423$
GeV\cite{databook}. There we use the effective values of the light
quark masses ($m_u$ and $m_d$) which can reproduce the hadron
contribution to the shift in the fine structure constant
$\alpha_{ew}(m_Z^2)$\cite{leger}. We take the fine structure
constant at the $Z^0$-pole as input parameter,
$\alpha_{ew}(m_Z^2)^{-1}|_{\overline{MS}}=127.918$\cite{databook}.
Then from Eq.(\ref{CT electric charge}) we get the counter-term of
the electric charge in $\overline {DR}$ scheme expressed as
\cite{count1, count2, eberl}
\begin{eqnarray}
\label{ecount}
   \delta Z_e &=&
   \frac{e^2}{6(4\pi)^2}
    \left\{ 4 \sum_f N_C^f e_f^2\left( \Delta+\log\frac{Q^2}{x_f^2} \right)+\sum_{\tilde{f}} \sum_{k=1}^2 N_C^f
    e_{f}^2 \left( \Delta+\log\frac{Q^2}{m^2_{\tilde{f}_k}} \right)     \right.       \nonumber  \\
  &&\left. + 4 \sum_{k=1}^2\left(\Delta+\log\frac{Q^2}{m^2_{\tilde{\chi}_k}}\right)
       +\sum_{k=1}^2\left( \Delta+\log\frac{Q^2}{m^2_{H_k^+}} \right)   \right.      \nonumber   \\
  &&\left.  - 22 \left(\Delta+\log\frac{Q^2}{m_W^2}\right)
  \right\},
\end{eqnarray}
where we take $x_f=m_Z$ when $m_f<m_Z$ and $x_t=m_t$. $Q_f$ is the
electric charge of (s)fermion and
$\Delta=2/\epsilon-\gamma+\log4\pi$. $N_C^f$ is color factor,
which equal to 1 and 3 for (s)leptons and (s)quarks, respectively.
It is obvious that there is a little discrepancy between our
electric charge counter-term expression( Eq.(\ref{ecount})) and
that in subsection 3.1 of Ref.\cite{eberl}.
\par
The MSSM parameters are determined by using FormCalc package with
following input parameters\cite{FeyArtsFormCalc}:
\par
(1) The input parameters for MSSM Higgs sector are the CP-odd mass
$M_{A^0}$ and $\tan \beta$ with the constraint $\tan \beta \geq
2.5$. The masses of the MSSM Higgs sector are fixed by taking into
account the significant radiative corrections.
\par
(2) The input parameters for the chargino and neutralino sector
are the gaugino mass parameters $M_1$, $M_2$ and the Higgsino-mass
parameter $\mu$. We adopt the grand unification theory(GUT)
relation $M_1 = (5/3)\tan^2 \theta_W M_2$ for simplification
\cite{mssm-2} and the gluino mass $m_{\tilde g}$ is evaluated by
$m_{\tilde g}={\alpha_s(Q)}/{\alpha_{ew}(m_Z)}\sin^2 \theta_W
M_2$.
\par
(3) For the sfermion sector, we assume
$M_{\tilde{Q}}=M_{\tilde{U}}=M_{\tilde{D}}=M_{\tilde{E}}=M_{\tilde{L}}=M_{SUSY}$
and take the soft trilinear couplings for sfermions $\tilde{q}$
and $\tilde{l}$ being equal, i.e., $A_q=A_l=A_f$.
\par
Except above SM and MSSM input parameters, we should have some
other parameters used in our numerical calculations, for example,
the QCD renormalization scale $Q$, the IR regularization parameter
$m_{\gamma}(m_g)$ and the soft cutoff $\Delta E_{\gamma,g} /E_b$.
In our following numerical calculations, we take the QCD
renormalization scale $Q$ to be $2m_{\tilde{f}_i}$, and set
$\Delta E_{\gamma, g} /E_b=10^{-3}$, $m_{\gamma,g}=10^{-5}$ GeV,
if there is no other statement. As we know, the final results
should be independent on IR regulator $m_{\gamma,g}$ and the
cutoff $\Delta E_{\gamma,g} /E_b$. For demonstration, we present
the dependence of the ${\cal O}(\alpha_{s})$ QCD corrections to
$\gamma \gamma \to \tilde{t}_1 \bar{\tilde{t_1}}(m_{\tilde{t}_1}=
148~GeV)$ in conditions of $\sqrt{\hat{s}}=500~GeV$ and $Set1$
parameters(see below) on the soft cutoff $\Delta E_{g} /E_b$ in
Fig.\ref{fig:softcutoff}. The full, dashed and dotted lines
correspond to $\Delta \hat{\sigma}^{QCD}_{vir+soft}$, $\Delta
\hat{\sigma}^{QCD}_{hard}$ and the total correction $\Delta
\hat{\sigma}^{QCD}$. As shown in this figure, the full ${\cal
O}(\alpha_{s})$ one-loop QCD correction $\Delta
\hat{\sigma}^{QCD}$ is independent of the soft cutoff $\Delta
E_g/E_b$ as $\Delta E_g /E_b$ running from $10^{-5}$ to $10^{-2}$,
although both $\Delta \hat{\sigma}^{QCD}_{vir+soft}$ and $\Delta
\hat{\sigma}^{QCD}_{hard}$ depend on cutoff strongly.

\begin{figure}[htb]
\centering
\includegraphics{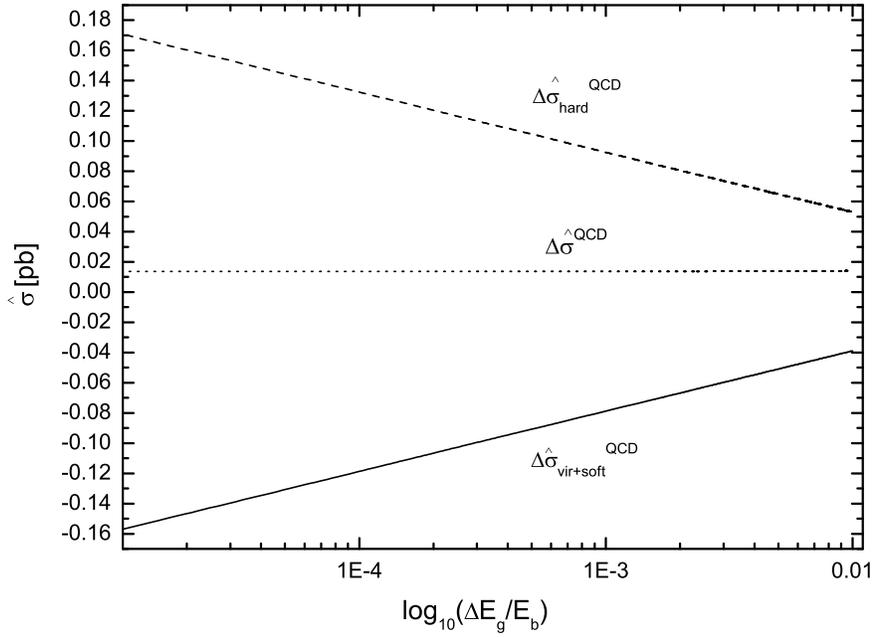}
\caption{The full ${\cal O}(\alpha_{s})$ QCD corrections to
$\gamma \gamma \to \tilde{t}_1 \bar{\tilde{t_1}}$ as a function of
the soft gluon cutoff $\Delta E_g /E_b$ in conditions of
$\sqrt{\hat{s}}=500~GeV$ and $Set1$ parameters. }
\label{fig:softcutoff}
\end{figure}
\par
In order to show and discuss the effects of the radiative
corrections to the subprocess of $\gamma \gamma \to \tilde{f}_i
\bar{\tilde{f_i}}$ quantitatively, we choose the following four
typical data sets:
\begin{itemize}
\item[$Set1$:] $\tan\beta=6$, $M_{A^0}=250$ GeV, $M_{SUSY}=200$ GeV, $\mu=800$ GeV,
$M_2=200$ GeV and $A_f=400$ GeV. \\
 Then we have $m_{\tilde{\tau}_{1,2}}=(185,223)$ GeV,
$m_{\tilde{t}_{1,2}}=(148,340)$ GeV and $m_{\tilde{b}_{1,2}}=(146,
250)$ GeV.
\item[$Set2$:] $\tan\beta=20$, $M_{A^0}=300$ GeV, $M_{SUSY}=400$ GeV, $\mu=1000$ GeV,
$M_2=200$ GeV and $A_f=-500$ GeV. \\
Then we have $m_{\tilde{\tau}_{1,2}}=(354,446)$ GeV,
$m_{\tilde{t}_{1,2}}=(304,533)$ GeV and
$m_{\tilde{b}_{1,2}}=(256,508)$ GeV.
\item[$Set3$:] $\tan\beta=30$, $M_{A^0}=300$ GeV, $M_{SUSY}=250$ GeV, $\mu=-200$ GeV,
$M_2=800$ GeV and $A_f=250$ GeV. \\
Then we have $m_{\tilde{\tau}_{1,2}}=(231,275)$ GeV,
$m_{\tilde{t}_{1,2}}=(215,368)$ GeV and
$m_{\tilde{b}_{1,2}}=(188,307)$ GeV.
\item[$Set4$:]
$\tan\beta=30$, $M_{A^0}=250$ GeV, $M_{SUSY}=200$ GeV, $\mu=200$
GeV,
$M_2=1000$ GeV and $A_f=300$ GeV. \\
Then we have $m_{\tilde{\tau}_{1,2}}=(179,228)$ GeV,
$m_{\tilde{t}_{1,2}}=(131,346)$ GeV and
$m_{\tilde{b}_{1,2}}=(124,263)$ GeV.
\end{itemize}
\par
With the input parameters $\tan\beta$, $M_{A^0}$, $M_{SUSY}$,
$\mu$, $M_2$ and $A_f$ in above data sets, all the masses of
supersymmetric particles can be obtained by using package
FormCalc. $Set1$(or $Set2$) is the case of gaugino-like with
small(or mediate) $\tan \beta$, but lighter(or heavier) sfermions,
while $Set3$ and $Set4$ are higgsino-like case with larger $\tan
\beta$.
\begin{figure}[hbtp]
\vspace*{-1cm} \epsfxsize = 8cm \epsfysize = 8cm
\epsfbox{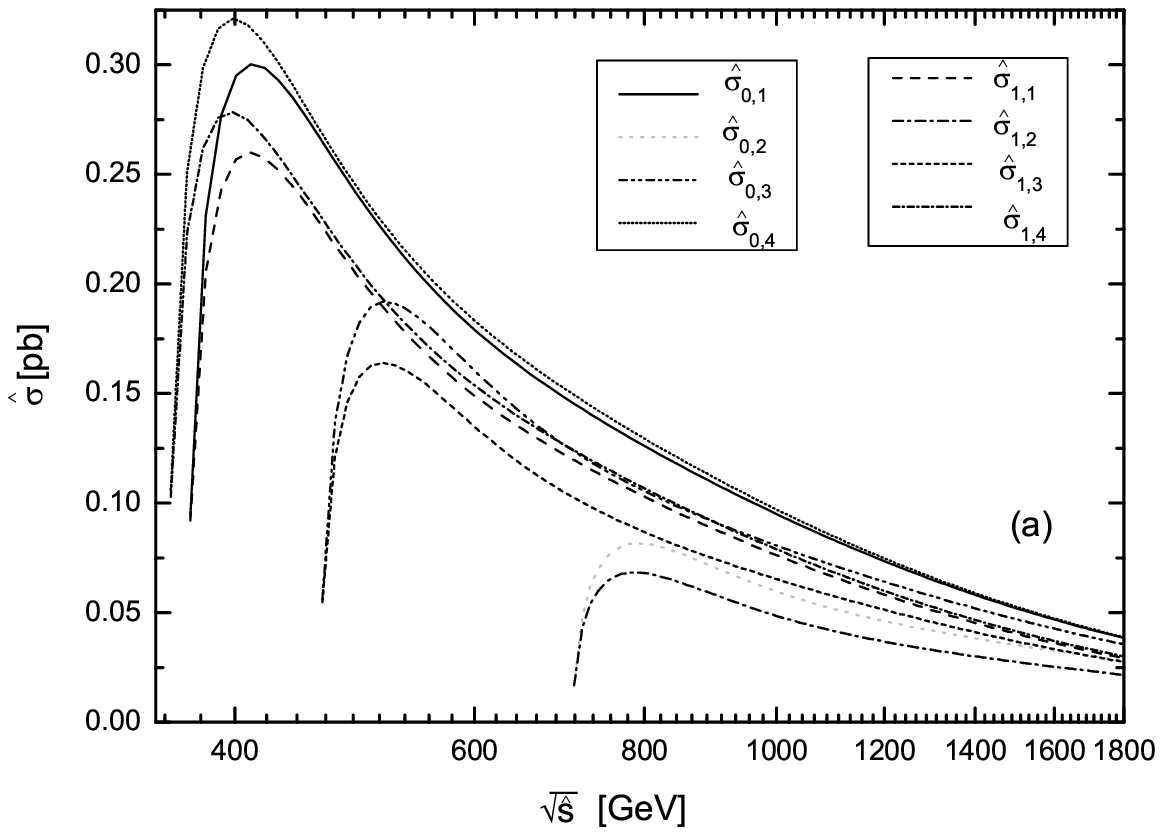} \epsfxsize = 8cm \epsfysize = 8cm
\epsfbox{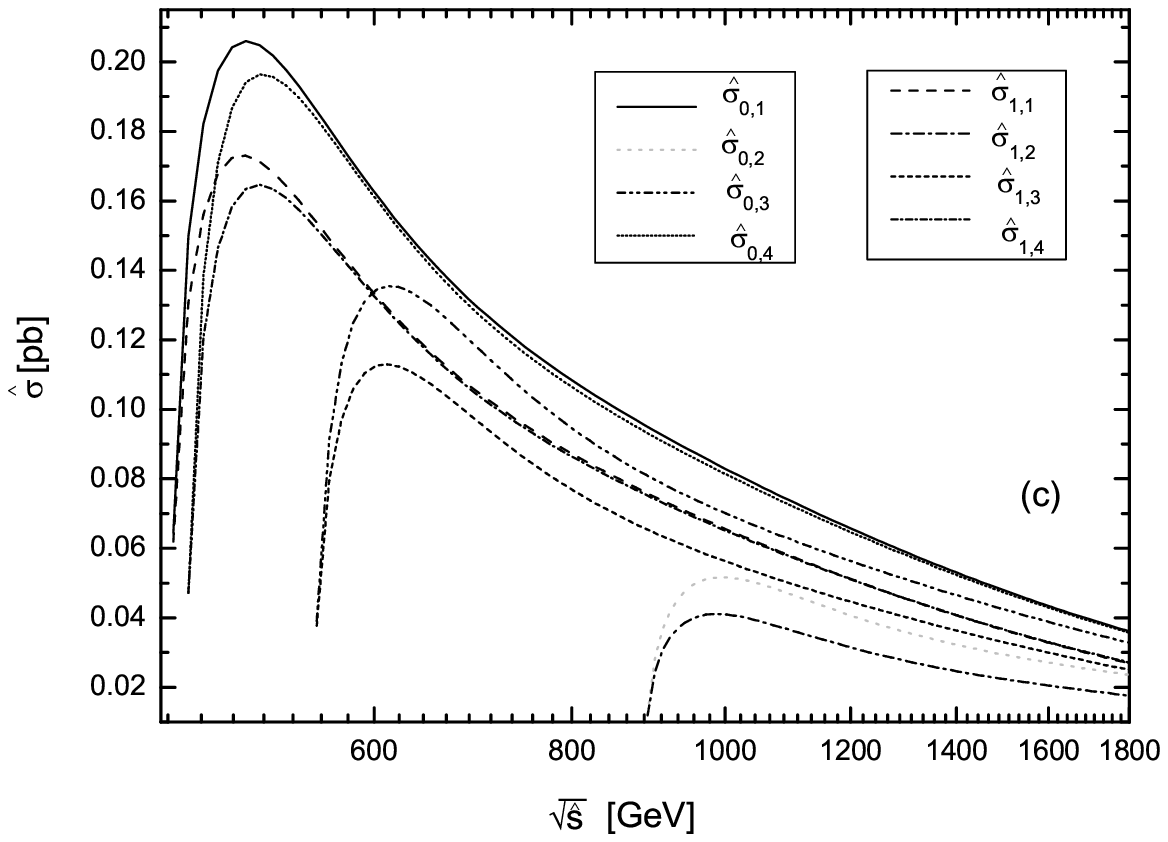} \epsfxsize = 8cm \epsfysize = 8cm
\epsfbox{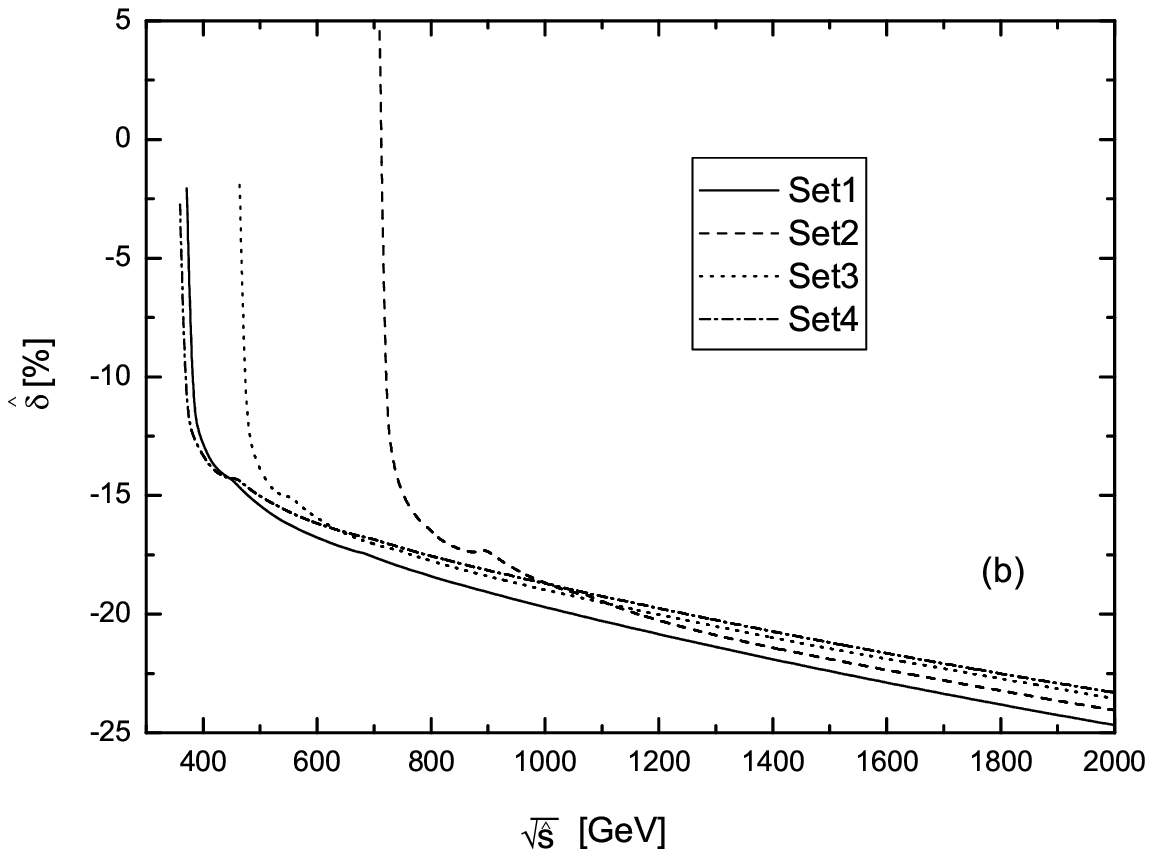} \epsfxsize = 8cm \epsfysize = 8cm
\epsfbox{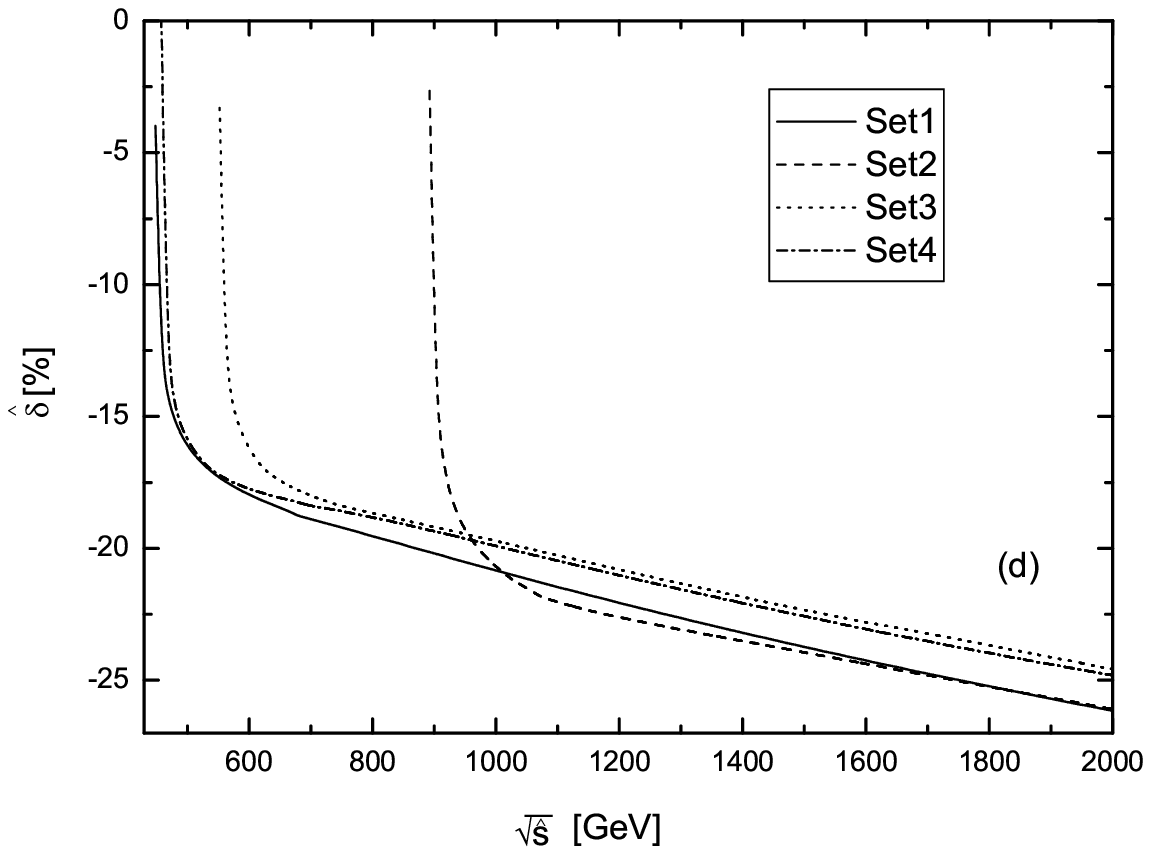} \vspace*{-0.3cm} \caption{ (a) The Born and
full ${\cal O}(\alpha_{ew})$ EW corrected cross sections for the
$\gamma \gamma \to \tilde{\tau}_1 \bar{\tilde{\tau_1}}$ subprocess
as the functions of c.m.s. energy of $\gamma \gamma$ collider
$\sqrt {\hat s}$ with four different data sets, respectively. (b)
The full ${\cal O}(\alpha_{ew})$ EW relative correction to the
$\gamma \gamma \to \tilde{\tau}_1 \bar{\tilde{\tau_1}}$
subprocess. The solid ,dashed, dotted and dash-dotted curves
correspond to four different data set cases, respectively. (c) The
Born and full ${\cal O}(\alpha_{ew})$ EW corrected cross sections
for the $\gamma \gamma \to \tilde{\tau}_2 \bar{\tilde{\tau_2}}$
subprocess as the functions of c.m.s. energy of $\gamma \gamma$
collider $\sqrt {\hat s}$ with four different data sets,
respectively. (d) The full ${\cal O}(\alpha_{ew})$ EW relative
correction to the $\gamma \gamma \to \tilde{\tau}_2
\bar{\tilde{\tau_2}}$ subprocess. The solid ,dashed, dotted and
dash-dotted curves correspond to four different data set cases,
respectively.}
\end{figure}
\par
The Born and the full ${\cal O}(\alpha_{ew})$ EW corrected cross
sections for the $\gamma \gamma \to \tilde{\tau}_1
\bar{\tilde{\tau_1}}$ subprocess as the functions of c.m.s. energy
of $\gamma \gamma$ collider with above four data sets are
displayed in Fig.5(a). There $\hat{\sigma}_{0,i}$'s are the Born
cross sections and $\hat{\sigma}_{1,i}$'s are the full ${\cal
O}(\alpha_{ew})$ EW corrected cross sections for the subprocess
$\gamma \gamma \to \tilde{\tau}_1 \bar{\tilde{\tau_1}}$. The
subscript $i$ goes from 1 to 4, which correspond to the data
$Set1$, $Set2$, $Set3$ and $Set4$ respectively. The ${\cal
O}(\alpha_{ew})$ EW corrected cross section with $Set 4$ can
achieve the maximal value 0.278 pb at the energy near the
threshold $\sqrt{\hat s}\sim 400$ GeV. When $\sqrt {\hat s}$
approaches to 1.5 TeV, the EW corrected cross section with $Set 2$
goes down to 27.5 fb, but it is still much larger than that for
the process of $e^+e^- \to \tilde{\tau}_1
\bar{\tilde{\tau}_1}$\cite{eberl, hollik} with the same input
parameters. In Fig.5(b), the relative ${\cal O}(\alpha_{ew})$ EW
corrections with the four data sets are depicted. As it can be
seen in this figure, the relative corrections $\hat \delta$ also
have their maximal values at the position near the threshold
energies and then decrease quantitatively with the increment of
$\sqrt {\hat s}$. When the c.m.s. energy $\sqrt {\hat s}$ goes
from the threshold value of $\tilde{\tau}_1 \bar{\tilde{\tau_1}}$
pair production to 2 TeV, the full ${\cal O}(\alpha_{ew})$ EW
corrections can enhance or reduce the Born cross section depending
on the colliding energy. At the position of colliding energy
$\sqrt {\hat s}$ = 2 TeV, the relative EW correction $\hat \delta$
can reach $-24.6\%$, $-24.1\%$, $-23.5\%$ and $-23.2\%$ for
$Set1$, $Set2$, $Set3$ and $Set4$ respectively. Fig.5(c) shows the
numerical results of the cross sections of $\gamma \gamma \to
\tilde{\tau}_2 \bar{\tilde{\tau_2}}$ subprocess both at the Born
level and one-loop level, as the functions of the colliding energy
$\sqrt{\hat{s}}$. Fig.5(d) displays the relative ${\cal
O}(\alpha_{ew})$ EW correction for $\tilde{\tau}_2
\bar{\tilde{\tau_2}}$-pair production as a function of
$\sqrt{\hat{s}}$. We find that the behavior of curves in Fig.5(c),
which correspond to the Born, the EW corrected cross
sections of $\tilde{\tau}_2 \bar{\tilde{\tau_2}}$ production,
are quite similar to those in Fig.5(a) for $\tilde{\tau}_1\bar{\tilde{\tau_1}}$
production. But the values of the cross sections for
$\tilde{\tau}_2 \bar{\tilde{\tau_2}}$ production are always
smaller due to the heavier mass of $\tilde{\tau}_2$, and can reach
0.173 pb near the threshold energy of $\tilde{\tau}_2
\bar{\tilde{\tau_2}}$ pair production in the case of $Set1$. The
magnitude of EW relative correction is about -26.1$\%$ or -24.5$\%$ at the
position of $\sqrt {\hat s}$ = 2 TeV, which is close to that of
$\gamma \gamma \to \tilde{\tau}_1 \bar{\tilde{\tau_1}}$
subprocess.
\begin{figure}[hbtp]
\vspace*{-1cm} \epsfxsize = 8cm \epsfysize = 8cm
\epsfbox{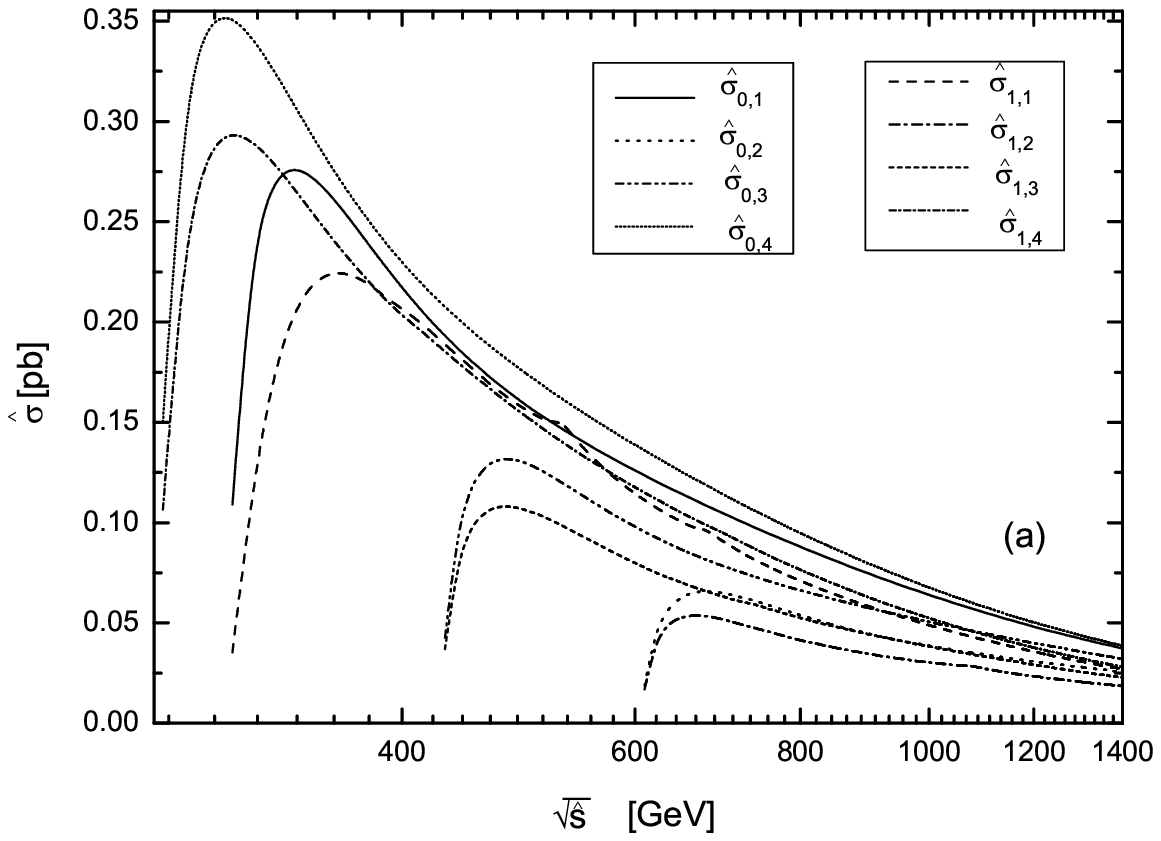} \epsfxsize = 8cm \epsfysize = 8cm
\epsfbox{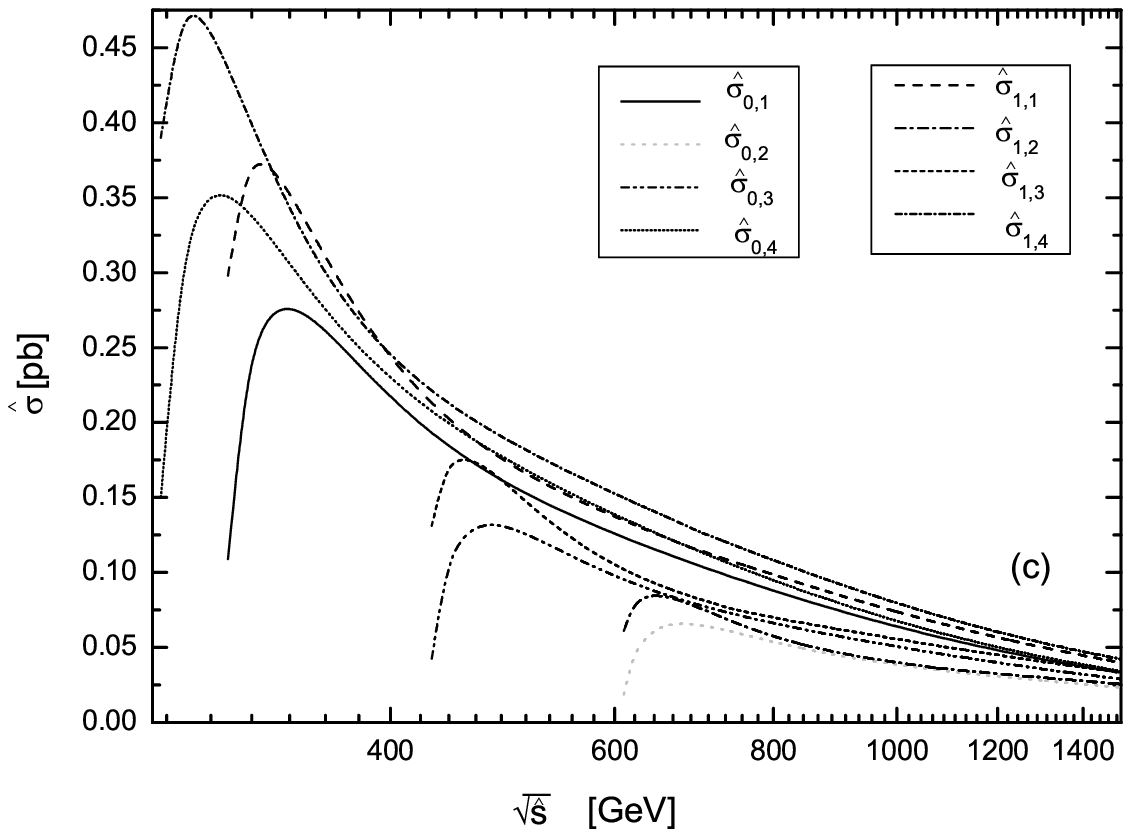} \epsfxsize = 8cm \epsfysize = 8cm
\epsfbox{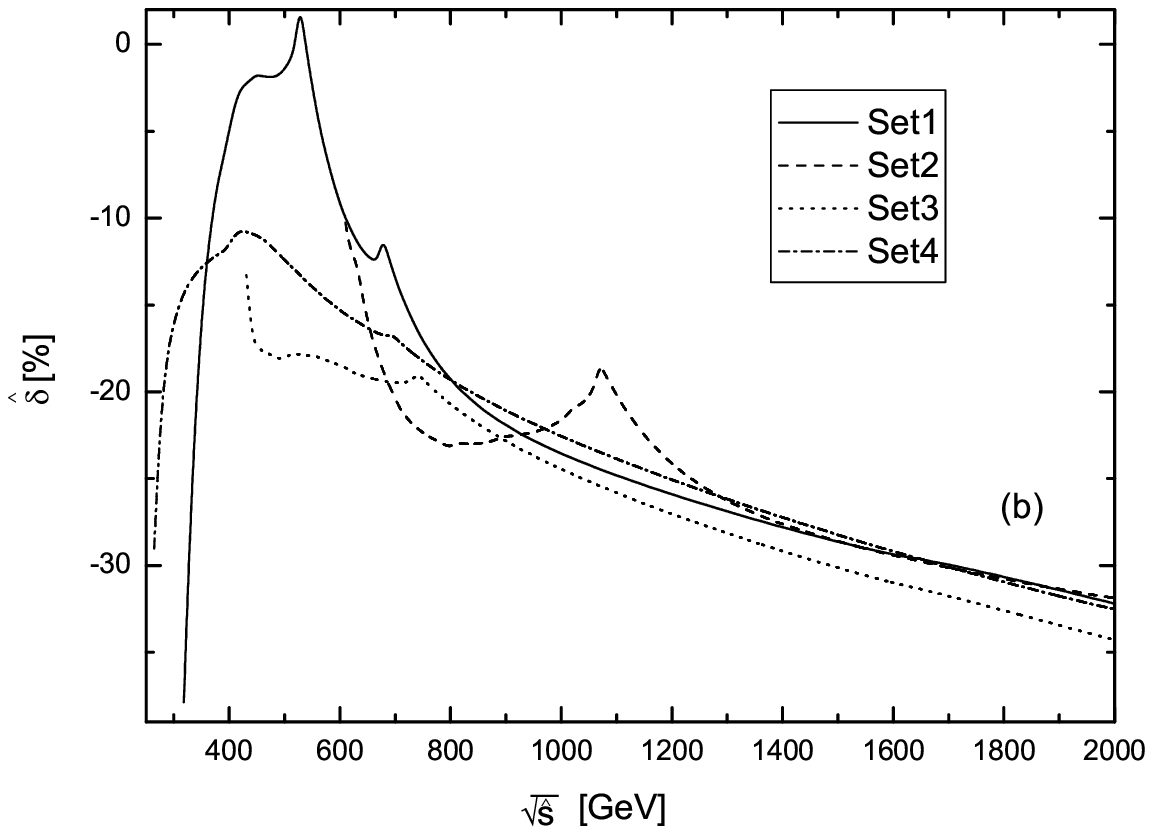} \epsfxsize = 8cm \epsfysize = 8cm
\epsfbox{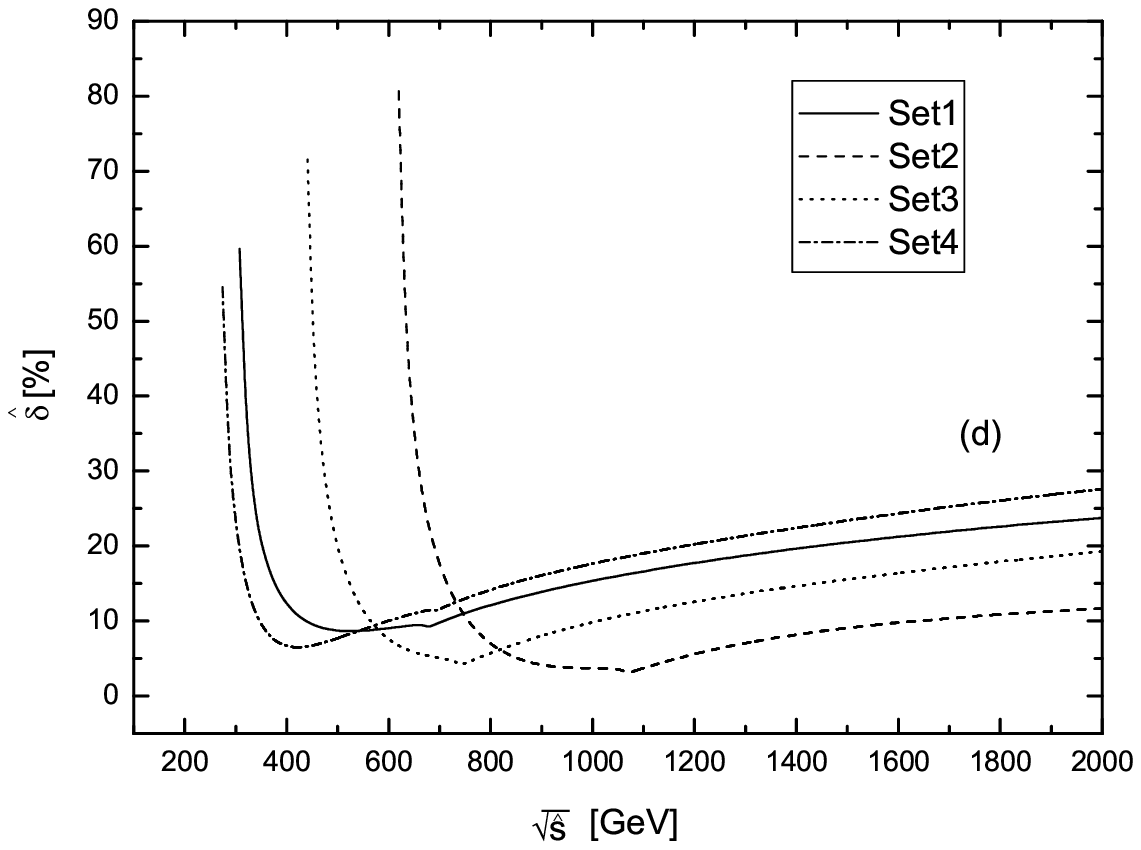} \vspace*{-0.3cm} \caption{ (a) The Born and
full ${\cal O}(\alpha_{ew})$ EW corrected cross sections for the
$\gamma \gamma \to \tilde t_1 \bar{\tilde{t_1}}$ subprocess as the
functions of c.m.s. energy of $\gamma \gamma$ collider $\sqrt
{\hat s}$ with four different data sets, respectively. (b) The
full ${\cal O}(\alpha_{ew})$ EW relative correction to $\gamma
\gamma \to \tilde t_1 \bar{\tilde{t_1}}$ subprocess. Four
different curves correspond to four different data sets,
respectively. (c) The Born and full ${\cal O}(\alpha_{s})$ QCD
corrected cross sections for the $\gamma \gamma \to \tilde t_1
\bar{\tilde{t_1}}$ subprocess as the functions of c.m.s. energy of
$\gamma \gamma$ collider $\sqrt {\hat s}$ with four different data
sets, respectively. (d) The full ${\cal O}(\alpha_{s})$ QCD
relative correction to $\gamma \gamma \to \tilde t_1
\bar{\tilde{t_1}}$ subprocess.}
\end{figure}
\begin{figure}[hbtp]
\vspace*{-1cm} \epsfxsize = 8cm \epsfysize = 8cm
\epsfbox{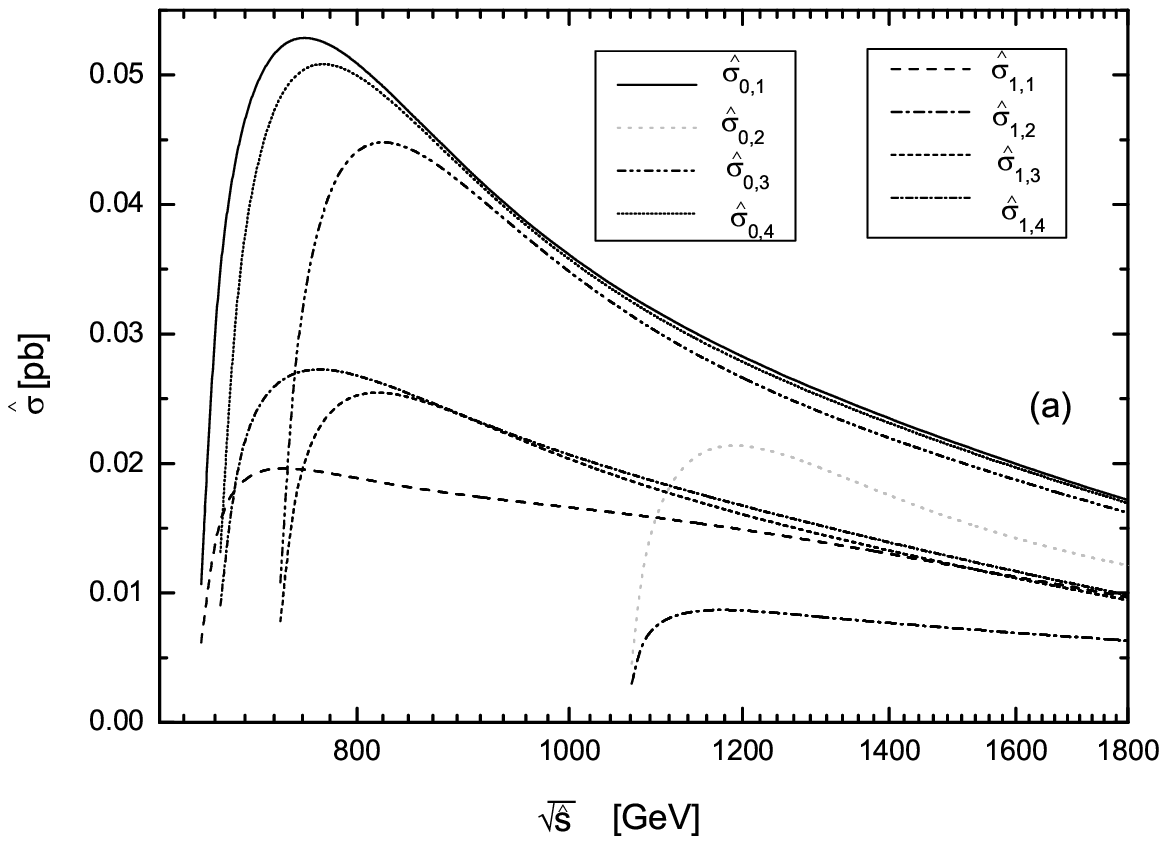} \epsfxsize = 8cm \epsfysize = 8cm
\epsfbox{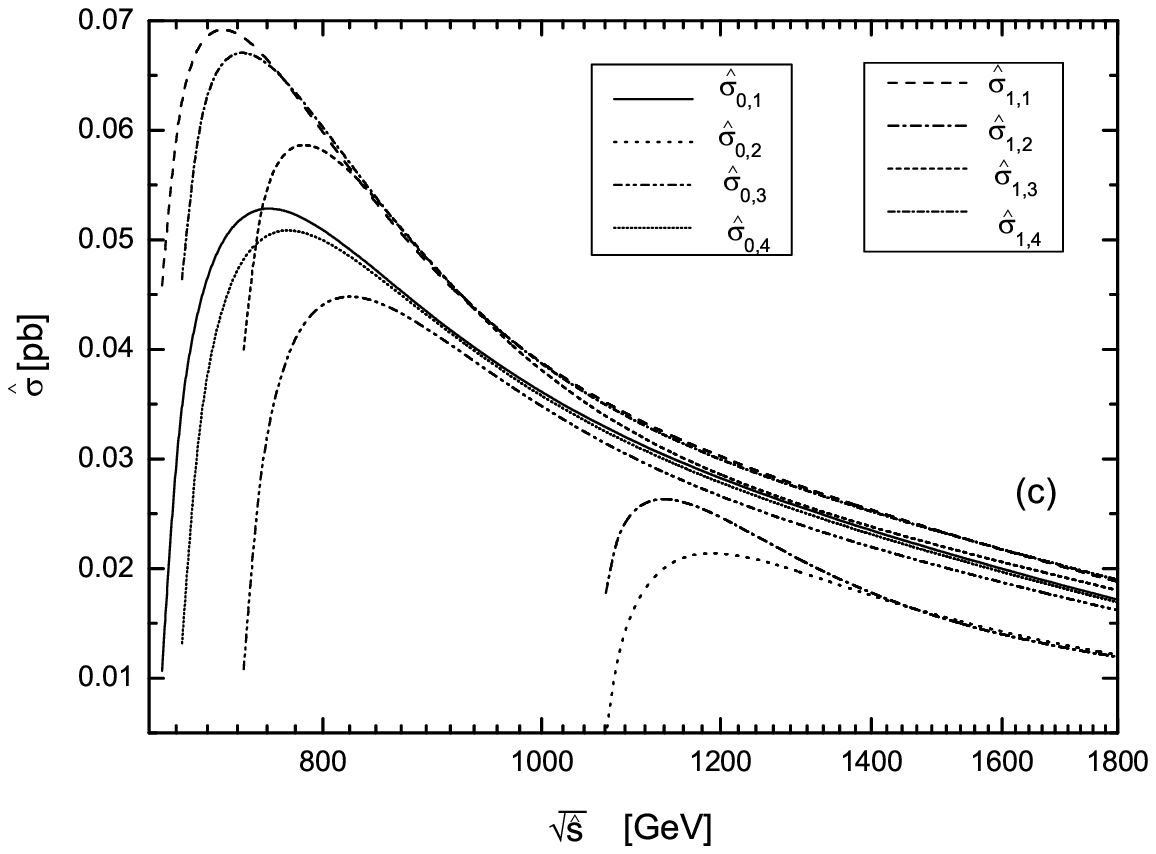} \epsfxsize = 8cm \epsfysize = 8cm
\epsfbox{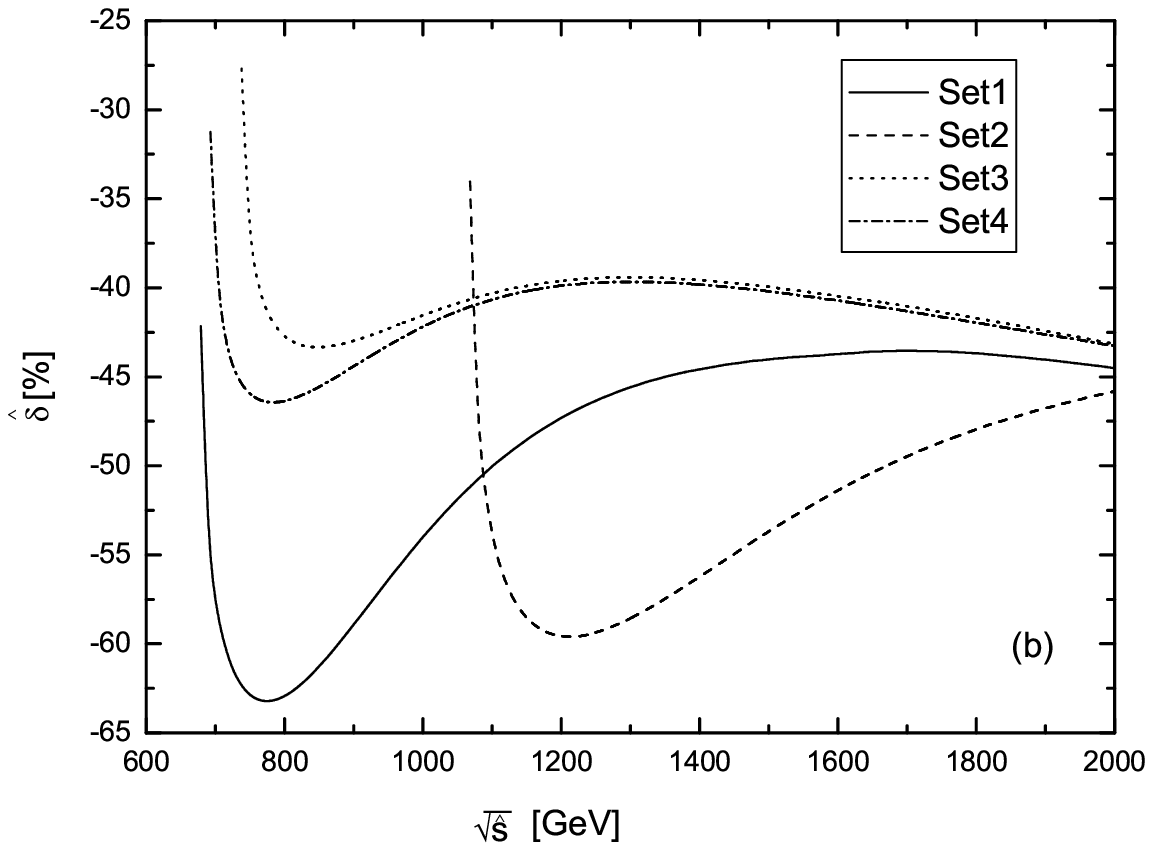} \epsfxsize = 8cm \epsfysize = 8cm
\epsfbox{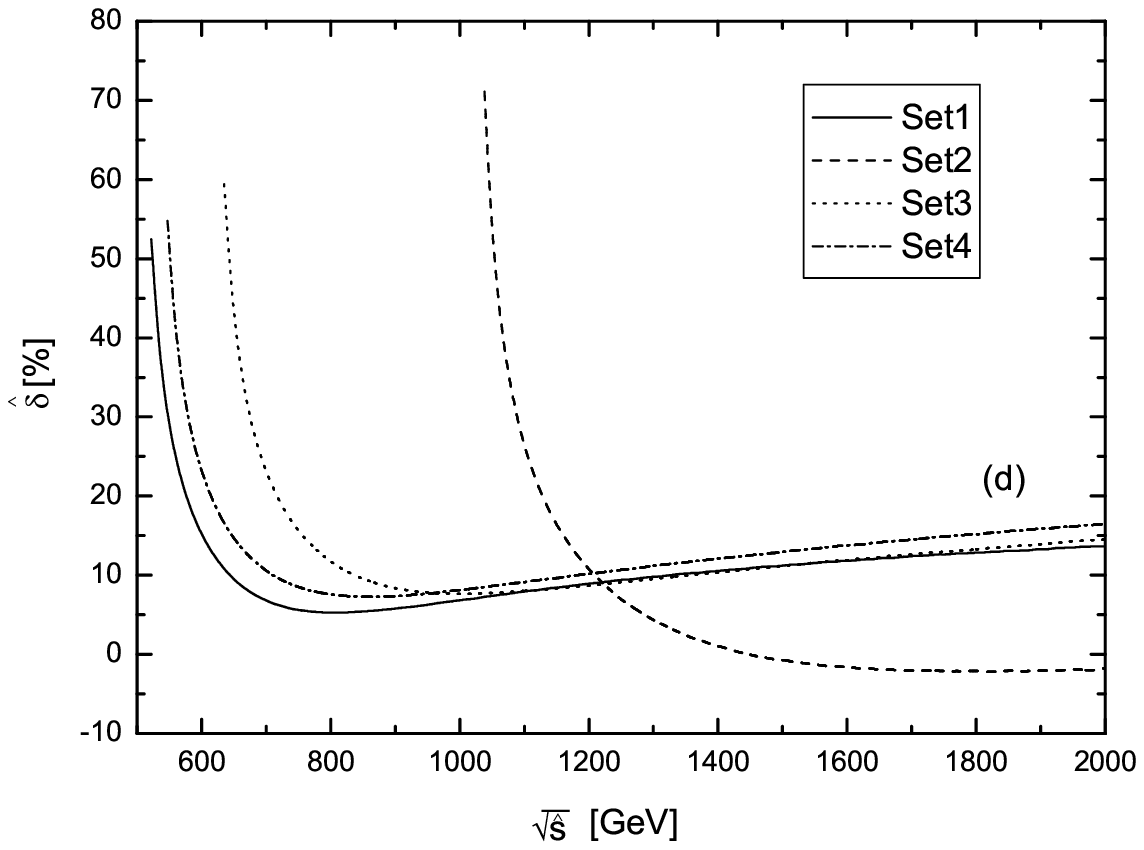} \vspace*{-0.3cm} \caption{ (a) The Born and
full ${\cal O}(\alpha_{ew})$ EW corrected cross sections for the
$\gamma \gamma \to \tilde t_2 \bar{\tilde{t_2}}$ subprocess as the
functions of c.m.s. energy of $\gamma \gamma$ collider $\sqrt
{\hat s}$ with four different data sets, respectively. (b) The
full ${\cal O}(\alpha_{ew})$ EW relative correction to $\gamma
\gamma \to \tilde t_2 \bar{\tilde{t_2}}$ subprocess. Four
different curves correspond to four different data sets,
respectively. (c) The Born and full ${\cal O}(\alpha_{s})$ QCD
corrected cross sections for the $\gamma \gamma \to \tilde t_2
\bar{\tilde{t_2}}$ subprocess as the functions of c.m.s. energy of
$\gamma \gamma$ collider $\sqrt {\hat s}$ with four different data
sets, respectively. (d) The full ${\cal O}(\alpha_{s})$ QCD
relative correction to $\gamma \gamma \to \tilde t_2
\bar{\tilde{t_2}}$ subprocess.}
\end{figure}
\par
In Fig.6(a) and Fig.6(c), we depict the full ${\cal
O}(\alpha_{ew})$ EW and ${\cal O}(\alpha_{s})$ QCD corrected cross
sections for the subprocess $\gamma \gamma \to \tilde{t}_1
\bar{\tilde{t_1}}$. Analogously, $\hat{\sigma}_{0,i}(i=1\cdots,4)$
mean the tree-level cross sections corresponding to the four input
data sets respectively, and $\hat {\sigma}_{1,i}$'s are the full
one-loop corrected cross sections. Fig.6(a) demonstrates that the
corresponding two curves for Born and ${\cal O}(\alpha_{ew})$ EW
corrected cross sections in the same condition of the input data
set, have the same line shape. While Fig.6(c) shows obviously that
the ${\cal O}(\alpha_{s})$ QCD corrections can be larger than the
${\cal O}(\alpha_{ew})$ EW corrections, especially near the
threshold. The EW and QCD relative corrections to $\tilde t_1
\bar{\tilde t}_1$ pair production subprocess are displayed in
Fig.6(b) and (d), respectively. From Fig.6(b), we can see that the
${\cal O}(\alpha_{ew})$ EW relative corrections to $\gamma \gamma
\to \tilde{t}_1 \bar{\tilde{t_1}}$ subprocess vary from positive
values to negative ones as $\sqrt {\hat s}$ running from the
threshold value to 2 TeV. The absolute value of the relative
correction $|\hat \delta ^{EW}|$ for $Set3$ can reach about
$34.2\%$ at the point of $\sqrt{\hat{s}}$ = 2 TeV and be over
$32\%$ near the threshold for $Set1$. Furthermore, with the same
input parameters as used in \cite{eberl}, for example, $Set2$, our
calculation shows that, when $\sqrt{\hat{s}}$ is between 1200 GeV
and 2000 GeV, the EW relative correction to $\gamma \gamma \to
\tilde{t}_1 \bar{\tilde{t_1}}$ subprocess is about $-24.1 \sim
-31.8\%$, while the EW relative correction to $e^+e^- \to
\tilde{t}_1 \bar{\tilde{t_1}}$ process is about $-10\%$
\cite{eberl}. We notice that on the curves in Fig.6(b) there are
some small spikes which are due to the resonance effects. For
example, the resonance effect at the position of $\sqrt {\hat s}
\sim 525$ GeV is caused by $\sqrt {\hat s} \sim 2m_{H^+}$ for
input data $Set1$, while the resonance effect at the position of
$\sqrt {\hat s} \sim 1066$ GeV is caused by $\sqrt {\hat s}\sim
2m_{\tilde{t}_2}$ for input data $Set2$. Furthermore, when we
observe Fig.6(d), which shows the relative QCD correction as a
function of c.m.s. energy $\sqrt {\hat s}$ for $\gamma \gamma \to
\tilde{t}_1 \bar{\tilde{t_1}}$, we can see that the values of
$\hat \delta ^{QCD}$ decrease rapidly to minimal values after
$\sqrt{\hat s}$ just goes up from the threshold energy and then
increase slowly to $23.8\%$, $11.4\%$, $19.3\%$ and $27.6\%$ at
the position of $\sqrt {\hat s}$ = 2 TeV for $Set1$, $Set2$,
$Set3$ and $Set4$, respectively.
\par
The results for subprocess $\gamma \gamma \to \tilde{t}_2
\bar{\tilde{t_2}}$ are drawn in Fig.7(a-d). The full ${\cal
O}(\alpha_{ew})$ EW and ${\cal O}(\alpha_{s})$ QCD corrected cross
sections are plotted in Fig.7(a) and Fig.7(c), respectively.
Comparing these two figures with Fig.6(a) and Fig.6(c), we can see
that the cross sections for the $\gamma \gamma \to \tilde{t}_2
\bar{\tilde{t_2}}$ subprocess are almost one order smaller than
those for the $\gamma \gamma \to \tilde{t}_1 \bar{\tilde{t_1}}$
subprocess quantitatively due to $m_{\tilde{t}_2} >
m_{\tilde{t}_1}$. The EW and QCD relative corrections to
subprocess $\gamma \gamma \to \tilde{t}_2 \bar{\tilde{t_2}}$ are
plotted in Fig.7(b) and Fig.7(d), respectively. From Fig.7(d), we
see that the values of QCD relative corrections are rather large
near the threshold, and at position of $\sqrt {\hat s}$ = 2 TeV
they are $10.2\%$, $-0.7\%$, $12.5\%$ and $13.2\%$ for data
$Set1$, $Set2$, $Set3$ and $Set4$ respectively, which are less
than the corresponding QCD relative corrections to the $\gamma
\gamma \to \tilde{t}_1 \bar{\tilde{t_1}}$ subprocess shown in
Fig.6(d). The absolute EW relative corrections are generally
larger than those of the absolute QCD relative corrections shown
in Fig.7(d), except in the threshold energy regions. The values of
EW relative corrections are about $-45\%$ for the two
gaugino-like data sets $Set 1$ and $Set 2$ and $-43\%$ for the two
higgsino-like data sets $Set 3$ and $Set 4$ at the position of
$\sqrt {\hat s}$ = 2 TeV. The EW relative corrections to $\gamma
\gamma \to \tilde{t}_2 \bar{\tilde{t_2}}$ cross section are
negative in the range of $\sqrt{\hat{s}}= 700 \sim 2000$ GeV with
all the four data sets and have the minimal values near the
position of $\sqrt {\hat s} = 800$ GeV for $Set 1$, $Set 3$ and
$Set 4$, and in the vicinity of $\sqrt {\hat s} \sim 1200$ GeV for
$Set 2$.
\begin{figure}[hbtp]
\vspace*{-1cm} \epsfxsize = 8cm \epsfysize = 8cm
\epsfbox{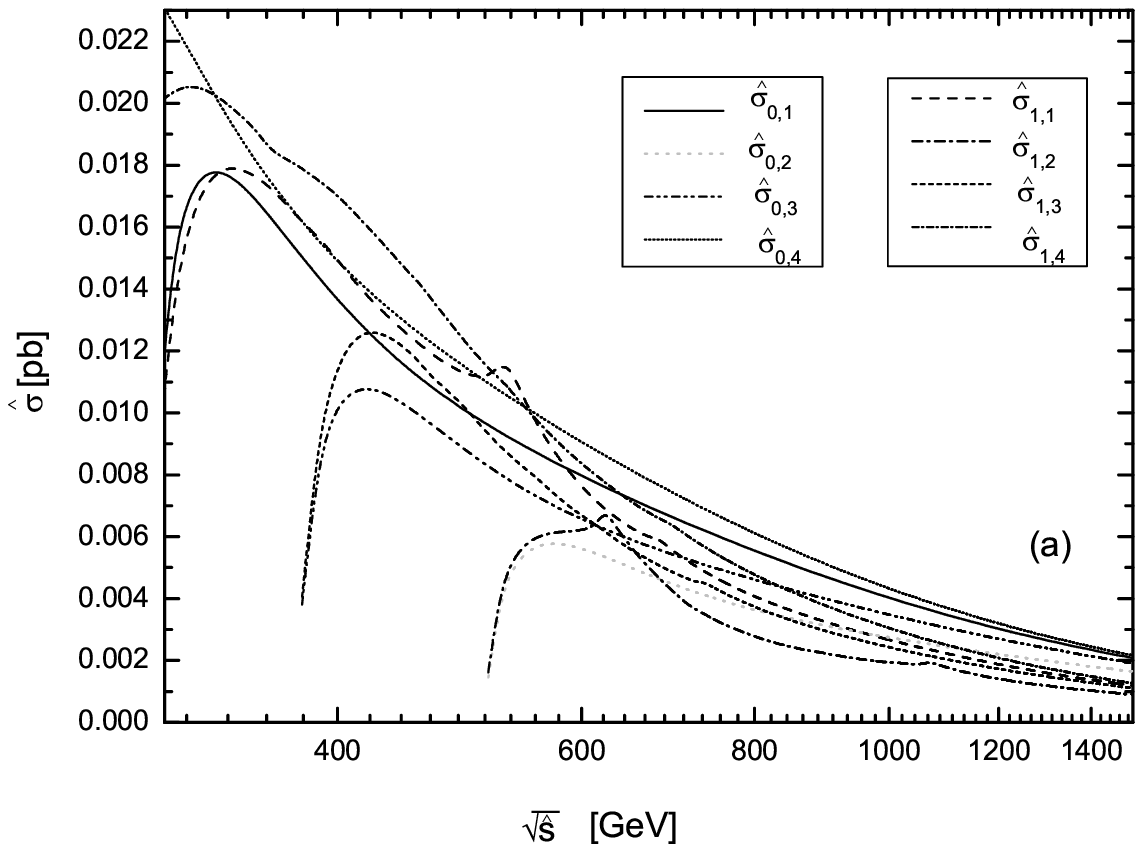} \epsfxsize = 8cm \epsfysize = 8cm
\epsfbox{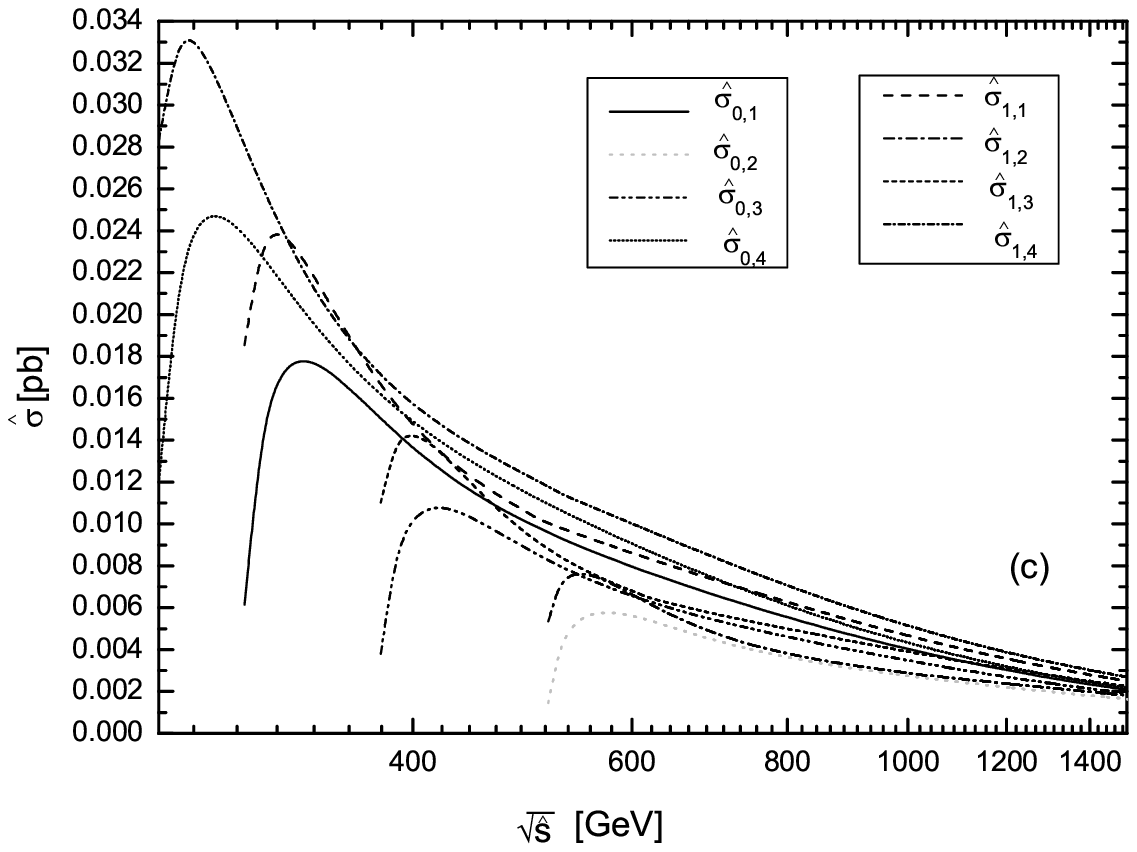} \epsfxsize = 8cm \epsfysize = 8cm
\epsfbox{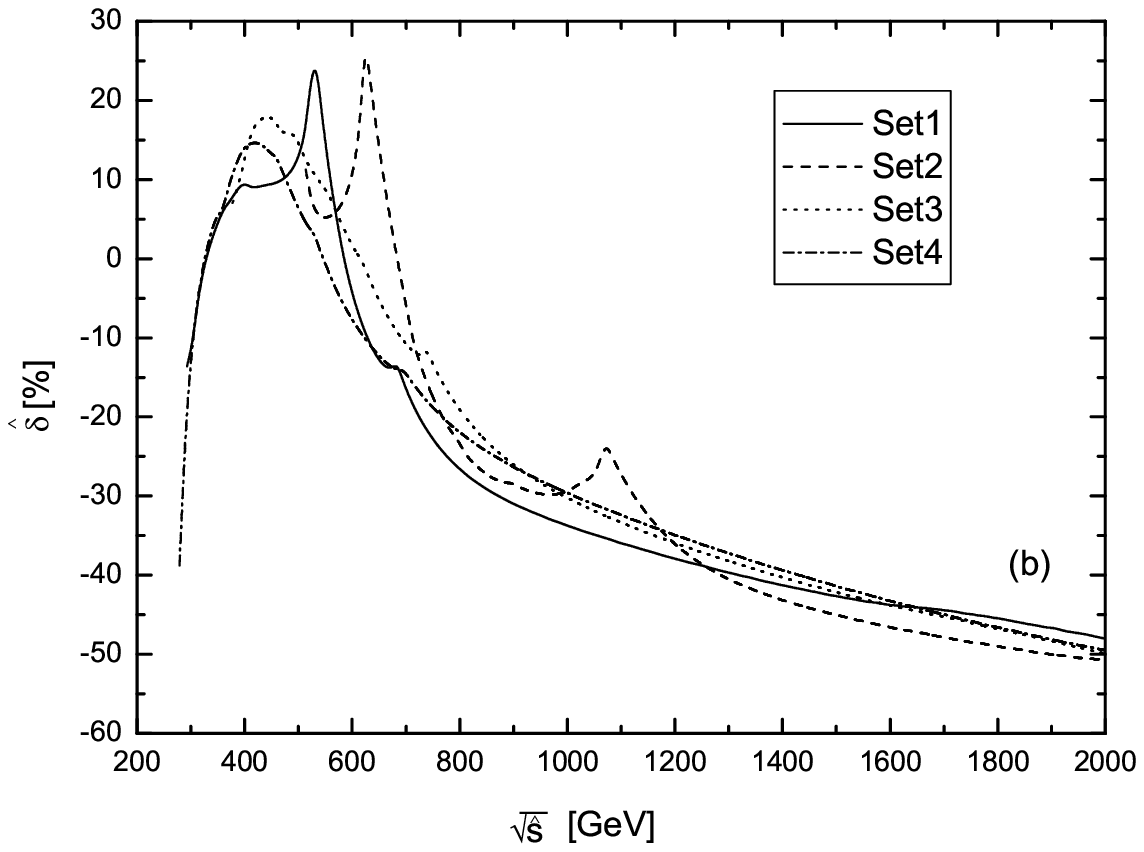} \epsfxsize = 8cm \epsfysize = 8cm
\epsfbox{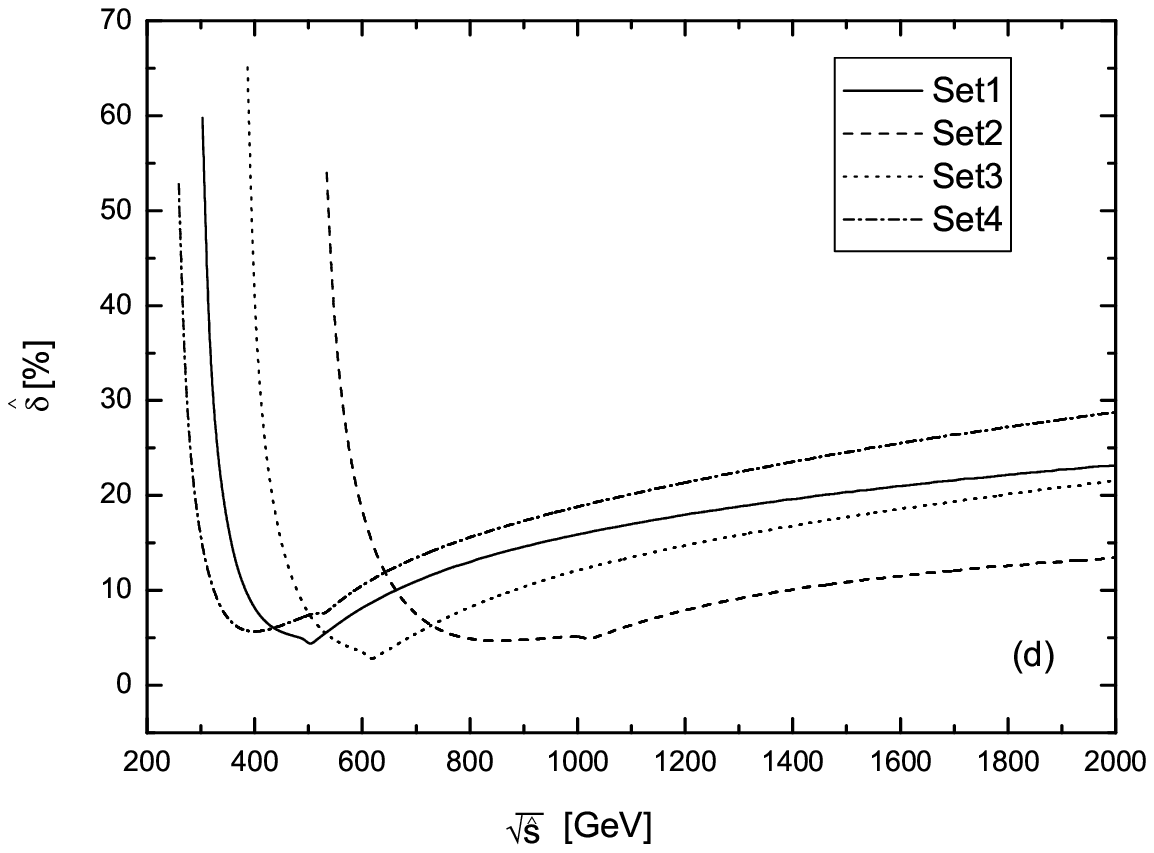} \vspace*{-0.3cm} \caption{ (a) The Born and
full ${\cal O}(\alpha_{ew})$ EW corrected cross sections for the
$\gamma \gamma \to \tilde b_1 \bar{\tilde{b_1}}$ subprocess as the
functions of c.m.s. energy $\sqrt {\hat s}$ with four data sets,
respectively. (b) The full one-loop ${\cal O}(\alpha_{ew})$ EW
relative correction to $\gamma \gamma \to \tilde b_1
\bar{\tilde{b_1}}$ subprocess. (c) The Born and full ${\cal
O}(\alpha_{s})$ QCD corrected cross sections for the $\gamma
\gamma \to \tilde b_1 \bar{\tilde{b_1}}$ subprocess as the
functions of c.m.s. energy $\sqrt {\hat s}$ with four data sets,
respectively. (d) The full ${\cal O}(\alpha_{s})$ QCD relative
correction to $\gamma \gamma \to \tilde b_1 \bar{\tilde{b_1}}$
subprocess. }
\end{figure}
\begin{figure}[hbtp]
\vspace*{-1cm} \epsfxsize = 8cm \epsfysize = 8cm
\epsfbox{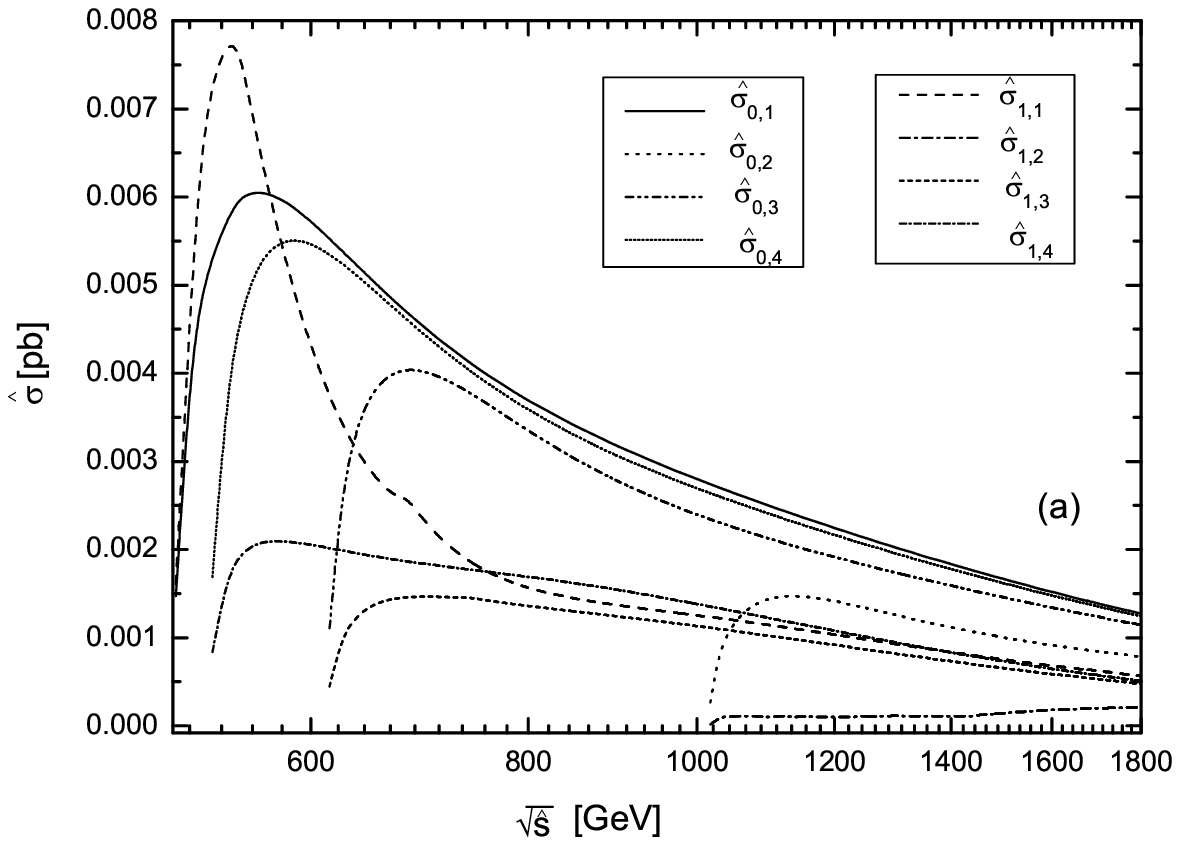} \epsfxsize = 8cm \epsfysize = 8cm
\epsfbox{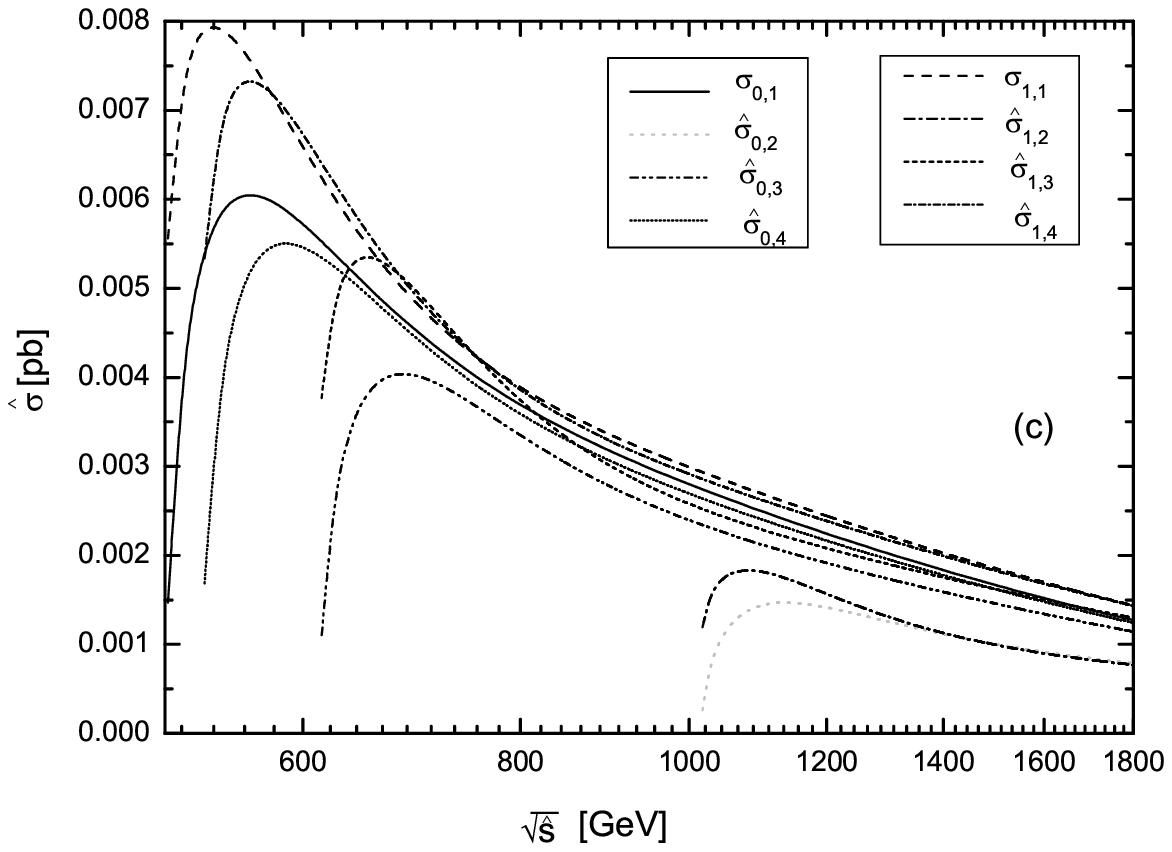} \epsfxsize = 8cm \epsfysize = 8cm
\epsfbox{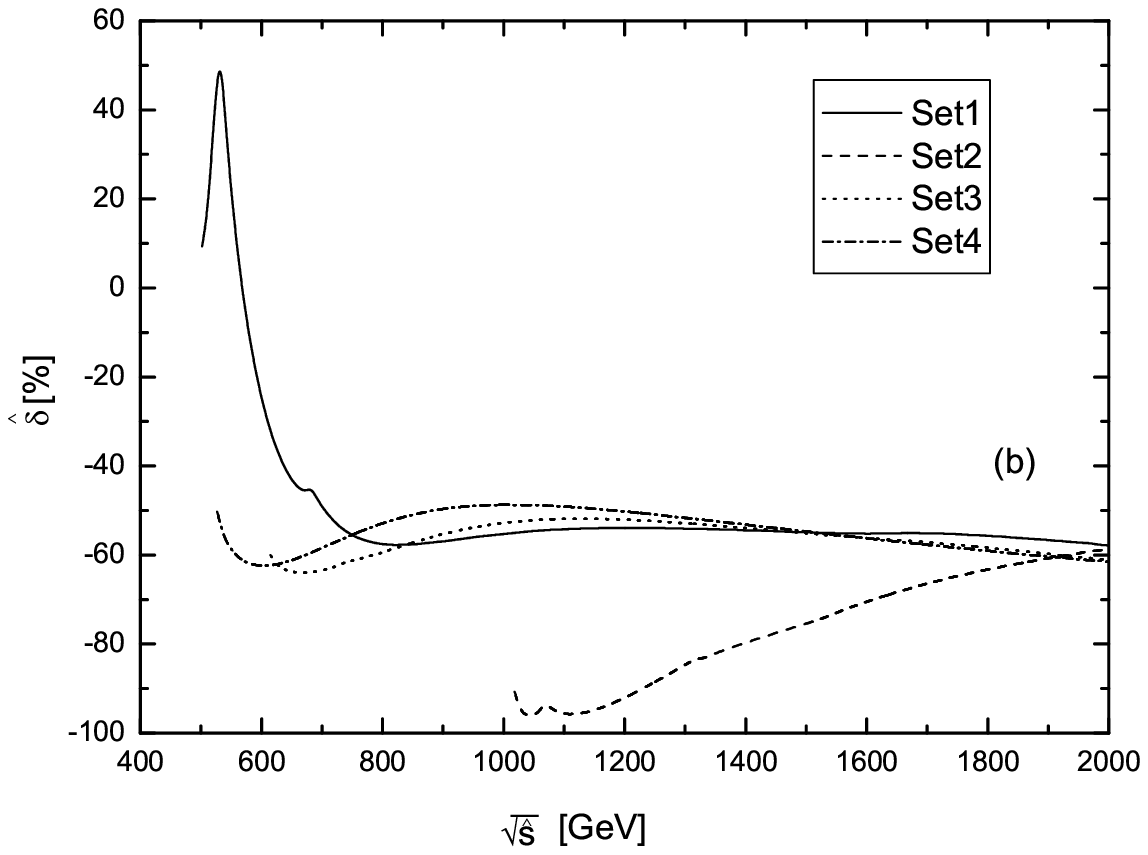} \epsfxsize = 8cm \epsfysize = 8cm
\epsfbox{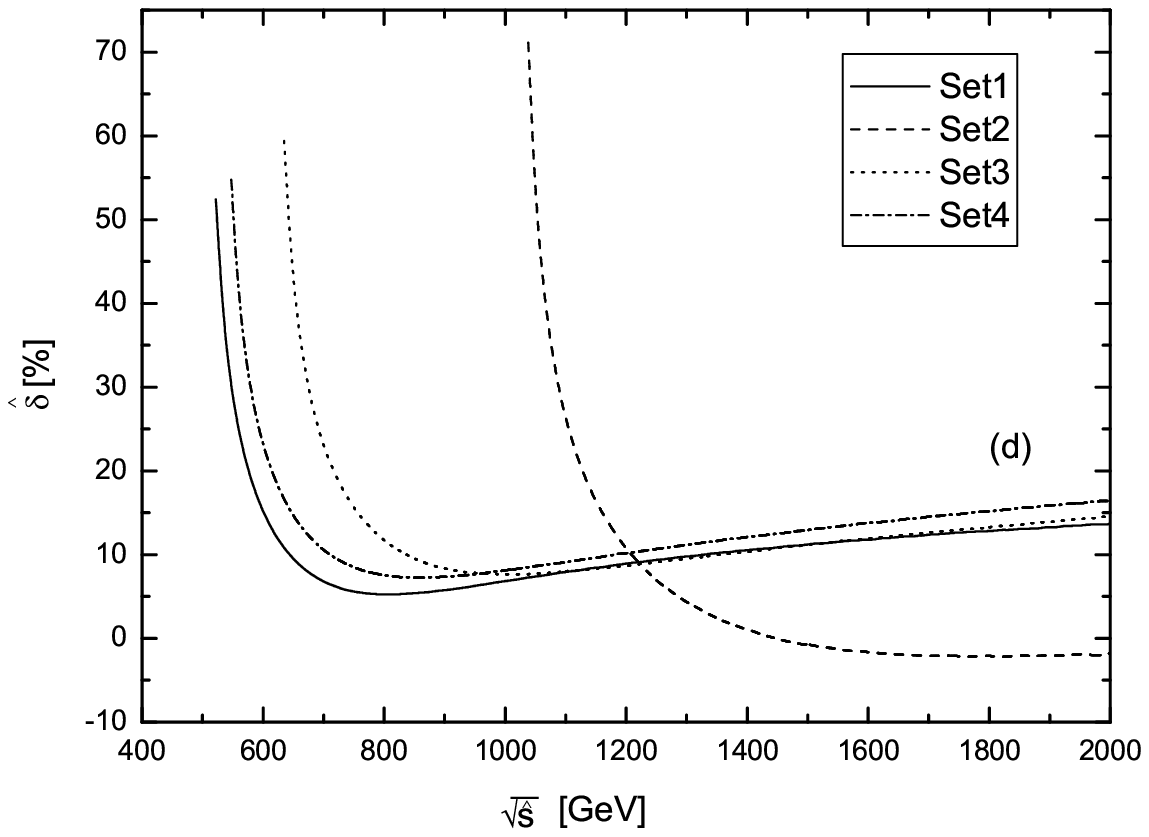} \vspace*{-0.3cm} \caption{ (a) The Born and
full ${\cal O}(\alpha_{ew})$ EW corrected cross sections for the
$\gamma \gamma \to \tilde b_2 \bar{\tilde{b_2}}$ subprocess as the
functions of c.m.s. energy $\sqrt {\hat s}$ with four data sets,
respectively. (b) The full ${\cal O}(\alpha_{ew})$ EW relative
corrections to $\gamma \gamma \to \tilde b_2 \bar{\tilde{b_2}}$
subprocess. (c) The Born and full ${\cal O}(\alpha_{s})$ QCD
corrected cross sections for the $\gamma \gamma \to \tilde b_2
\bar{\tilde{b_2}}$ subprocess as the functions of c.m.s. energy
$\sqrt {\hat s}$ with four data sets, respectively. (d) The full
${\cal O}(\alpha_{s})$ QCD relative correction to $\gamma \gamma
\to \tilde b_2 \bar{\tilde{b_2}}$ subprocess.}
\end{figure}
\par
We also show the $\tilde{b}_i\bar{\tilde{b}}_i,~(i=1,2)$ pair
productions in Fig.8 and Fig.9. Fig.8(a) is plotted for the Born
and full ${\cal O}(\alpha_{ew})$ EW corrected cross sections of
the $\gamma \gamma \to \tilde{b}_1 \bar{\tilde{b}}_1$ subprocess
as the functions of $\sqrt{\hat s}$ with four data sets
respectively, and Fig.8(c) for the Born and full ${\cal
O}(\alpha_{s})$ QCD corrected cross sections. Fig.8(b) and
Fig.8(d) display the EW and QCD relative corrections,
respectively. These figures show that the behaviors of the curves
for $\tilde b_1 \bar{\tilde b}_1$ pair production are similar to
those for $\tilde t_1 \bar{\tilde{t}_1}$ pair production.
Comparing Fig.6(b) with Fig.8(b), we notice that the full ${\cal
O}(\alpha_{ew})$ EW relative correction to $\gamma \gamma \to
\tilde{b}_1 \bar{\tilde{b}}_1$ subprocess is larger than that to
the $\gamma \gamma \to \tilde{t}_1 \bar{\tilde{t}_1}$ subprocess.
The maximum absolute value of the former can reach $50.7\%$ for
$Set2$ at the position of $\sqrt {\hat s} \sim 2000$ GeV, which is
even larger than the QCD correction to $\gamma \gamma \to
\tilde{b}_1 \bar{\tilde{b}}_1$. Again, all of the small spikes
appearing on the curves of Fig.8(a-b) are due to the resonance
effects, such as the spikes at the positions of $\sqrt {\hat s}
\sim 525$ GeV for $Set 1$ and $\sqrt {\hat s} \sim 621$ GeV for
$Set2$ are caused by $\sqrt {\hat s} \sim 2m_{H^+}$, while
condition of $\sqrt {\hat s} \sim 2m_{\tilde{t}_2}$ leads to the
spike at the position of $\sqrt {\hat s} \sim 1066$ GeV for
$Set2$. In Fig.8(d), the solid, dashed, dotted and dash-dotted
lines correspond to the QCD relative corrections with parameter
scenarios $Set1$, $Set2$, $Set3$ and $Set4$, respectively.
Although Fig.8(d) demonstrates that the the QCD relative
corrections in the region near the threshold energy of the
$\tilde{b}_1\bar{\tilde{b}}_1$ pair production are extremely
large, these values are untrustworthy due to the non-perturbative
QCD effects. The values of the relative ${\cal O}(\alpha_{s})$ QCD
corrections at the position of $\sqrt {\hat s} = 2000$ GeV, are
$23.1\%$, $13.4\%$, $21.5\%$ and $28.7\%$, for $Set1$, $Set2$,
$Set3$ and $Set4$ respectively.
\par
The full ${\cal O}(\alpha_{ew})$ EW and ${\cal O}(\alpha_{s})$ QCD
corrected cross sections for the $\gamma \gamma \to \tilde{b}_2
\bar{\tilde{b}}_2$ subprocess are depicted in Fig.9(a) and
Fig.9(c) separately, while their corresponding relative
corrections are plotted in Fig.9(b) and Fig.9(d) respectively.
Although the line-shapes in Fig.9(a) and fig.9(c) are similar with
the corresponding ones in Fig.8(a) and (c) for the $\tilde{b}_1
\bar{\tilde{b_1}}$ pair production, the values of the corrected
cross section in Fig.9(a) and Fig.9(c) are much smaller due to
$m_{\tilde{b}_2}>m_{\tilde{b}_1}$. Nevertheless, Fig.9(b) shows
that when $\sqrt{\hat s}$ is large enough, the absolute EW
relative corrections to $\gamma \gamma \to \tilde{b}_2
\bar{\tilde{b}}_2$ approach about $50\%$ or beyond for all four
data sets. The peak at the position of $\sqrt {\hat s} \sim 525$
GeV for $Set1$ in Fig.9(b) comes from the resonance effect of
$\sqrt{\hat s} \sim 2m_{H^+}$. We find also from Fig.9(d) that the
absolute QCD relative corrections to the $\gamma \gamma \to
\tilde{b}_2 \bar{\tilde{b}}_2$ subprocess are generally comparable
with the EW corrections or even smaller than EW ones, especially
in the large colliding energy region.
\begin{figure}[hbtp]
\vspace*{-1cm} \epsfxsize = 8cm \epsfysize = 8cm
\epsfbox{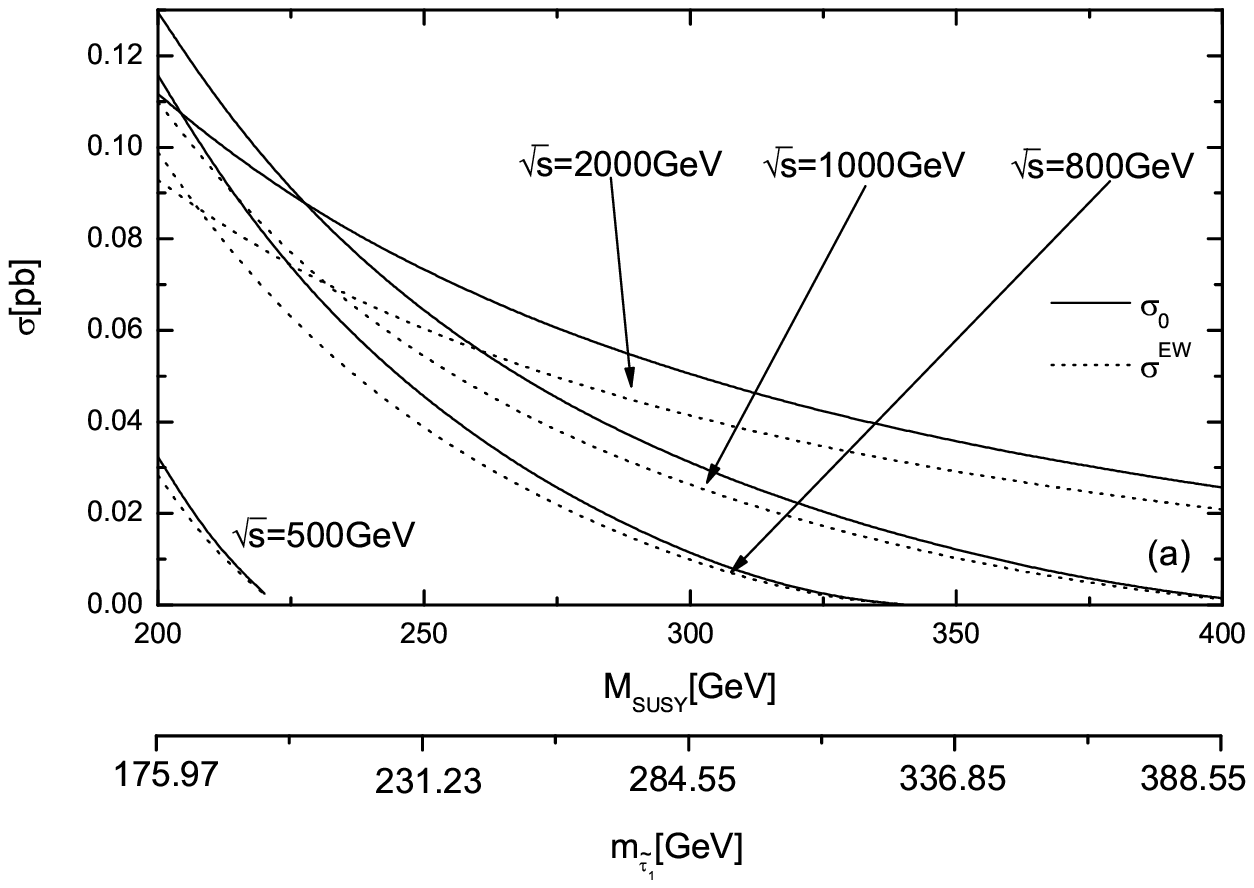} \epsfxsize = 8cm \epsfysize = 8cm
\epsfbox{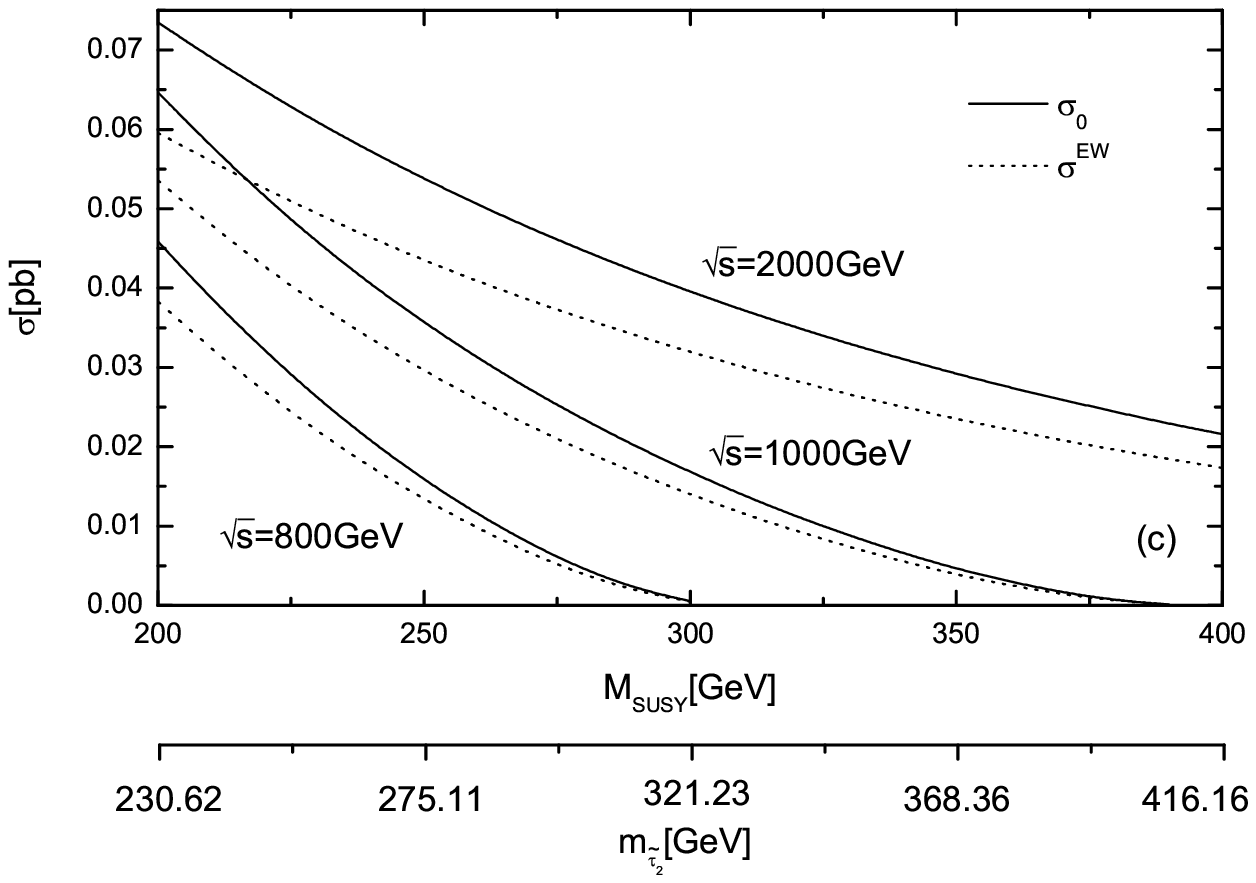} \epsfxsize = 8cm \epsfysize = 8cm
\epsfbox{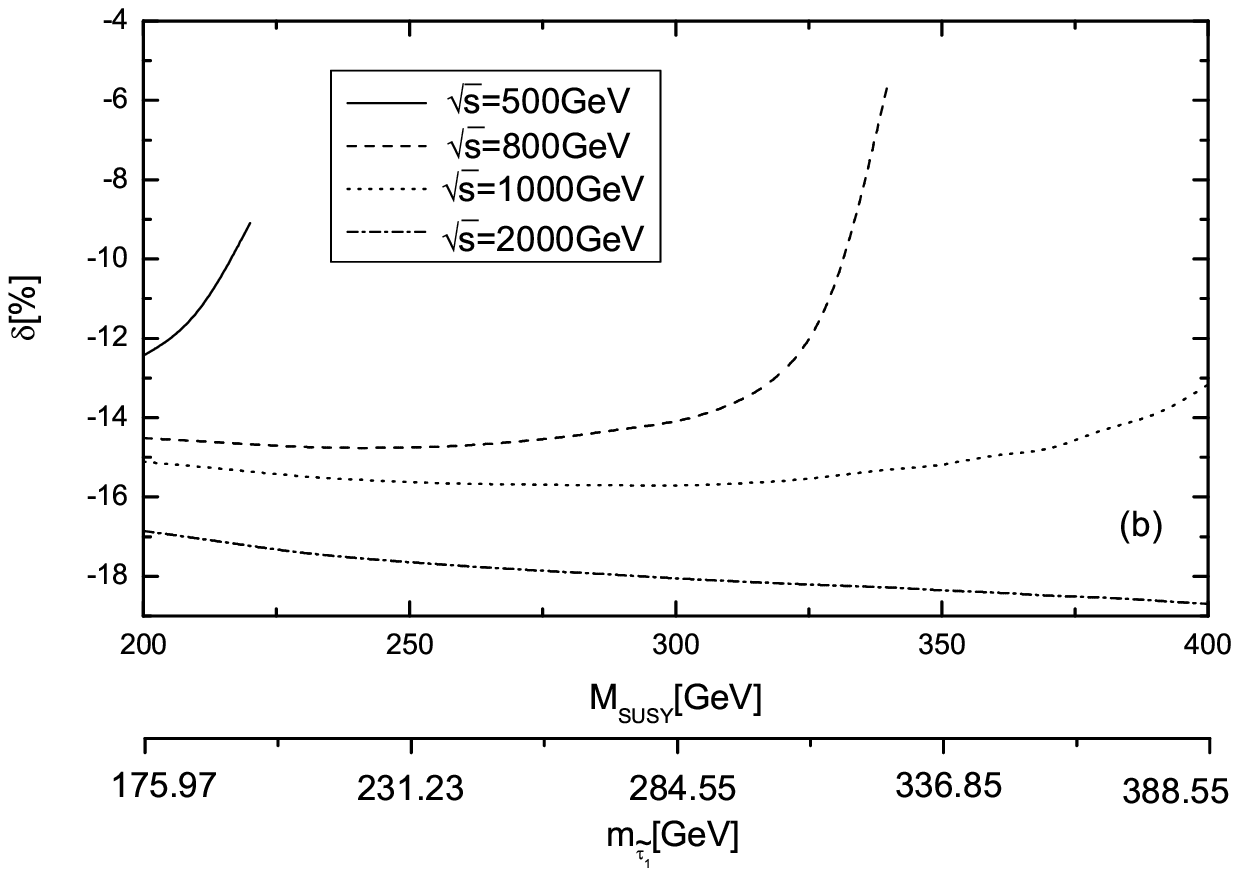} \epsfxsize = 8cm \epsfysize = 8cm
\epsfbox{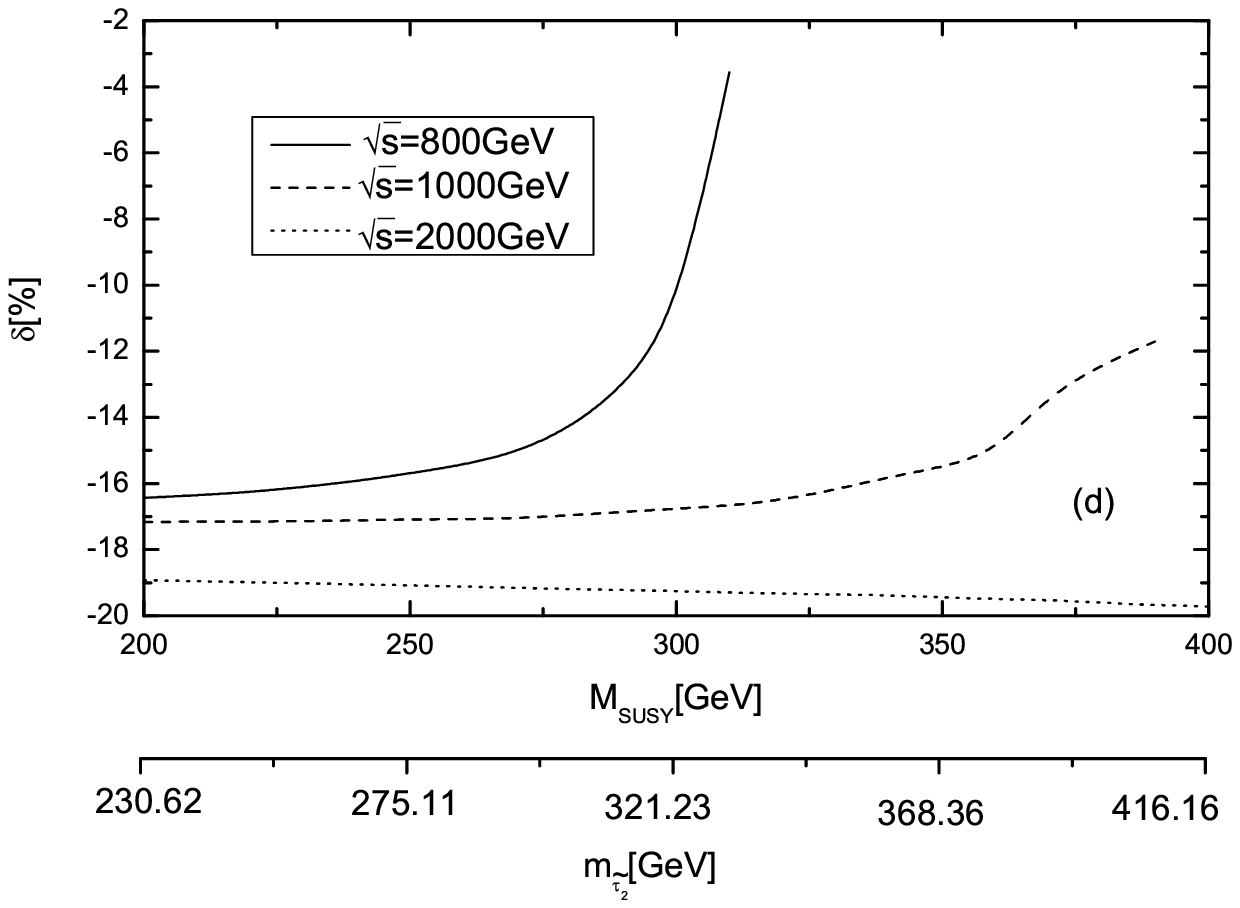} \epsfxsize = 8cm \epsfysize = 8cm
\vspace*{-0.3cm} \caption{ (a) The Born and full ${\cal
O}(\alpha_{ew})$ EW corrected cross sections for the $e^+e^- \to
\gamma \gamma \to \tilde{\tau}_1 \bar{\tilde{\tau_1}}$ process as
functions of the soft-breaking sfermion mass $M_{SUSY}$ with
$\sqrt s$ = 500 GeV, 800GeV, 1000 GeV, 2000 GeV, respectively. (b)
The full ${\cal O}(\alpha_{ew})$ EW relative corrections to the
$e^+e^- \to \gamma \gamma \to \tilde{\tau}_1 \bar{\tilde{\tau_1}}$
process as the functions of $M_{SUSY}$ with $\sqrt s$ = 500 GeV,
800 GeV, 1000 GeV, 2000 GeV, respectively. (c) The Born and full
${\cal O}(\alpha_{ew})$ EW corrected cross sections for the
$e^+e^- \to \gamma \gamma \to \tilde{\tau}_2 \bar{\tilde{\tau_2}}$
process as functions of the soft-breaking sfermion mass $M_{SUSY}$
with $\sqrt s$ = 500 GeV, 800GeV, 1000 GeV, 2000 GeV,
respectively. (d) The full ${\cal O}(\alpha_{ew})$ EW relative
corrections to the $e^+e^- \to \gamma \gamma \to \tilde{\tau}_2
\bar{\tilde{\tau_2}}$ process as the functions of $M_{SUSY}$ with
$\sqrt s$ = 500 GeV, 800 GeV, 1000 GeV, 2000 GeV, respectively. }
\end{figure}
\par
In the following discussion, we present some numerical results
about the parent process $e^+e^- \to \gamma \gamma \to \tilde{f}_i
\bar{\tilde{f_i}}(f=\tau,t,b,~i=1,2)$. For convenience, we denote
the cross sections of the parent process $e^+e^- \to \gamma \gamma
\to \tilde{f}_i \bar{\tilde{f_i}}$ containing the ${\cal
O}(\alpha_{ew})$ EW and ${\cal O}(\alpha_{s})$ QCD corrections as
$$
\sigma^{EW} = \sigma_0 + \Delta \sigma^{EW} = \sigma_0 (1+
\delta^{EW}), ~ ~ ~ \sigma^{QCD} = \sigma_0 + \Delta \sigma^{QCD}
= \sigma_0 (1+ \delta^{QCD})
$$
where $\delta^{EW}$ and $\delta^{QCD}$ are the ${\cal
O}(\alpha_{ew})$ EW and ${\cal O}(\alpha_{s})$ QCD relative
correction respectively. In the following numerical calculations,
we take the input parameters of higgsino-like data $Set3$, but let
$M_{SUSY}$ running from 200 Gev to 400 GeV. Fig.10(a) and
Fig.10(c) show the Born and full ${\cal O}(\alpha_{ew})$ EW
corrected cross sections for the $e^+e^- \to \gamma \gamma \to
\tilde{\tau}_1 \bar{\tilde{\tau_1}}$ and $e^+e^- \to \gamma \gamma
\to \tilde{\tau}_2 \bar{\tilde{\tau_2}}$ process as the functions
of the soft-breaking sfermion mass $M_{SUSY}$. In Fig.10(a), the
solid and dashed curves correspond to Born and full ${\cal
O}(\alpha_{ew})$ EW corrected cross sections for $\sqrt {s}$ = 500
GeV, 800 GeV, 1000 GeV, 2000 GeV, respectively . It is obvious
that all the curves for the Born and EW corrected cross sections
decrease rapidly as $M_{SUSY}$ going up from 200 to 400 GeV, but
the damping decrement is getting smaller with the increment of the
colliding energy $\sqrt{s}$. We can read from Fig.10(a) that the
values of the EW corrected cross sections decrease from 28.3 fb,
98.9 fb, 110 fb and 92.8 fb to 2.25 fb, 0.1 fb, 1.28 fb and 20.8
fb for $\sqrt s$ = 500 GeV, 800 GeV, 1000 GeV, 2000 GeV
respectively, when $M_{SUSY}$ increases from 200 GeV to 400 GeV.
In Fig.10(c), the curves show the Born and ${\cal O}(\alpha_{ew})$
EW corrected cross sections for $e^+e^- \to \gamma \gamma \to
\tilde{\tau}_2 \bar{\tilde{\tau_2}}$ process with $\sqrt s$ = 800
GeV, 1000 GeV, 2000 GeV respectively. All the curves have the
analogous tendency as the curves in Fig.10(a). The values of the
corrected cross sections in Fig.10(c) are smaller than those for
$e^+e^- \to \gamma \gamma \to \tilde{\tau}_1 \bar{\tilde{\tau_1}}$
in Fig.10(a). The EW relative corrections to the $e^+e^- \to
\gamma \gamma \to \tilde{\tau}_1 \bar{\tilde{\tau_1}}$ process as
the functions of $M_{SUSY}$ are depicted in Fig.10(b) for $\sqrt s$
= 500 GeV, 800 GeV, 1000 GeV and 2000 GeV. From this figure, we
can see that in the range of $M_{SUSY}$ = 200 GeV to 400 GeV,
this relative correction can reach $-5.46\%$ at the position of
$M_{SUSY}$ = 340 GeV when we take $\sqrt s$ = 800 GeV. If we take
$e^+e^-$ colliding energy $\sqrt s$ = 2 TeV, we can get $-18.69\%$
relative correction to $e^+e^- \to \gamma \gamma \to
\tilde{\tau}_1 \bar{\tilde{\tau_1}}$ process when $M_{SUSY}$ = 400
GeV. Fig.10(d) displays the EW relative corrections to $e^+e^- \to
\gamma \gamma \to \tilde{\tau}_2 \bar{\tilde{\tau_2}}$ process. We
can see from this figure that the numerical values of these
relative corrections increase rapidly from $-16.43\%$ to $-3.56\%$
when $M_{SUSY}$ goes up from 200 GeV to 310 GeV for $\sqrt s$ =
800 GeV, and from $-17.16\%$ to $-11.7\%$ when $M_{SUSY}$ increases
from 200 GeV to 390 GeV for $\sqrt s$ = 1000 GeV. But for $\sqrt
s$ = 2000 GeV the value of $\delta^{EW}$ is relative stable and
keeps in the range of [-18.93$\%$, -19.72$\%$] as $M_{SUSY}$ varying
from 200 GeV to 400 GeV.
\begin{figure}[hbtp]
\vspace*{-1cm} \epsfxsize = 8cm \epsfysize = 8cm
\epsfbox{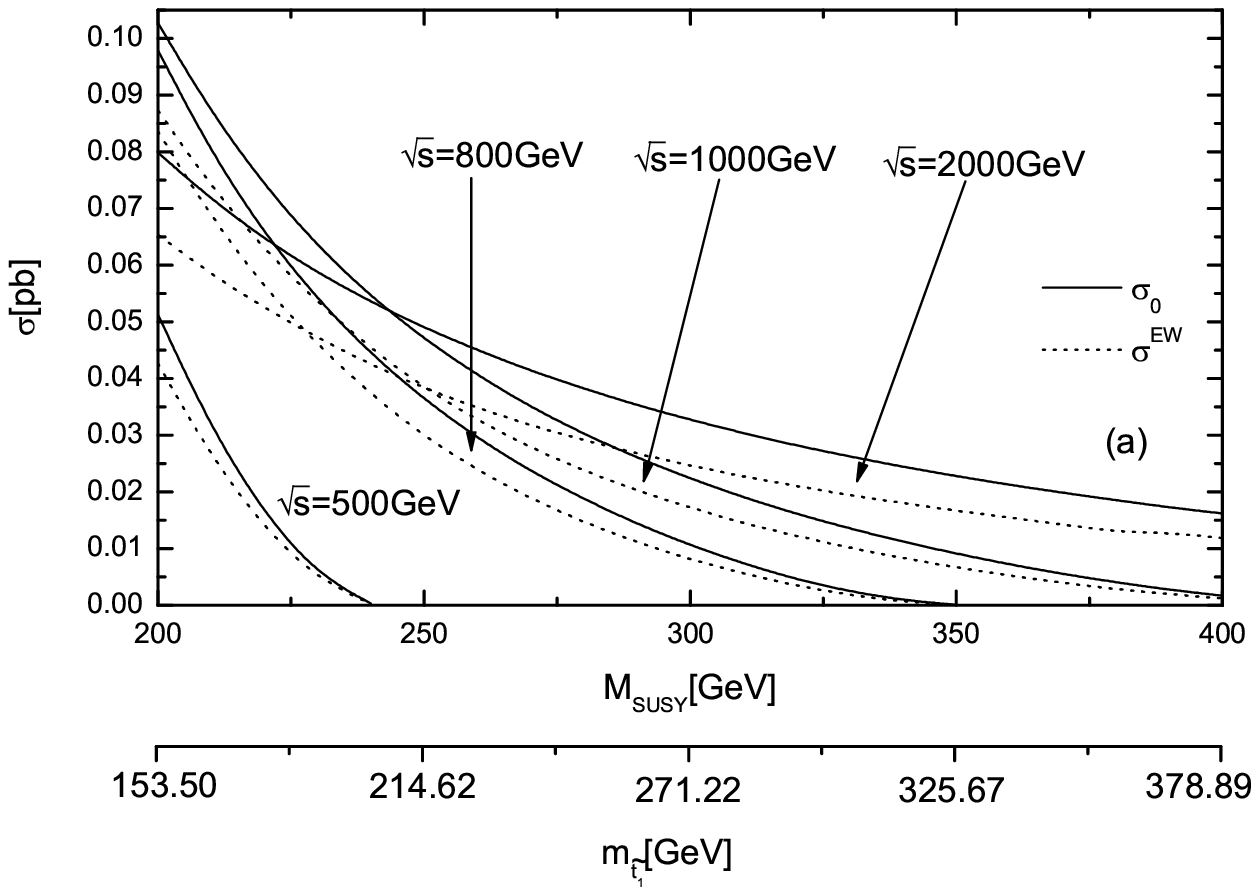} \epsfxsize = 8cm \epsfysize = 8cm
\epsfbox{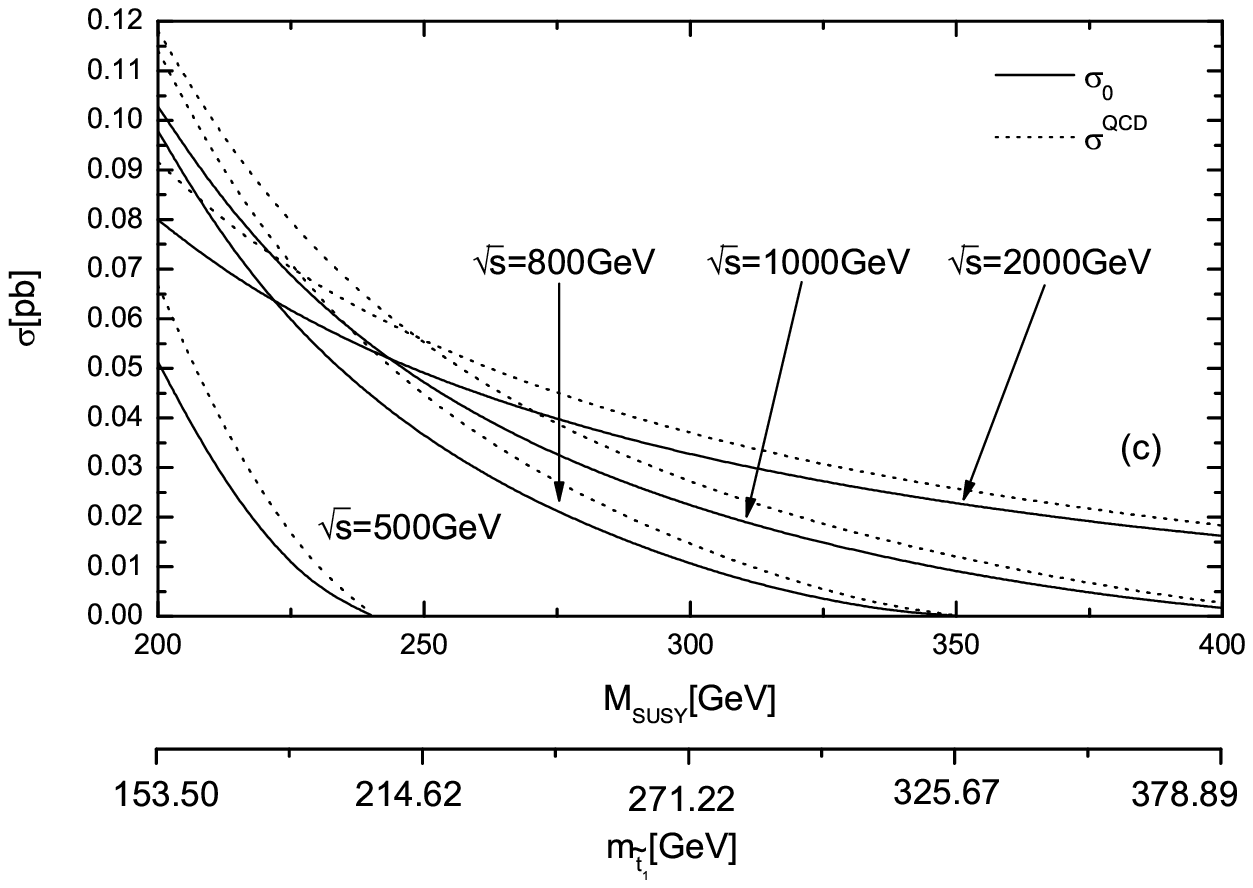} \epsfxsize = 8cm \epsfysize = 8cm
\epsfbox{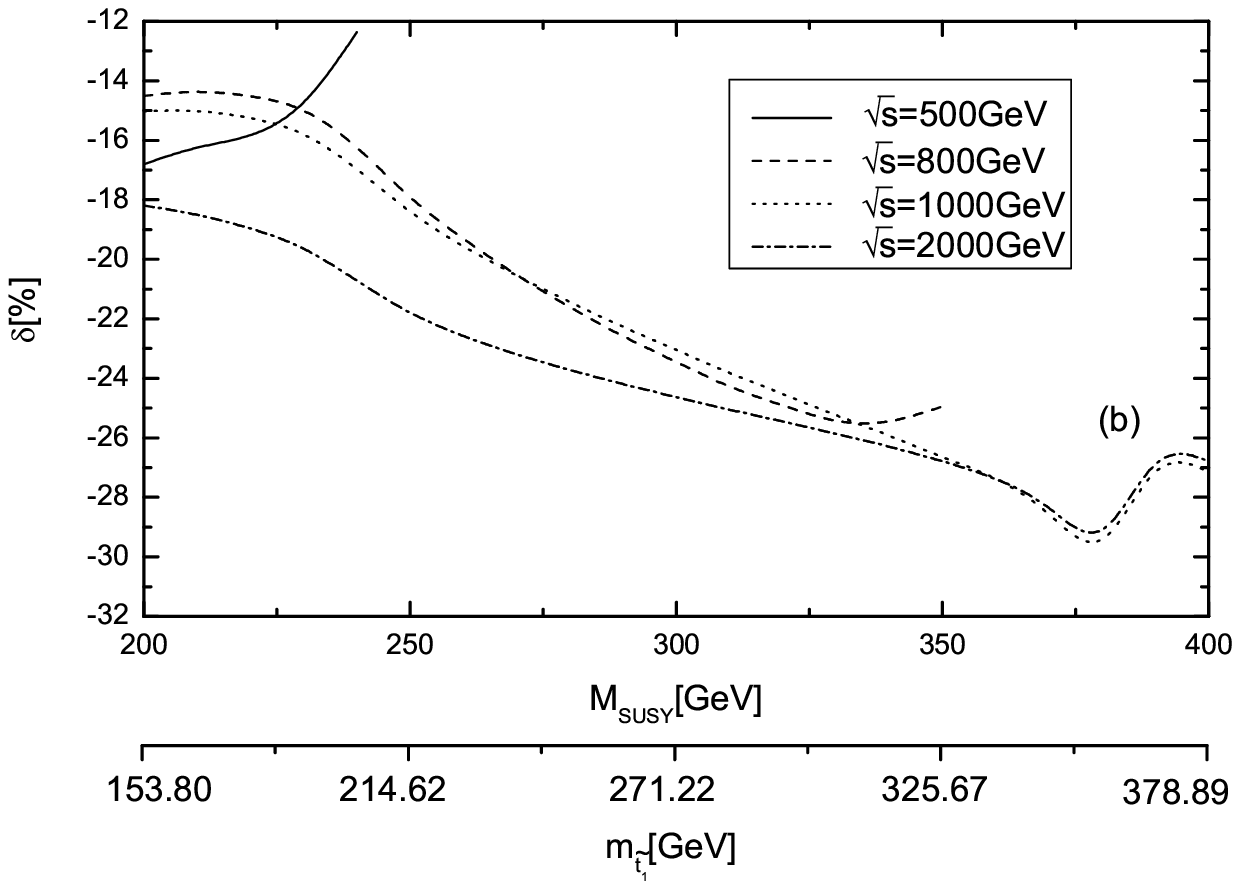} \epsfxsize = 8cm \epsfysize = 8cm
\epsfbox{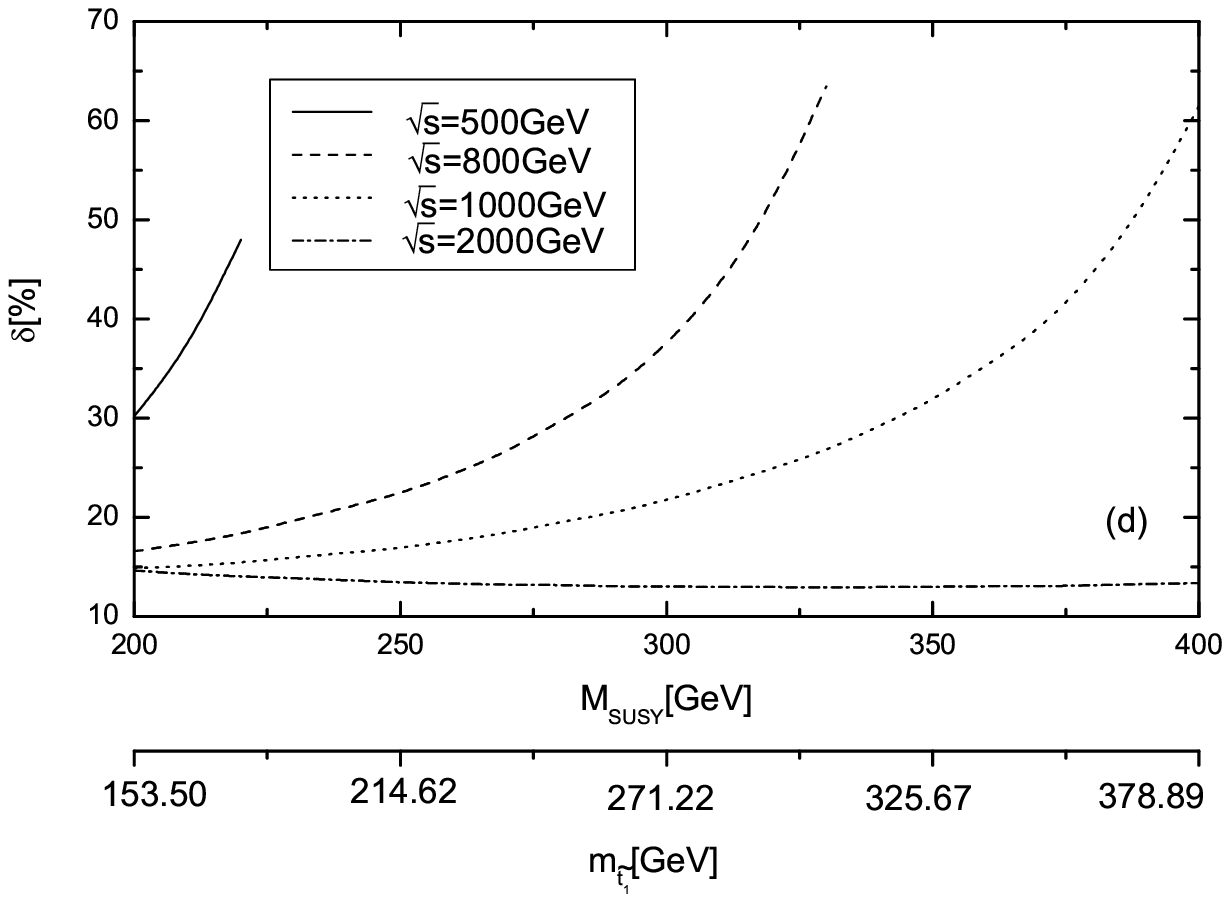} \vspace*{-0.3cm} \caption{ (a) The Born and
full ${\cal O}(\alpha_{ew})$ EW corrected cross sections for the
$e^+e^- \to \gamma \gamma \to \tilde{t}_1 \bar{\tilde{t_1}}$
process as the functions of the soft-breaking sfermion mass
$M_{SUSY}$ with $\sqrt s$ = 500, 800, 1000, 2000 GeV,
respectively. (b) The full ${\cal O}(\alpha_{ew})$ EW relative
corrections to the $e^+e^- \to \gamma \gamma \to \tilde{t}_1
\bar{\tilde{t_1}}$ process as the functions of $M_{SUSY}$ with
$\sqrt s$ = 500 GeV, 800 GeV, 1000 GeV, 2000 GeV, respectively.
(c) The Born and full ${\cal O}(\alpha_{s})$ QCD corrected cross
sections for the $e^+e^- \to \gamma \gamma \to \tilde{t}_1
\bar{\tilde{t_1}}$ process as the functions of the soft-breaking
sfermion mass $M_{SUSY}$ with $\sqrt s$ = 500, 800, 1000, 2000
GeV, respectively. (d) The full ${\cal O}(\alpha_{s})$ QCD
relative corrections to the $e^+e^- \to \gamma \gamma \to
\tilde{t}_1 \bar{\tilde{t_1}}$ process as the functions of
$M_{SUSY}$ with $\sqrt s$ = 500 GeV, 800 GeV, 1000 GeV, 2000 GeV,
respectively.}
\end{figure}
\begin{figure}[hbtp]
\vspace*{-1cm} \epsfxsize = 8cm \epsfysize = 8cm
\epsfbox{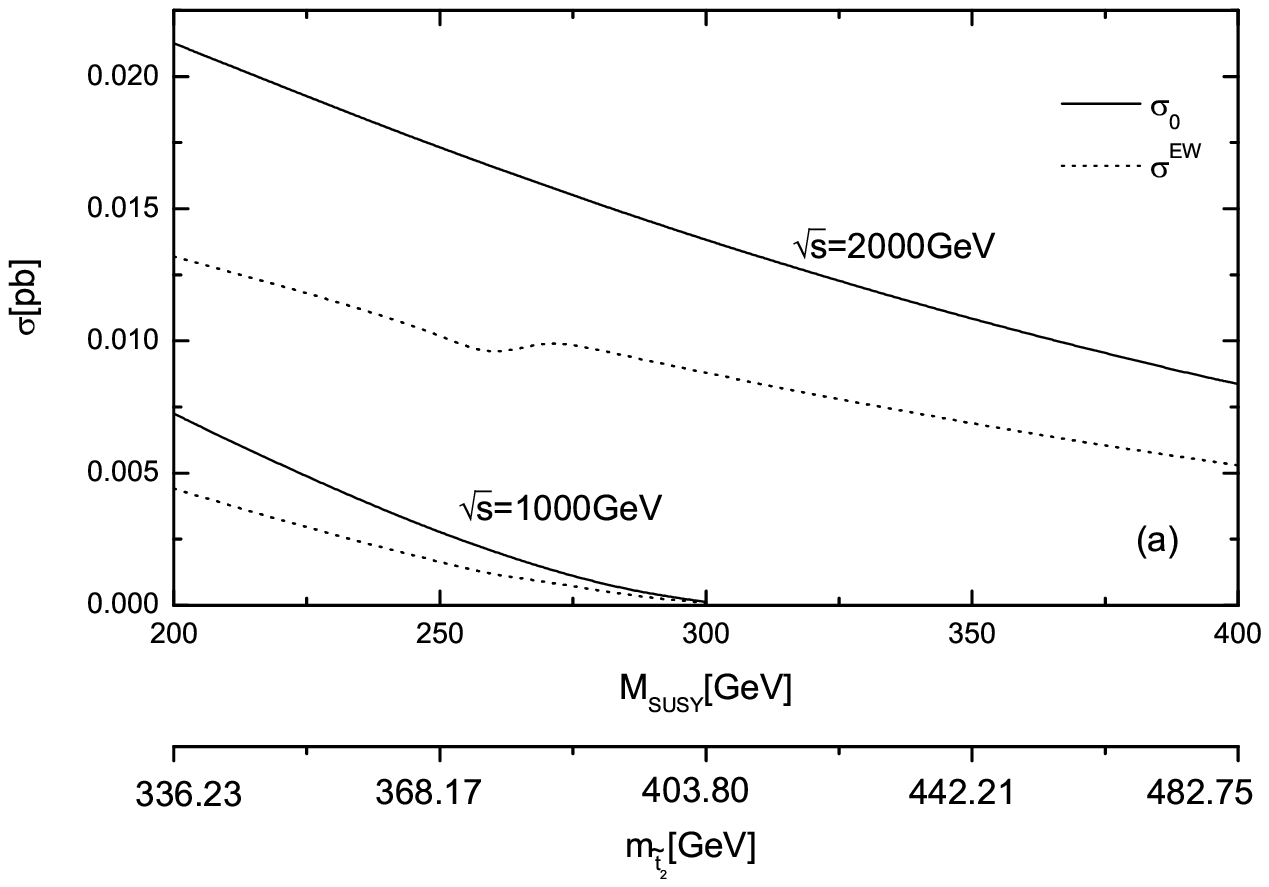} \epsfxsize = 8cm \epsfysize = 8cm
\epsfbox{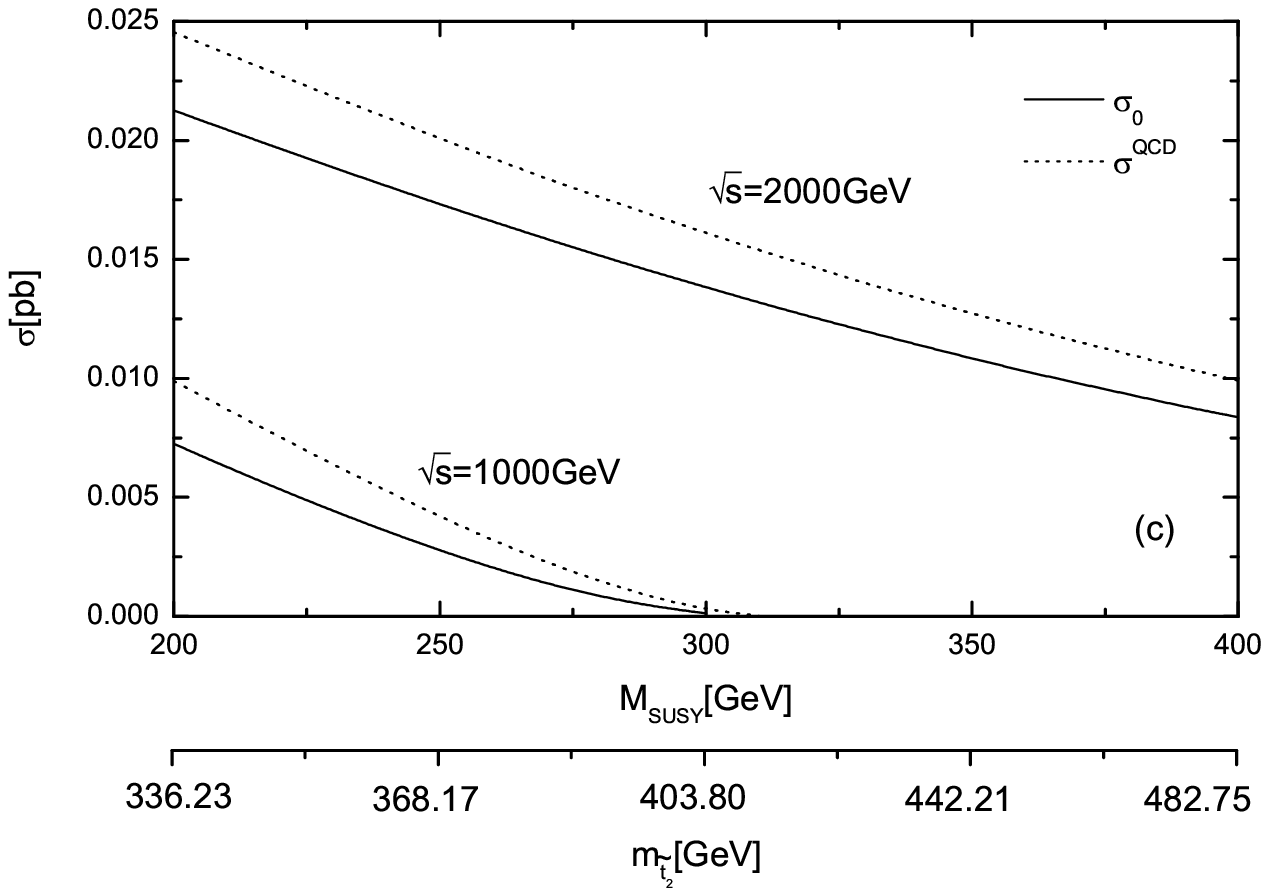} \epsfxsize = 8cm \epsfysize = 8cm
\epsfbox{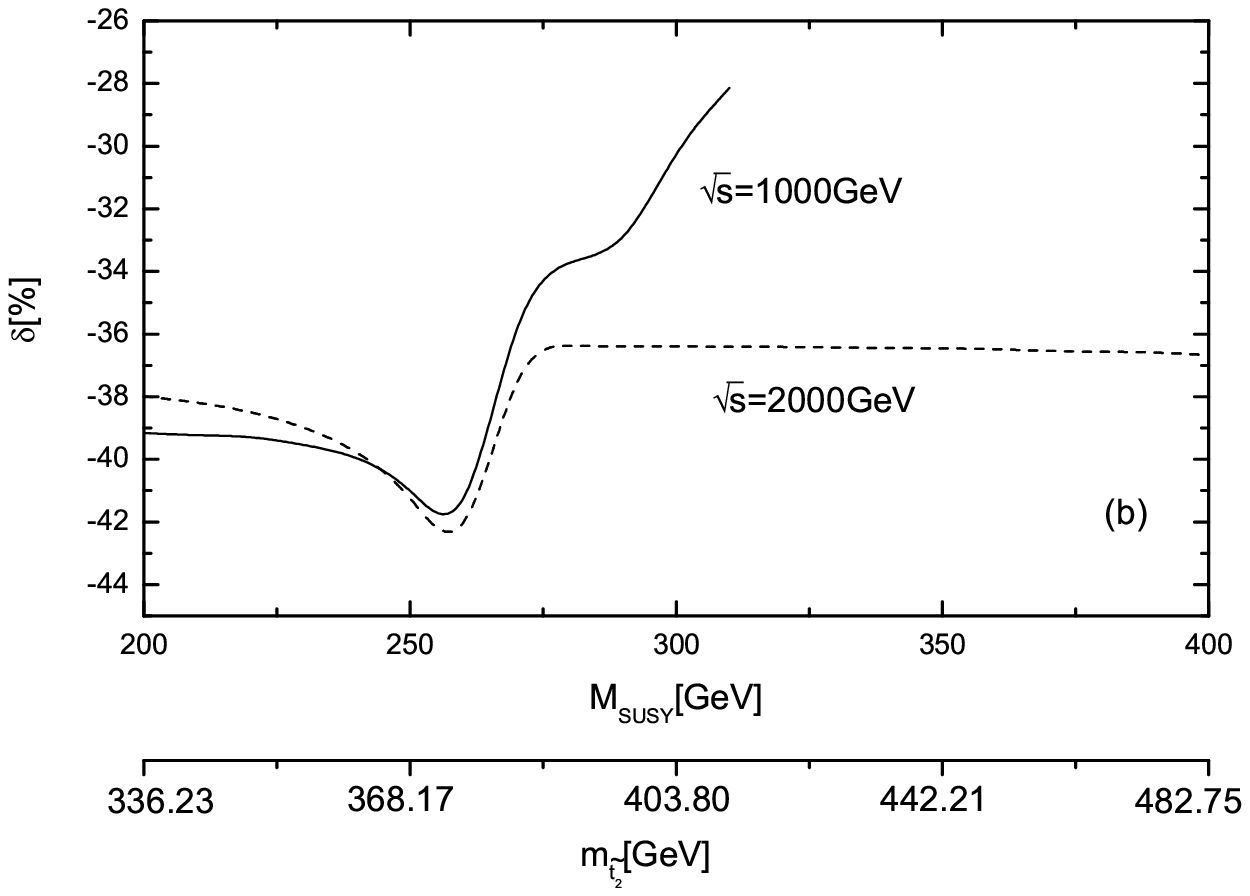} \epsfxsize = 8cm \epsfysize = 8cm
\epsfbox{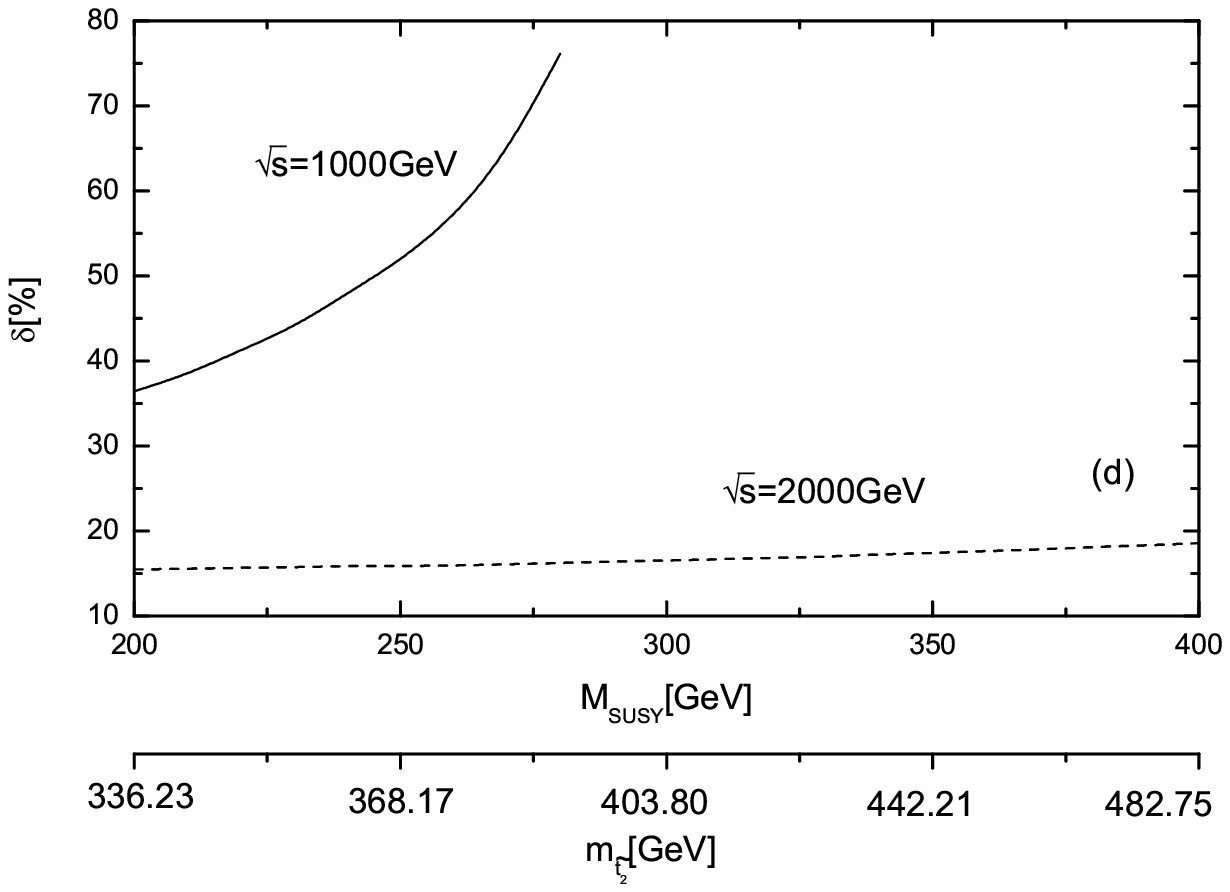} \vspace*{-0.3cm} \caption{ (a) The Born and
full ${\cal O}(\alpha_{ew})$ EW corrected cross sections for the
$e^+e^- \to \gamma \gamma \to \tilde{t}_2 \bar{\tilde{t_2}}$
process as the functions of the soft-breaking sfermion mass
$M_{SUSY}$ with $\sqrt s$ = 1000 GeV, 2000 GeV, respectively. (b)
The full ${\cal O}(\alpha_{ew})$ EW relative corrections to the
$e^+e^- \to \gamma \gamma \to \tilde{t}_2 \bar{\tilde{t_2}}$
process as the functions of $M_{SUSY}$ with $\sqrt s$ = 1000 GeV,
2000 GeV, respectively. (c) The Born and full ${\cal
O}(\alpha_{s})$ QCD corrected cross sections for the $e^+e^- \to
\gamma \gamma \to \tilde{t}_2 \bar{\tilde{t_2}}$ process as
functions of the soft-breaking sfermion mass $M_{SUSY}$ with
$\sqrt s$ = 1000 GeV, 2000 GeV, respectively. (d) The full ${\cal
O}(\alpha_{s})$ QCD relative corrections to the $e^+e^- \to \gamma
\gamma \to \tilde{t}_2 \bar{\tilde{t_2}}$ process as the functions
of $M_{SUSY}$ with $\sqrt s$ = 1000 GeV, 2000 GeV, respectively.}
\end{figure}
\par
The numerical results for the process $e^+e^- \to \gamma \gamma
\to \tilde{t}_1 \bar{\tilde{t_1}}$ are plotted in Fig.11.
Fig.11(a) and Fig.11(c) display the Born and full one-loop EW and
QCD corrected cross sections as the functions of $M_{SUSY}$ with
$\sqrt s$ = 500 GeV, 800 GeV, 1000 GeV, 2000 GeV, respectively. As
we expect, the curves in Fig.11(a) for the cross sections in Born
approximation and at ${\cal O}(\alpha_{ew})$ EW one-loop level,
have some similar behaviors with those for the
$\tilde{\tau}_1\bar{\tilde \tau}_1$ production process shown in
Fig.10(a). We can find from Fig.11(a) that the full ${\cal
O}(\alpha_{ew})$ EW corrected cross sections decrease from 87.3 fb
and 65.3 fb to 1.21 fb and 11.8 fb when $M_{SUSY}$ goes from 200
GeV to 400 GeV for $\sqrt s$ = 1000 GeV, 2000 GeV, respectively.
In Fig11(c) the solid and dotted curves correspond to the Born and
one-loop QCD corrected cross sections versus $M_{SUSY}$ with
$\sqrt s$ = 500 GeV, 800 GeV, 1000 GeV, 2000 GeV, respectively.
From this figure, we can see that the value of the QCD corrected
cross section reaches 118 fb for $\sqrt{s}$ = 1000 GeV with our
chosen parameters. In order to study the EW and QCD radiative
corrections more clearly, we plot the EW and QCD relative
corrections to the $e^+e^- \to \gamma \gamma \to \tilde{t}_1
\bar{\tilde{t_1}}$ process in Fig.11(b) and Fig.11(d). In
Fig.11(b), the resonance effect at the position of $M_{SUSY}$ =
386 GeV is due to the condition of $m_{\tilde{t}_1} \sim m_t +
m_{\tilde{\chi}^0_1}$. Fig.11(b) shows the EW relative corrections
for $\sqrt s$ = 1000 GeV and 2000 GeV can reach $-27.15\%$ and $-26.78\%$
at the position of $M_{SUSY}$ = 400 GeV. From Fig.11(d) we find the
curves for $\sqrt s$ = 500 GeV, 800 GeV and 1000 GeV go up fleetly with the
increment of $M_{SUSY}$, but for the curve of $\sqrt s$ = 2000 GeV
the relative correction is almost stable, varying in the range of
[$14.6\%$, $13.3\%$].
\par
The results for $e^+e^- \to \gamma \gamma \to \tilde{t}_2
\bar{\tilde{t_2}}$ are represented in Fig.12. Fig.12(a) shows the
plot of the Born and full one-loop EW corrected cross sections
versus $M_{SUSY}$. Fig.12(b) describes the EW relative corrections
as the functions of $M_{SUSY}$. The QCD corrected cross sections
and QCD relative corrections are plotted in Fig.12(c) and
Fig.12(d), respectively. In Fig.12(a) the EW corrected cross
section of $e^+e^- \to \gamma \gamma \to \tilde{t}_2
\bar{\tilde{t_2}}$ process decreases from 13.2 fb to 5.3 fb for
$\sqrt s$ = 2000 GeV, when $M_{SUSY}$ increases from 200 GeV to
400 GeV. At the position of $M_{SUSY}$ = 266 GeV in Fig.12(a),
there is a dithering on the curve of $\sqrt s$ = 2000 GeV, which
is due to resonance effect of $m_{\tilde{t}_2} \sim m_t +
m_{\tilde{\chi}^0_2}$. The resonance effect on the EW relative
correction curves at the position near $M_{SUSY}$ = 266 GeV is
also shown in Fig.12(b). In Fig.12(d), the QCD relative
corrections for $\sqrt s$ = 1000 GeV are rather larger and vary in
the region between 36.4$\%$ and 76.1$\%$ with the increment of
$M_{SUSY}$, but the QCD relative corrections for $\sqrt s$ = 2000
GeV are smaller, and have the values in the range between 15.5$\%$
and 18.6$\%$.
\begin{figure}[hbtp]
\vspace*{-1cm} \epsfxsize = 8cm \epsfysize = 8cm
\epsfbox{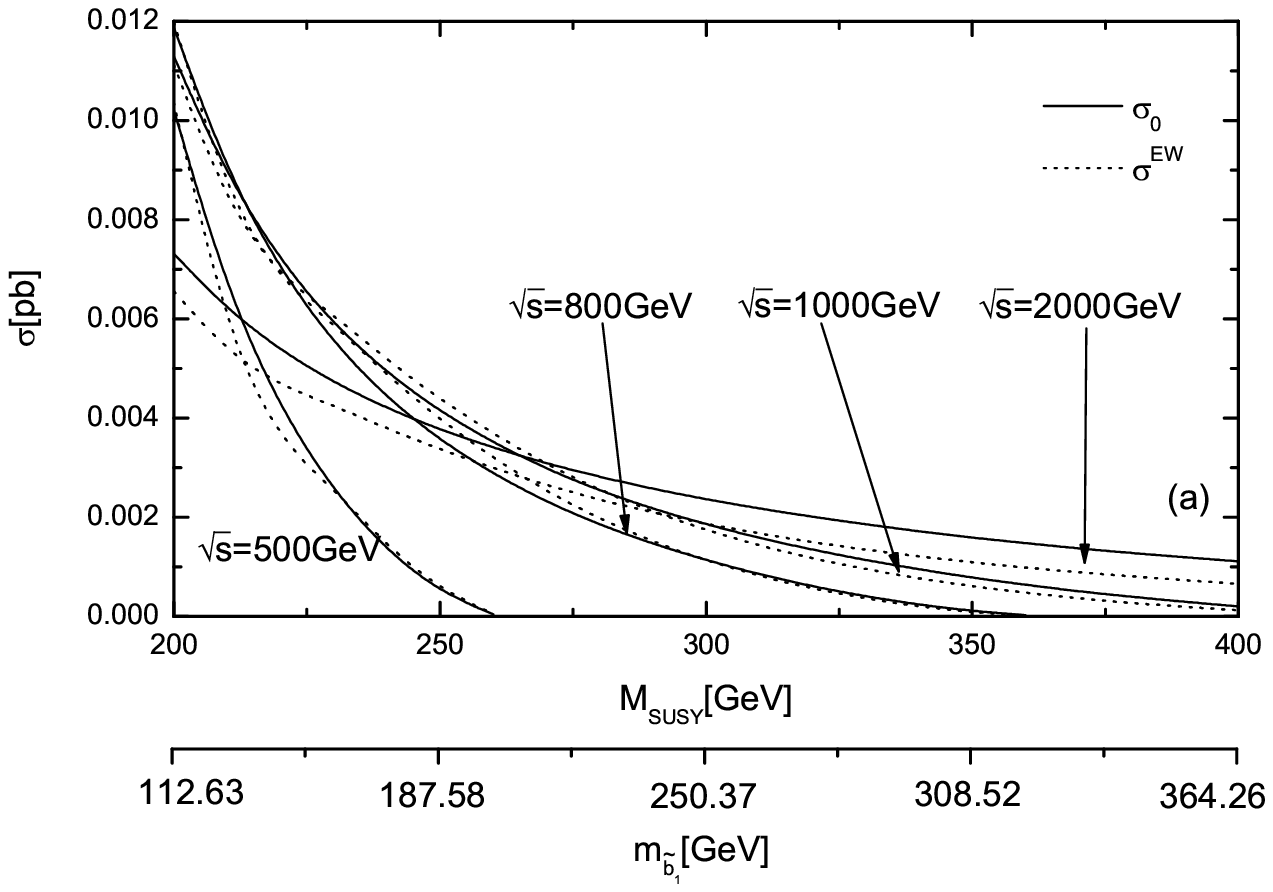} \epsfxsize = 8cm \epsfysize = 8cm
\epsfbox{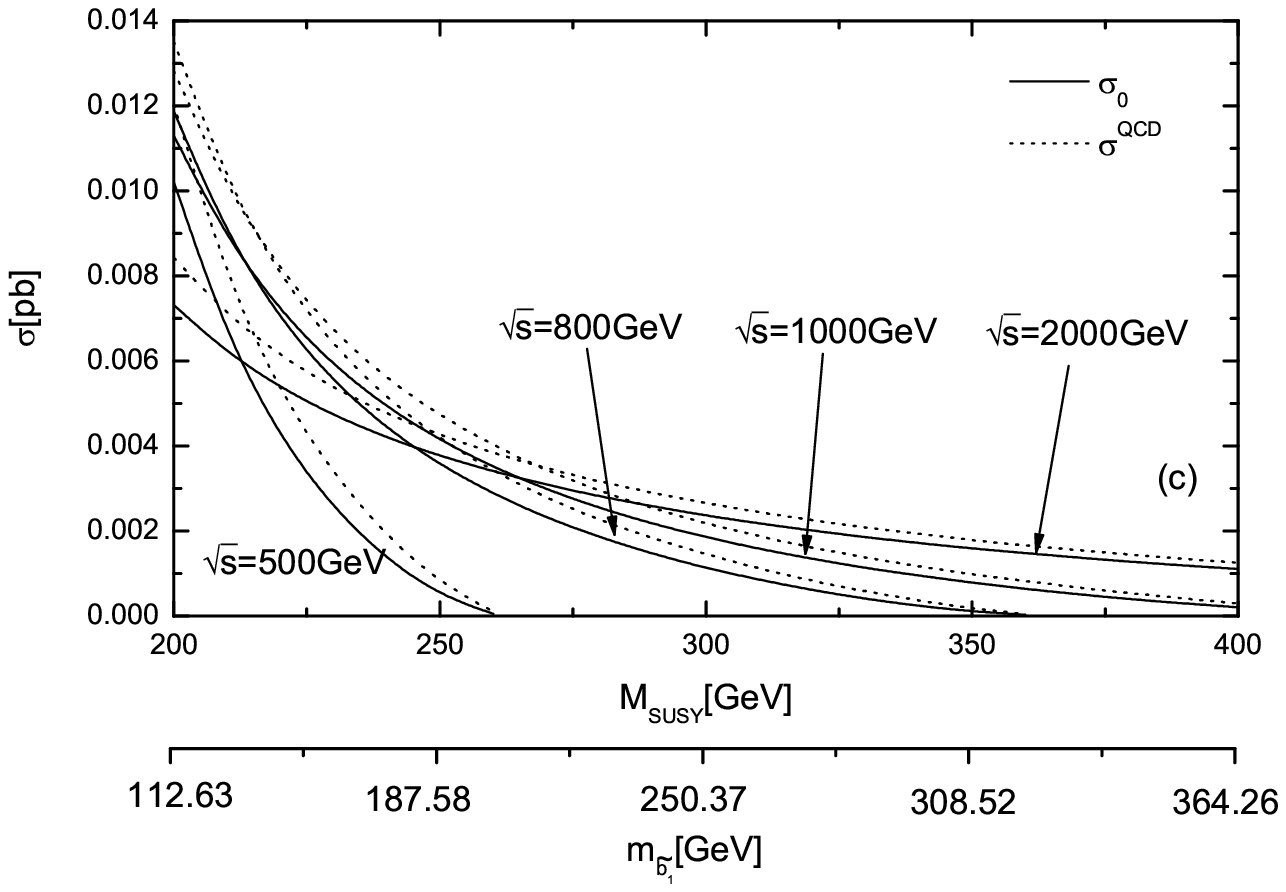} \epsfxsize = 8cm \epsfysize = 8cm
\epsfbox{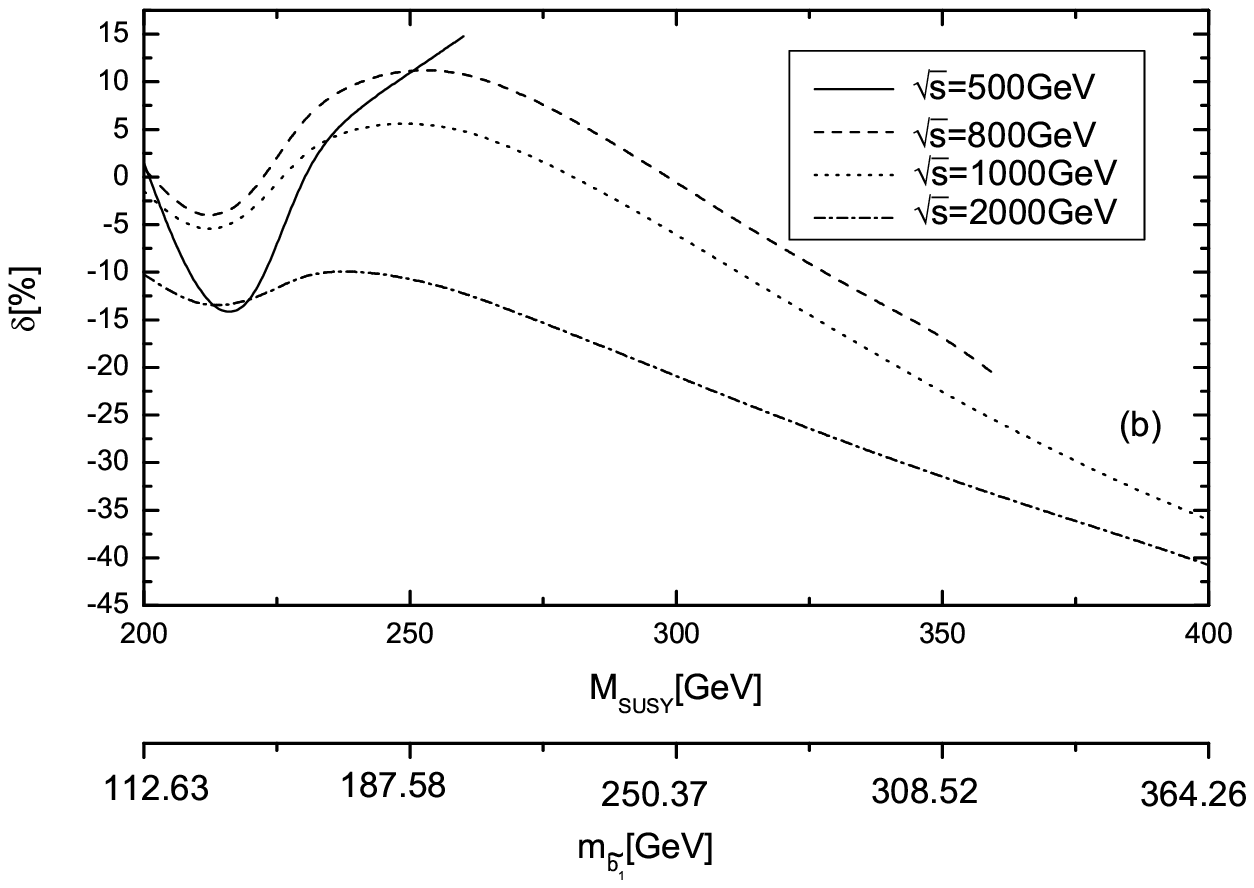} \epsfxsize = 8cm \epsfysize = 8cm
\epsfbox{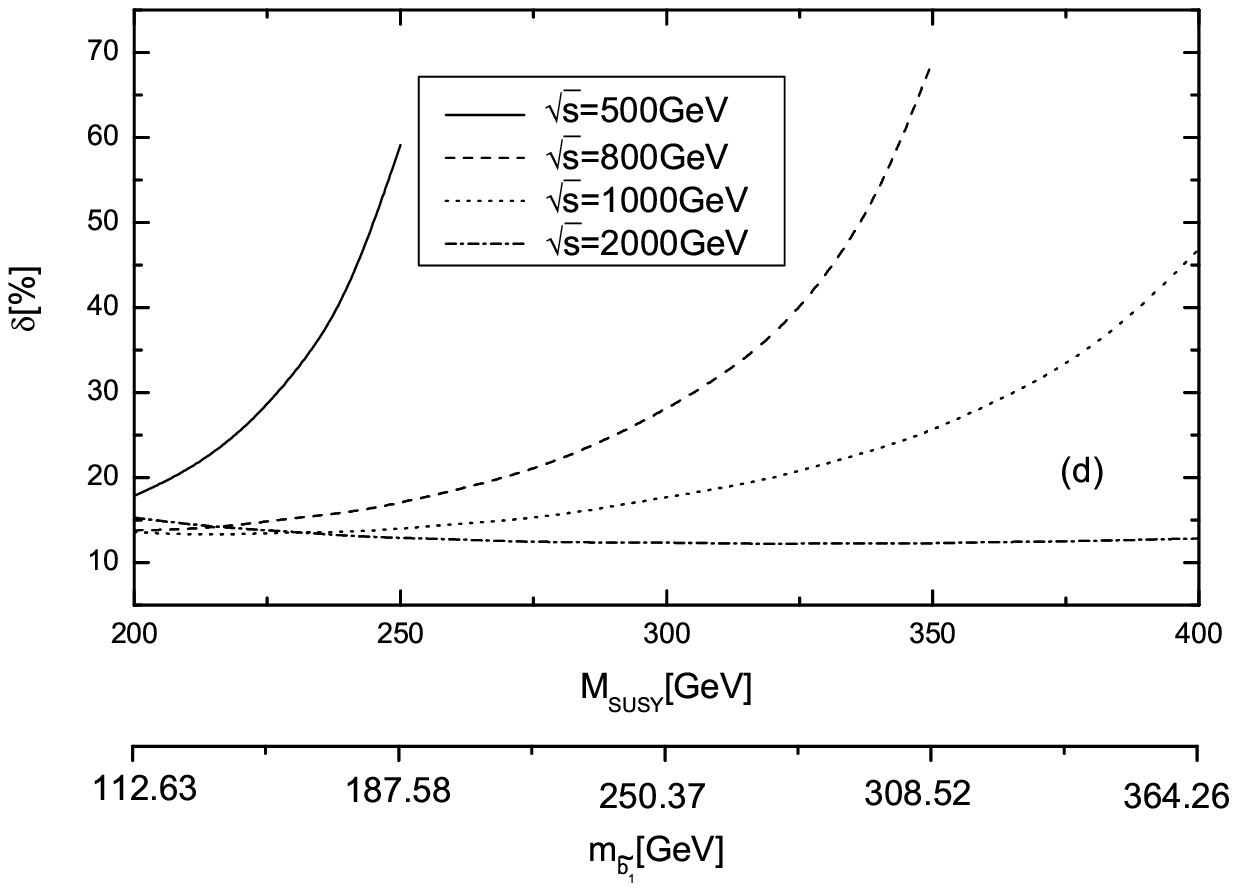} \vspace*{-0.3cm} \caption{ (a) The Born and
full ${\cal O}(\alpha_{ew})$ EW corrected cross sections for the
$e^+e^- \to \gamma \gamma \to \tilde{b}_1 \bar{\tilde{b_1}}$
process as the functions of the soft-breaking sfermion mass
$M_{SUSY}$ with $\sqrt s$ = 500 GeV, 800 GeV, 1000 GeV, 2000 GeV,
respectively. (b) The full ${\cal O}(\alpha_{ew})$ EW relative
corrections to the $e^+e^- \to \gamma \gamma \to \tilde{b}_1
\bar{\tilde{b_1}}$ process as the functions of $M_{SUSY}$ with
$\sqrt s$ = 500 GeV, 800 GeV, 1000 GeV, 2000 GeV, respectively.
(c) The Born and full ${\cal O}(\alpha_{s})$ QCD corrected cross
sections for the $e^+e^- \to \gamma \gamma \to \tilde{b}_1
\bar{\tilde{b_1}}$ process as the functions of the soft-breaking
sfermion mass $M_{SUSY}$ with $\sqrt s$ = 500 GeV, 800 GeV, 1000
GeV, 2000 GeV, respectively. (d) The full ${\cal O}(\alpha_{s})$
QCD relative corrections to the $e^+e^- \to \gamma \gamma \to
\tilde{b}_1 \bar{\tilde{b_1}}$ process as the functions of
$M_{SUSY}$ with $\sqrt s$ = 500 GeV, 800 GeV, 1000 GeV, 2000 GeV,
respectively.}
\end{figure}
\begin{figure}[hbtp]
\vspace*{-1cm} \epsfxsize = 8cm \epsfysize = 8cm
\epsfbox{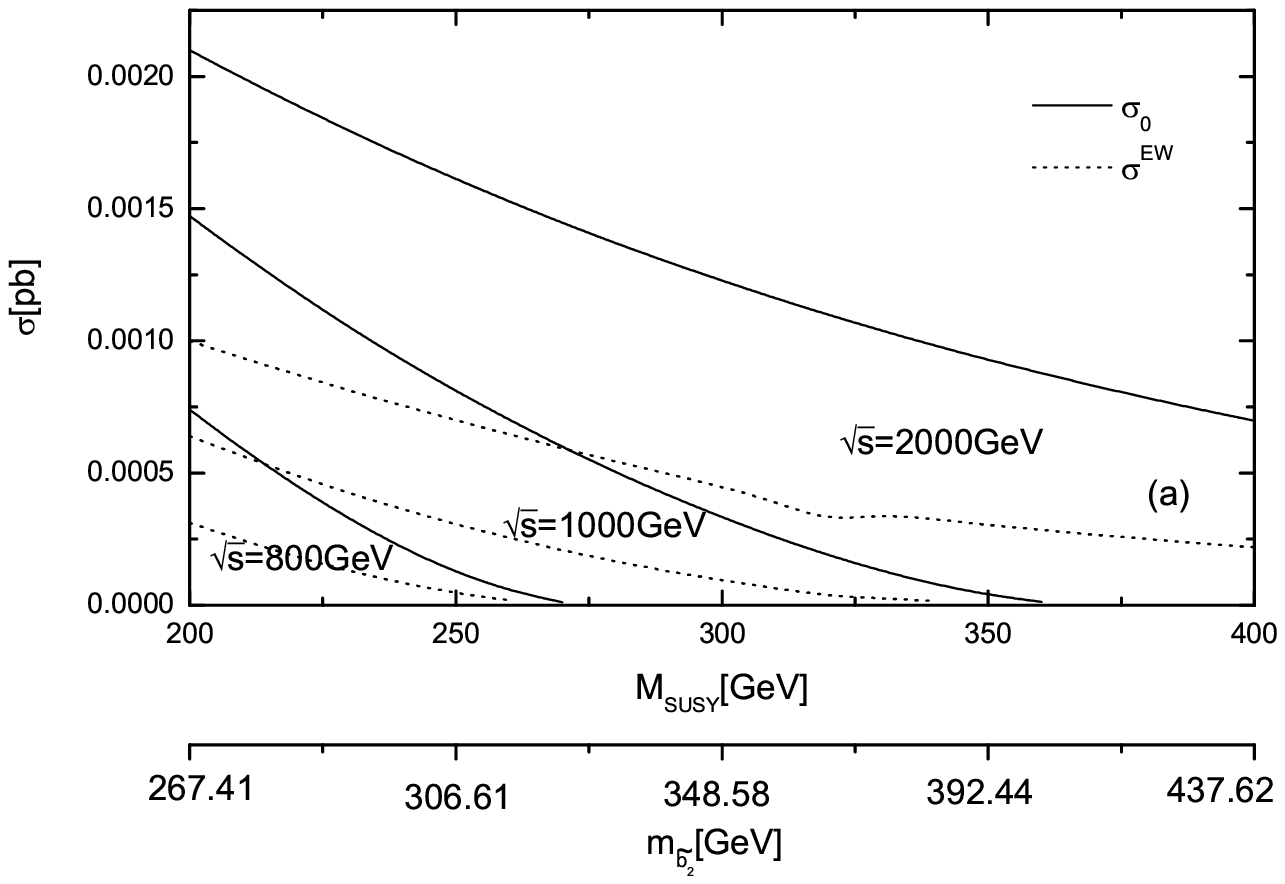} \epsfxsize = 8cm \epsfysize = 8cm
\epsfbox{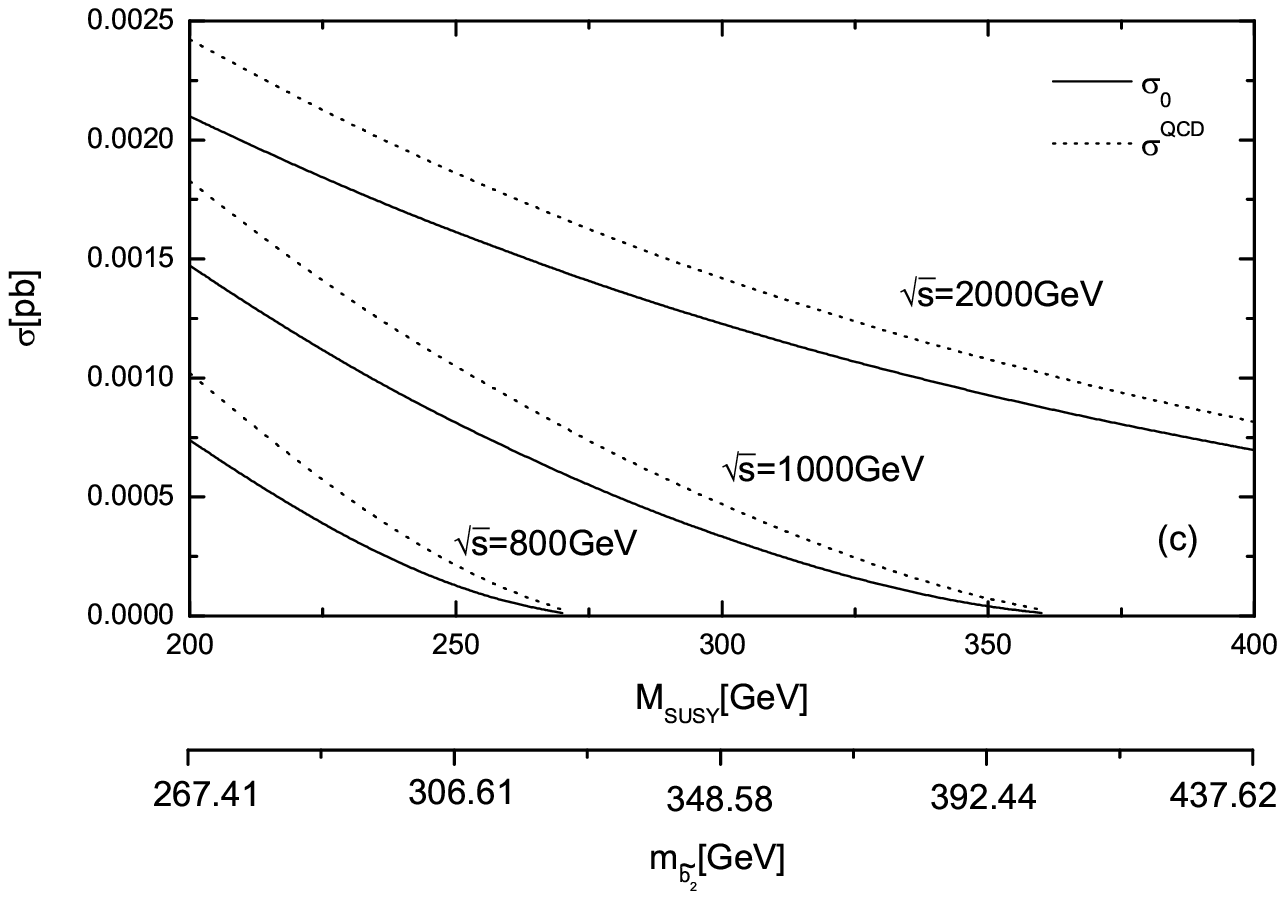} \epsfxsize = 8cm \epsfysize = 8cm
\epsfbox{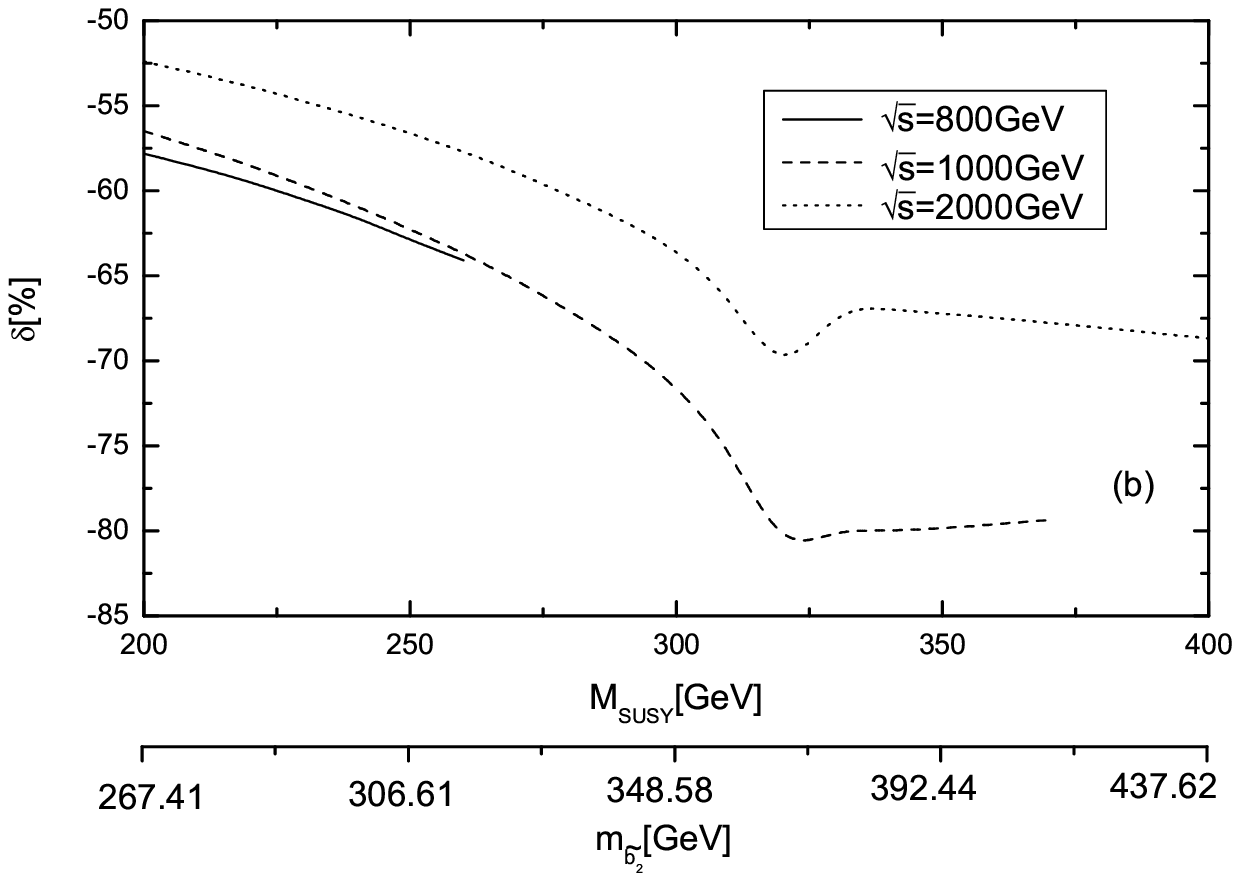} \epsfxsize = 8cm \epsfysize = 8cm
\epsfbox{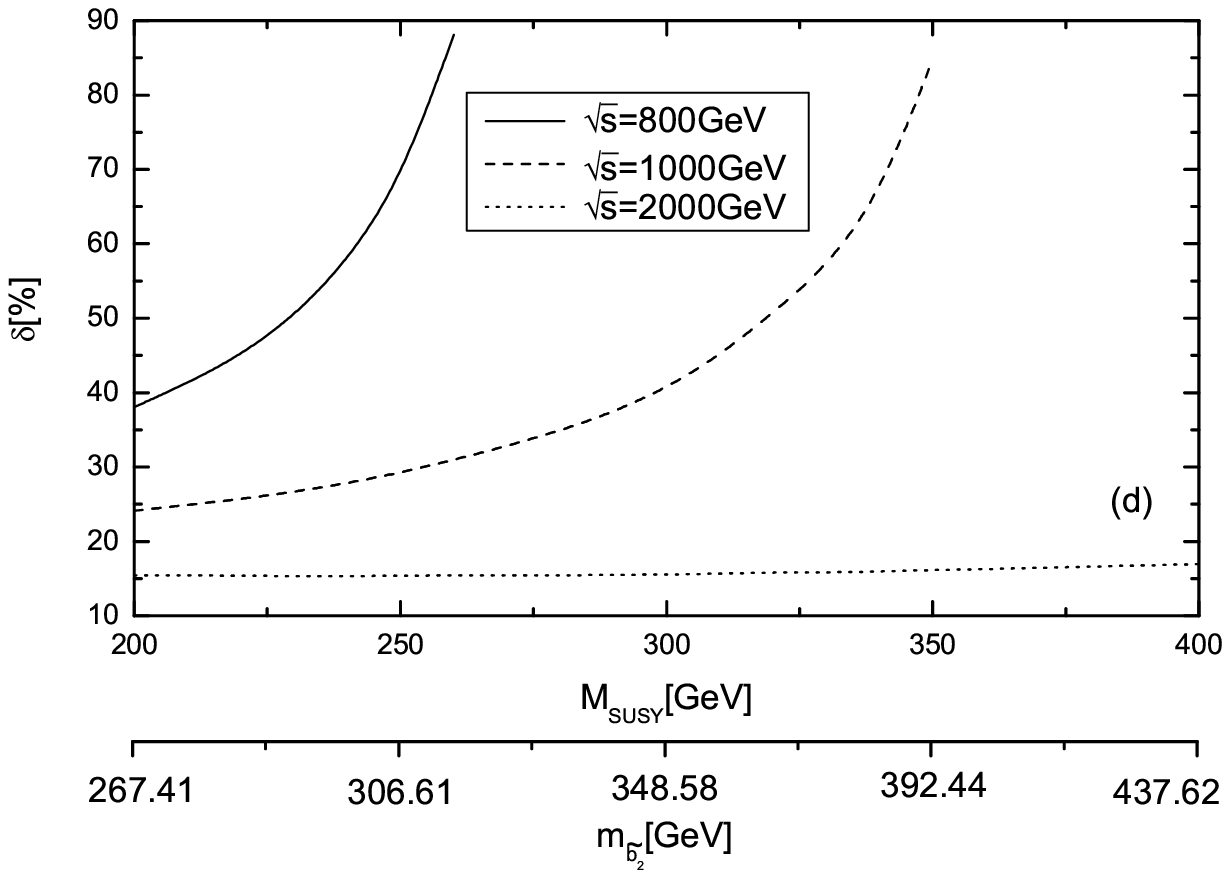} \vspace*{-0.3cm} \caption{ (a) The Born and
full ${\cal O}(\alpha_{ew})$ EW corrected cross sections for the
$e^+e^- \to \gamma \gamma \to \tilde{b}_2 \bar{\tilde{b_2}}$
process as the functions of the soft-breaking sfermion mass
$M_{SUSY}$ with $\sqrt s$ = 800 GeV, 1000 GeV, 2000 GeV,
respectively. (b) The full ${\cal O}(\alpha_{ew})$ EW relative
corrections to the $e^+e^- \to \gamma \gamma \to \tilde{b}_2
\bar{\tilde{b_2}}$ process as the functions of $M_{SUSY}$ with
$\sqrt s$ = 800 GeV, 1000 GeV, 2000 GeV, respectively. (c) The
Born and full ${\cal O}(\alpha_{s})$ QCD corrected cross sections
for the $e^+e^- \to \gamma \gamma \to \tilde{b}_2
\bar{\tilde{b_2}}$ process as the functions of the soft-breaking
sfermion mass $M_{SUSY}$ with $\sqrt s$ = 800 GeV, 1000 GeV, 2000
GeV, respectively. (d) The full ${\cal O}(\alpha_{s})$ QCD
relative corrections to the $e^+e^- \to \gamma \gamma \to
\tilde{b}_2 \bar{\tilde{b_2}}$ process as the functions of
$M_{SUSY}$ with $\sqrt s$ = 800 GeV, 1000 GeV, 2000 GeV,
respectively.}
\end{figure}
\par
We also present the results of the $e^+e^- \to \gamma \gamma \to
\tilde{b}_1 \bar{\tilde{b}}_1$ process in Fig.13. Fig.13(a) and
Fig.13(c) show the Born and full one-loop EW and QCD corrected
cross sections, respectively. In Fig.13(a) and Fig.13(c) we find
that all the curves of cross sections at the Born level and
involving EW and QCD one-loop contributions, decrease with the
increment of the $M_{SUSY}$. For example, when $M_{SUSY}$ varies
from 200 to 400 GeV, the two curves of the cross sections
including full one-loop EW corrections for $\sqrt s$ = 1000 GeV
and 2000 GeV in Fig.13(a), goes down from 11.1 fb and 6.6 fb to 0.13
fb and 0.65 fb respectively. While the two curves in Fig.13(c),
which represent the cross sections including QCD corrections for
$\sqrt s$ = 1000 GeV and 2000 GeV, decrease from 12.8 fb and 8.4
fb to 0.3 fb and 1.3 fb, respectively. Fig.13(b) shows the full
one-loop EW relative corrections to the $e^+e^- \to \gamma \gamma
\to \tilde{b}_1 \bar{\tilde{b}}_1$ process, as the functions of
$M_{SUSY}$ for $\sqrt s$ = 500, 800, 1000, 2000 GeV respectively.
We see that there occur the resonance effects on each curve at the
position of $m_{\tilde b_1} \sim m_{H^+}-m_{\tilde t_1} \sim 133$
GeV (corresponding to $M_{SUSY}\sim 212$ GeV). Fig.13(d) displays
the full one-loop QCD relative corrections as the functions of
$M_{SUSY}$ with $\sqrt s$ = 500, 800, 1000, 2000 GeV,
respectively. The QCD relative corrections can be rather larger
for $\sqrt s $ = 500 GeV and 800 GeV, and can reach 59$\%$ and
69$\%$ at the positions of $M_{SUSY}$ = 250 GeV and 350 GeV,
respectively.
\par
Finally, we present the Born and full one-loop EW and QCD
corrections to $e^+e^- \to \gamma \gamma \to \tilde{b}_2
\bar{\tilde{b}}_2$ process in Fig.14. Comparing Fig.14(a) with
Fig.13(a), we can see that the Born and full one-loop EW corrected
cross sections for $\tilde{b}_2 \bar{\tilde{b}}_2$ pair production
are smaller than the corresponding ones for $\tilde{b}_1
\bar{\tilde{b_1}}$ pair production because of
$m_{\tilde{b}_2}>m_{\tilde{b}_1}$. But the EW relative corrections
to $e^+e^- \to \gamma \gamma \to \tilde{b}_2 \bar{\tilde{b}}_2$
process are rather large and can be comparable with their QCD
corrections as shown in Fig.14(b) and Fig.14(d). We can also find
that when $M_{SUSY}$ = 200 GeV, the EW relative corrections are
-57.8$\%$ (for $\sqrt s$ = 800 GeV), -56.5$\%$ (for $\sqrt s$ =
1000 GeV) and -52.4$\%$ (for $\sqrt s$ = 2000 GeV), and when
$M_{SUSY}$ = 200 GeV the QCD relative corrections are 38.1$\%$ and
24.1$\%$ for $\sqrt s$ = 800 GeV and 1000 GeV respectively, but
for $\sqrt s$ = 2000 GeV the QCD relative correction varies in the
a small range of [15.4$\%$, 16.9$\%$].

\vskip 5mm
\section{Summery}
In this paper, we have calculated the full ${\cal O}(\alpha_{ew})$
EW and ${\cal O}(\alpha_{s})$ QCD contributions to the third
generation scalar fermion
($\tilde{\tau}_i,\tilde{t}_i,\tilde{b}_i,i=1,2$) pair production
in $\gamma\gamma$ collision at an $e^+e^-$ collider. The
calculation of the radiative corrections was carried out
analytically and numerically. The numerical results were discussed
in conditions of both gaugino-like and higgsino-like input
parameter scenarios. Our investigation shows that the full ${\cal
O}(\alpha_{ew})$ EW relative corrections to both subprocesses and
parent processes are typically of the order 10$\sim$30$\%$, and
the EW relative corrections to the squark pair production can be
comparable with the ${\cal O}(\alpha_{s})$ QCD contributions in
some parameter space, especially in high $\gamma\gamma$ and
$e^+e^-$ colliding energy regions. For example, the EW relative
corrections can reach -50$\%$ and -40.76$\%$ at
$\sqrt{\hat{s}}(\sqrt{s})$ = 2000 GeV to the subprocess $\gamma
\gamma \to \tilde{b}_1 \bar{\tilde{b}}_1$ and its parent process
$e^+e^- \to \gamma \gamma \to \tilde{b}_1 \bar{\tilde{b_1}}$,
respectively. We find the full ${\cal O}(\alpha_{s})$ QCD
corrections to these squark pair production subprocesses are also
large under our input data sets, for example, the QCD relative
correction is 28.7$\%$ for $\gamma \gamma \to \tilde{b}_1
\bar{\tilde{b_1}}$ subprocess with $\sqrt{s}=$ 2000 GeV and $Set4$
parameters. In conclusion, our numerical results have indicated
that the full ${\cal O}(\alpha_{ew})$ EW corrections to $e^+e^-
\to \gamma\gamma \to \tilde{f}_i
\bar{\tilde{f}}_i~(f=t,b,\tau,i=1,2)$ processes and ${\cal
O}(\alpha_{s})$ QCD corrections to $e^+e^- \to \gamma\gamma \to
\tilde{q}_i \bar{\tilde{q}}_i~(q=t,b,i=1,2)$ processes, give
substantial contributions in some parameter space. Therefore,
these radiative corrections cannot be neglected in considering the
third generation sfermion pair productions in $\gamma\gamma$
collision mode at future linear colliders.

\vskip 5mm \noindent{\large\bf Acknowledgments:}

This work was supported in part by the National Natural Science
Foundation of China and special fund sponsored by China Academy of
Science.

\vskip 10mm
\begin{flushleft} {\bf Figure Captions} \end{flushleft}
\par
{\bf Figure 1} The lowest order Feynman diagrams for the subprocess
$\gamma \gamma \to \tilde{f}_i \bar{\tilde{f_i}} ~ (f= \tau, t,
b)$.
\par
{\bf Figure 2} The real photon emission diagrams for the process
$\gamma \gamma \to \tilde{f}_i \bar{\tilde{f_i}} \gamma ~ (f=
\tau, t, b)$.
\par
{\bf Figure 3} The real gluon emission diagrams for the process
$\gamma \gamma \to \tilde{q}_i \bar{\tilde{q_i}} g ~ (q= t, b)$.
\par
{\bf Figure 4} The full ${\cal O}(\alpha_{s})$ QCD corrections to
$\gamma \gamma \to \tilde{t}_1 \bar{\tilde{t_1}}$ as a function of
the soft gluon cutoff $\Delta E_g /E_b$ in conditions of
$\sqrt{\hat{s}}=500~GeV$ and $Set1$ parameters.
\par
{\bf Figure 5(a)} The Born and full ${\cal O}(\alpha_{ew})$ EW
corrected cross sections for the $\gamma \gamma \to \tilde{\tau}_1
\bar{\tilde{\tau_1}}$ subprocess as the functions of c.m.s. energy
of $\gamma \gamma$ collider $\sqrt {\hat s}$ with four different
data sets, respectively.
\par
{\bf Figure 5(b)}The full ${\cal O}(\alpha_{ew})$ EW relative
correction to the $\gamma \gamma \to \tilde{\tau}_1
\bar{\tilde{\tau_1}}$ subprocess. The solid ,dashed, dotted and
dash-dotted curves correspond to four different data set cases,
respectively.
\par
{\bf Figure 5(c)}The Born and full ${\cal O}(\alpha_{ew})$ EW
corrected cross sections for the $\gamma \gamma \to \tilde{\tau}_2
\bar{\tilde{\tau_2}}$ subprocess as the functions of c.m.s. energy
of $\gamma \gamma$ collider $\sqrt {\hat s}$ with four different
data sets, respectively.
\par
{\bf Figure 5(d)} The full ${\cal O}(\alpha_{ew})$ EW relative
correction to the $\gamma \gamma \to \tilde{\tau}_2
\bar{\tilde{\tau_2}}$ subprocess. The solid ,dashed, dotted and
dash-dotted curves correspond to four different data set cases,
respectively.
\par
{\bf Figure 6(a)} The Born and full ${\cal O}(\alpha_{ew})$ EW
corrected cross sections for the $\gamma \gamma \to \tilde t_1
\bar{\tilde{t_1}}$ subprocess as the functions of c.m.s. energy of
$\gamma \gamma$ collider $\sqrt {\hat s}$ with four different data
sets, respectively.
\par
{\bf Figure 6(b)} The full ${\cal O}(\alpha_{ew})$ EW relative
correction to $\gamma \gamma \to \tilde t_1 \bar{\tilde{t_1}}$
subprocess. Four different curves correspond to four different
data sets, respectively.
\par
{\bf Figure 6(c)} The Born and full ${\cal O}(\alpha_{s})$ QCD
corrected cross sections for the $\gamma \gamma \to \tilde t_1
\bar{\tilde{t_1}}$ subprocess as the functions of c.m.s. energy of
$\gamma \gamma$ collider $\sqrt {\hat s}$ with four different data
sets, respectively. .
\par
{\bf Figure 6(d)}  The full ${\cal O}(\alpha_{s})$ QCD
relative correction to $\gamma \gamma \to \tilde t_1
\bar{\tilde{t_1}}$ subprocess.
\par
{\bf Figure 7(a)} The Born and full ${\cal O}(\alpha_{ew})$ EW
corrected cross sections for the $\gamma \gamma \to \tilde t_2
\bar{\tilde{t_2}}$ subprocess as the functions of c.m.s. energy of
$\gamma \gamma$ collider $\sqrt {\hat s}$ with four different data
sets, respectively.
\par
{\bf Figure 7(b)} The full ${\cal O}(\alpha_{ew})$ EW relative
correction to $\gamma \gamma \to \tilde t_2 \bar{\tilde{t_2}}$
subprocess. Four different curves correspond to four different
data sets, respectively.
\par
{\bf Figure 7(c)} The Born and full ${\cal O}(\alpha_{s})$ QCD
corrected cross sections for the $\gamma \gamma \to \tilde t_2
\bar{\tilde{t_2}}$ subprocess as the functions of c.m.s. energy of
$\gamma \gamma$ collider $\sqrt {\hat s}$ with four different data
sets, respectively.
\par
{\bf Figure 7(d)} The full ${\cal O}(\alpha_{s})$ QCD
relative correction to $\gamma \gamma \to \tilde t_2
\bar{\tilde{t_2}}$ subprocess.
\par
{\bf Figure 8(a)} The Born and full ${\cal O}(\alpha_{ew})$ EW
corrected cross sections for the $\gamma \gamma \to \tilde b_1
\bar{\tilde{b_1}}$ subprocess as the functions of c.m.s. energy
$\sqrt {\hat s}$ with four data sets, respectively.
\par
{\bf Figure 8(b)} The full one-loop ${\cal
O}(\alpha_{ew})$ EW relative correction to $\gamma \gamma \to
\tilde b_1 \bar{\tilde{b_1}}$ subprocess.
\par
{\bf Figure 8(c)} The Born and full ${\cal O}(\alpha_{s})$ QCD
corrected cross sections for the $\gamma \gamma \to \tilde b_1
\bar{\tilde{b_1}}$ subprocess as the functions of c.m.s. energy
$\sqrt {\hat s}$ with four data sets, respectively.
\par
{\bf Figure 8(d)} The full ${\cal O}(\alpha_{s})$ QCD relative correction
to $\gamma \gamma \to \tilde b_1 \bar{\tilde{b_1}}$ subprocess.
\par
{\bf Figure 9(a)} The Born and full ${\cal O}(\alpha_{ew})$ EW
corrected cross sections for the $\gamma \gamma \to \tilde b_2
\bar{\tilde{b_2}}$ subprocess as the functions of c.m.s. energy
$\sqrt {\hat s}$ with four data sets, respectively.
\par
{\bf Figure 9(b)} The full ${\cal
O}(\alpha_{ew})$ EW relative corrections to $\gamma \gamma \to
\tilde b_2 \bar{\tilde{b_2}}$ subprocess.
\par
{\bf Figure 9(c)} The Born and full ${\cal O}(\alpha_{s})$ QCD
corrected cross sections for the $\gamma \gamma \to \tilde b_2
\bar{\tilde{b_2}}$ subprocess as the functions of c.m.s. energy
$\sqrt {\hat s}$ with four data sets, respectively.
\par
{\bf Figure 9(d)} The full ${\cal O}(\alpha_{s})$ QCD relative correction
to $\gamma \gamma \to \tilde b_2 \bar{\tilde{b_2}}$ subprocess.
\par
{\bf Figure 10(a)} The Born and full ${\cal
O}(\alpha_{ew})$ EW corrected cross sections for the $e^+e^- \to
\gamma \gamma \to \tilde{\tau}_1 \bar{\tilde{\tau_1}}$ process as
functions of the soft-breaking sfermion mass $M_{SUSY}$ with
$\sqrt s$ = 500 GeV, 800GeV, 1000 GeV, 2000 GeV, respectively.
\par
{\bf Figure 10(b)} The full ${\cal O}(\alpha_{ew})$ EW relative
corrections to the $e^+e^- \to \gamma \gamma \to \tilde{\tau}_1
\bar{\tilde{\tau_1}}$ process as the functions of $M_{SUSY}$ with
$\sqrt s$ = 500 GeV, 800 GeV, 1000 GeV, 2000 GeV, respectively.
\par
{\bf Figure 10(c)} he Born and full ${\cal O}(\alpha_{ew})$ EW
corrected cross sections for the $e^+e^- \to \gamma \gamma \to
\tilde{\tau}_2 \bar{\tilde{\tau_2}}$ process as functions of the
soft-breaking sfermion mass $M_{SUSY}$ with $\sqrt s$ = 500 GeV,
800GeV, 1000 GeV, 2000 GeV, respectively.
\par
{\bf Figure 10(d)} The full ${\cal O}(\alpha_{ew})$ EW relative
corrections to the $e^+e^- \to \gamma \gamma \to \tilde{\tau}_2
\bar{\tilde{\tau_2}}$ process as the functions of $M_{SUSY}$ with
$\sqrt s$ = 500 GeV, 800 GeV, 1000 GeV, 2000 GeV, respectively.
\par
{\bf Figure 11(a)} The Born and full ${\cal O}(\alpha_{ew})$ EW
corrected cross sections for the $e^+e^- \to \gamma \gamma \to
\tilde{t}_1 \bar{\tilde{t_1}}$ process as the functions of the
soft-breaking sfermion mass $M_{SUSY}$ with $\sqrt s$ = 500, 800,
1000, 2000 GeV, respectively.
\par
{\bf Figure 11(b)} The full ${\cal O}(\alpha_{ew})$ EW relative
corrections to the $e^+e^- \to \gamma \gamma \to \tilde{t}_1
\bar{\tilde{t_1}}$ process as the functions of $M_{SUSY}$ with
$\sqrt s$ = 500 GeV, 800 GeV, 1000 GeV, 2000 GeV, respectively.
\par
{\bf Figure 11(c)} The Born and full ${\cal O}(\alpha_{s})$ QCD
corrected cross sections for the $e^+e^- \to \gamma \gamma \to
\tilde{t}_1 \bar{\tilde{t_1}}$ process as the functions of the
soft-breaking sfermion mass $M_{SUSY}$ with $\sqrt s$ = 500, 800,
1000, 2000 GeV, respectively.
\par
{\bf Figure 11(d)} The full ${\cal O}(\alpha_{s})$ QCD relative
corrections to the $e^+e^- \to \gamma \gamma \to \tilde{t}_1
\bar{\tilde{t_1}}$ process as the functions of $M_{SUSY}$ with
$\sqrt s$ = 500 GeV, 800 GeV, 1000 GeV, 2000 GeV, respectively.
\par
{\bf Figure 12(a)} The Born and full ${\cal O}(\alpha_{ew})$ EW
corrected cross sections for the $e^+e^- \to \gamma \gamma \to
\tilde{t}_2 \bar{\tilde{t_2}}$ process as the functions of the
soft-breaking sfermion mass $M_{SUSY}$ with $\sqrt s$ = 1000 GeV,
2000 GeV, respectively.
\par
{\bf Figure 12(b)} The full ${\cal O}(\alpha_{ew})$ EW relative
corrections to the $e^+e^- \to \gamma \gamma \to \tilde{t}_2
\bar{\tilde{t_2}}$ process as the functions of $M_{SUSY}$ with
$\sqrt s$ = 1000 GeV, 2000 GeV, respectively.
\par
{\bf Figure 12(c)} The Born and full ${\cal O}(\alpha_{s})$ QCD
corrected cross sections for the $e^+e^- \to \gamma \gamma \to
\tilde{t}_2 \bar{\tilde{t_2}}$ process as the functions of the
soft-breaking sfermion mass $M_{SUSY}$ with $\sqrt s$ = 1000 GeV,
2000 GeV, respectively.
\par
{\bf Figure 12(d)} The full ${\cal O}(\alpha_{s})$ QCD relative
corrections to the $e^+e^- \to \gamma \gamma \to \tilde{t}_2
\bar{\tilde{t_2}}$ process as the functions of $M_{SUSY}$ with
$\sqrt s$ = 1000 GeV, 2000 GeV, respectively.
\par
{\bf Figure 13(a)} The Born and full ${\cal O}(\alpha_{ew})$ EW
corrected cross sections for the $e^+e^- \to \gamma \gamma \to
\tilde{b}_1 \bar{\tilde{b_1}}$ process as the functions of the
soft-breaking sfermion mass $M_{SUSY}$ with $\sqrt s$ = 500 GeV,
800 GeV, 1000 GeV, 2000 GeV, respectively.
\par
{\bf Figure 13(b)} The full ${\cal O}(\alpha_{ew})$ EW relative
corrections to the $e^+e^- \to \gamma \gamma \to \tilde{b}_1
\bar{\tilde{b_1}}$ process as the functions of $M_{SUSY}$ with
$\sqrt s$ = 500 GeV, 800 GeV, 1000 GeV, 2000 GeV, respectively.
\par
{\bf Figure 13(c)} The Born and full ${\cal O}(\alpha_{s})$ QCD
corrected cross sections for the $e^+e^- \to \gamma \gamma \to
\tilde{b}_1 \bar{\tilde{b_1}}$ process as the functions of the
soft-breaking sfermion mass $M_{SUSY}$ with $\sqrt s$ = 500 GeV,
800 GeV, 1000 GeV, 2000 GeV, respectively.
\par
{\bf Figure 13(d)} The full ${\cal O}(\alpha_{s})$ QCD relative
corrections to the $e^+e^- \to \gamma \gamma \to \tilde{b}_1
\bar{\tilde{b_1}}$ process as the functions of $M_{SUSY}$ with
$\sqrt s$ = 500 GeV, 800 GeV, 1000 GeV, 2000 GeV, respectively.
\par
{\bf Figure 14(a)} The Born and full ${\cal O}(\alpha_{ew})$ EW
corrected cross sections for the $e^+e^- \to \gamma \gamma \to
\tilde{b}_2 \bar{\tilde{b_2}}$ process as the functions of the
soft-breaking sfermion mass $M_{SUSY}$ with $\sqrt s$ = 800 GeV,
1000 GeV, 2000 GeV, respectively.
\par
{\bf Figure 14(b)} The full ${\cal O}(\alpha_{ew})$ EW relative
corrections to the $e^+e^- \to \gamma \gamma \to \tilde{b}_2
\bar{\tilde{b_2}}$ process as the functions of $M_{SUSY}$ with
$\sqrt s$ = 800 GeV, 1000 GeV, 2000 GeV, respectively.
\par
{\bf Figure 14(c)} The Born and full ${\cal O}(\alpha_{s})$ QCD
corrected cross sections for the $e^+e^- \to \gamma \gamma \to
\tilde{b}_2 \bar{\tilde{b_2}}$ process as the functions of the
soft-breaking sfermion mass $M_{SUSY}$ with $\sqrt s$ = 800 GeV,
1000 GeV, 2000 GeV, respectively.
\par
{\bf Figure 14(d)} The full ${\cal O}(\alpha_{s})$ QCD relative
corrections to the $e^+e^- \to \gamma \gamma \to \tilde{b}_2
\bar{\tilde{b_2}}$ process as the functions of $M_{SUSY}$ with
$\sqrt s$ = 800 GeV, 1000 GeV, 2000 GeV, respectively.

\vskip 5mm
\begin{thebibliography}{s25}
\bibitem{ellis} J. Ellis and S. Rudaz, Phys. Lett. B128(1983)248; J.F. Gunion and H.E. Haber,
             Nucl. Phys. B272(1986)1.
\bibitem{NLC} C. Adolphsen $et$ $al$., (International Study Group Collaboration),
            `International study group progress report on linear collider development',
        SLAC-R-559 and KEK-REPORT-2000-7 (April, 2000).
\bibitem{JLC} N. Akasaka $et$ $al$.,`JLC design study', KEK-REPORT-97-1.
\bibitem{TESLA} R. Brinkmann, K. Flottmann, J. Rossbach, P. Schmuser, N. Walker and H. Weise(editor),
                `TESLA: The superconducting electron positron linear collider with an
                 integrated X-ray laser laboratory. Technical design report, Part
                 2: The Accelerator', DESY-01-11 (March, 2001).
\bibitem{CLIC} `A 3 TeV $e^+e^-$ Linear Collider Based on
                CLIC Technology', G. Guignard(editor), CERN-2000-008.
\bibitem{zerwas} A. Freitas, A. von Manteuffel, P. M. Zerwas, Eur.Phys.J. C34 (2004) 487-512.
\bibitem{huitu} K. Huitu, J. Maalampi, M. Raidal, Phys.Lett. B328 (1994) 60-66.
\bibitem{howard} Howard Baer, B. W. Harris, Mary Hall Reno, Phys.Rev. D57 (1998) 5871-5874.
\bibitem{freitas} A. Freitas, D. J. Miller, eConf C010630 (2001) E3061.
\bibitem{hikasa} K.I. Hikasa and M. Kobayashi, Phys. Rev. D36(1987)724.
\bibitem{been} W. Beenakker, R. Hopker and P.M. Zerwas, Phys. Lett. B349(1995)463;
             A. Arhrib, M. Capdequi-Peyranere and A. Djouadi, Phys. Rev D52(1995)1404;
             H. Eberl, A. Bartl and W. Majeroto, Nucl.Phys. B472(1996)481, hep-ph/9603206.
\bibitem{han} C.H. Chang, L. Han, W.G. Ma and Z.H. Yu, Nucl.Phys. B515 (1998) 15-33.
\bibitem{eberl} K. Kova\v{r}\'{\i}k, C. Weber, H. Eberl, W. Majerotto, Phys.Lett. B591 (2004) 242-254.
\bibitem{hollik} A. Arhrib, W. Hollik, JHEP 0404 (2004) 073.
\bibitem{a1}  I.F. Ginzbyrg, G.L. Kotkin, V.G. Serbo and V.I. Telnov, Pis'ma ZHETF
              34 (1981)514; Nucl. Instr. Methods 205 (1983)47.
\bibitem{a2} Frank Cuypers (MPI Munich), talk presented at the International Europhysics
             Conference on High Energy Physics, Brussels,
             27/7-2/8,1995, MPI-PhT/95-93, hep-ph/9509400;
\bibitem{s14} G.'t Hooft and M. Veltman, Nucl. Phys. {\bf B153}, 365 (1979).
\bibitem{sd} I.F. Ginzburg, G.L. Kotkin, V.G. Serbo and V.I. Telnov, Nucl. Instr. Methods
             205 (1983) 47; L. Han, C.G. Hu, C.S. Li and W.G. Ma, Phys. Rev. D54(1996)2363.
\bibitem{COMS} A. Denner, Fortschr. Phys. {\bf 41}, 307 (1993).
\bibitem{FA3} T. Hahn, Comp. Phys. Commun. {\bf 140}, 418 (2001).
\bibitem{PSS} W.T. Giele and E.W.N. Glover,
              Phys. Rev. {\bf D46}, 1980 (1992);
          W.T. Giele, E.W. Glover and D.A. Kosower,
          Nucl. Phys. {\bf B403}, 633 (1993);
          S. Keller and E. Laenen, Phys. Rev. {\bf D59}, 114004 (1999).
\bibitem{Velt} G.'t Hooft and M. Veltman, Nucl. Phys. {\bf B153}, 365 (1979).
\bibitem{Com1} I. Ginzburg, G. Kotkin, V. Serbo and V. Telnov,
Pizma ZhETF, {\bf 34} (1981) 514; JETP Lett. {\bf 34} (1982) 491.
Preprint INP 81-50, 1981, Novosibirsk.
\bibitem{Com2} I. Ginzburg, G. Kotkin, V. Serbo and V. Telnov,
Nucl. Instr. \& Meth. {\bf 205} (1983) 47, Preprint INP 81-102,
1991, Novosibirsk.
\bibitem{Com3} I. Ginzburg, G. Kotkin, S. Panfil, V. Serbo and V.
Telnov, Nucl. Instr. \& Meth. {\bf 219} (1984) 5.
\bibitem{photon para} K. Cheung, Phys.Rev. {\bf D47}(1993)3750.
\bibitem{rb} R. Blankenbecler and S.D.Drell, Phys. Rev. Lett. 61(1988)2324;
             F. Halzen, C.S. Kim and M.L. Stong, Phys. Lett. B274(1992)489;
             M. Drees and R.M. Godbole, Phys. Lett. 67(1991)1189.
\bibitem{vt} V. Telnov, Nucl. Instr. Methods A294(1990)72.
\bibitem{phospec} V. Telnov, Nucl. Instrum. Methods Phys. Res. {\bf A294}(1990)72;
            L. Ginzburg, G. Kotkin and H. Spiesberger, Fortschr. Phys. {\bf 34}(1986)687.
\bibitem{mssm-2} J. F. Gunion, H. E. Haber, Nucl. Phys. {\bf B272} (1986) 1.
\bibitem{databook} S. Eidelman, {\it et al.,} Phys. Lett. {\bf B592}(2004)1.
\bibitem{leger} F. Jegerlehner, DESY 01-029, hep-ph/0105283.
\bibitem{count1}C. Weber, H. Eberl, W. Majerotto, Phys. Lett. {\bf B572}(2003) 56, hep-ph/0305250.
\bibitem{count2}H. Eberl, M. Kincel, W. Majerotto and Y. Yamada, Nucl. Phys. {\bf B625}(2002) 372, hep-ph/0111303.
\bibitem{FeyArtsFormCalc}Thomas Hahn and Christian Schappacher,Comput.Phys.Commun. 143(2002)54-68, hep-ph/0105349.
\end{thebibliography}
\end{document}